\documentclass[reprint,10pt,prx,aps, amsmath,amssymb,superscriptaddress,floatfix]{revtex4-2}

\usepackage{placeins}
\usepackage{float}
\usepackage{graphicx} 
\graphicspath{ {./images/} }
\usepackage{siunitx}
\usepackage{braket}
\usepackage{xcolor}
\usepackage{hyperref}

\allowdisplaybreaks

\begin{document}
\title{
All-to-all connectivity via adiabatic ZZ interactions in a superconducting flux qubit processor
}
\author{Elena Lupo}
\affiliation{Institute for Quantum Computing Analytics (PGI-12) Forschungszentrum J\"ulich, 52425 J\"ulich, Germany}
\affiliation{School of Mathematics and Physics, University of Surrey, Guildford GU2 7XH, United Kingdom}

\author{Nguyen H. Le}
\affiliation{Qolumbus SAS, 78 Boulevard de la République, 92100 Boulogne-Billancourt, France}

\author{Eran Ginossar}
\affiliation{School of Mathematics and Physics, University of Surrey, Guildford GU2 7XH, United Kingdom}

\author{Michael Stern}
\affiliation{Department of Physics, Technion – Israel Institute of Technology, Haifa 3200003, Israel}

\begin{abstract}
Scalable superconducting quantum processors would benefit from high qubit connectivity, yet existing architectures connect neighbouring qubits via tunable couplers, which introduce substantial wiring and control overhead. 
Here, we propose a baseband-driven  adiabatic ZZ gate that entangles any selected pair in a register of flux qubits coupled via a common cavity. The qubits stay dc-biased at their flux sweet spot. Strong drives applied selectively through local flux lines, at frequencies far below the qubit transitions, adiabatically tilt the quantization axes and generate an effective longitudinal ZZ interaction. The strength of the coupling is tunable and controlled by the amplitudes and relative phase of the drives rather than by any resonance condition.
Numerical simulations give coherent gate errors below $10^{-4}$ over broad ranges of drive frequency, with comparable fidelities for two simultaneous gates in a four-qubit register. The protocol is robust against frequency fluctuations and remains effective even when many qubits are coupled to the cavity.

\end{abstract}

\date{\today}

\maketitle

\section{Introduction}
\begin{figure}[t]
    \centering
    \includegraphics[width=\linewidth]{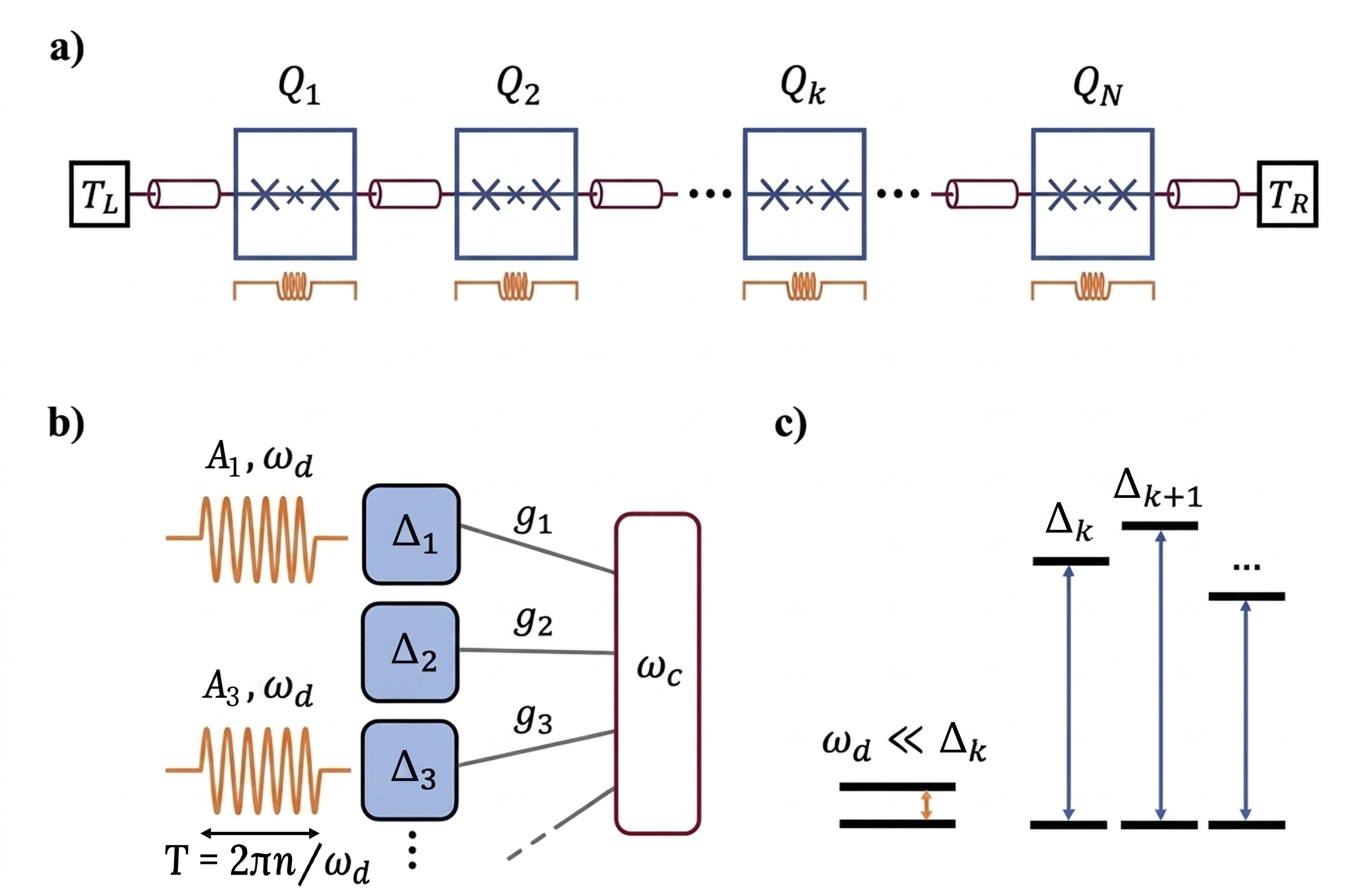}
\caption{\label{fig1_schemes} All-to-all flux qubit architecture and  entangling gate: (a) quantum processor consisting of $N$ gradiometric flux qubits, $Q_1,\dots,Q_N$, galvanically coupled to a common superconducting resonator, each addressed through a local flux line. At the optimal operating point, the qubit–cavity coupling is predominantly transverse, while the gradiometric symmetry suppresses direct magnetic-dipole coupling between qubits. (b) Schematic of the proposed two-qubit gate in a multiqubit system comprising two-level qubits transversely coupled to a common bosonic (cavity) mode. A selected pair is entangled by applying slow drives with matching frequencies $\omega_d$ and phases, and with a commensurate gate duration of $T=2\pi n/\omega_d$,  $n\in\mathbb{N}$; (c)  Involved frequencies'
scales: the qubits operate in the gigahertz range, whereas the
drive frequencies are typically a few hundred megahertz, such
that $\omega_{d} \ll \Delta_{k}$, and the driven qubits approximately follow
their instantaneous eigenstates in an adiabatic process.}
\end{figure}

The connectivity of a quantum processor plays a central role in determining its computational performance and scalability. Highly connected architectures allow distant qubits to interact directly, avoiding costly chains of operations \cite{herbert2020depth} thereby reducing the depth of
compiled circuits and limiting the accumulation of gate
errors and decoherence during execution.  High connectivity is particularly important for recently developed quantum low-density parity-check (QLDPC) codes \cite{gottesmanConstantOverhead2014,breuckmannQuantumLDPCCodes2021,panteleevAsymptoticallyGood2022}, and can substantially reduce the physical-qubit overhead of fault-tolerant quantum computation compared to surface codes \cite{cain2026shor,zhao2026ultra}. A natural route to such connectivity is to couple many qubits to a common bosonic mode, such as a collective motional mode in trapped-ion systems or a shared cavity mode in superconducting- and spin-qubit architectures.
\\\indent
For superconducting architectures, such connectivity is difficult to achieve with transmons because of several limitations.  Their weak anharmonicity leads to frequency crowding and spectral collisions as the processor grows \cite{Koch2007,Schutjens2013}. In addition, virtual transitions through higher levels generate residual parasitic
$ZZ$ interactions between coupled qubits sharing a common cavity \cite{zhaoSuppressionStaticZZ2021,kandalaHighFidelityCNOT2021,magesanEffectiveHamiltonian2020}. This quantum crosstalk has motivated architectures based on tunable couplers \cite{yanTunableCouplingScheme2018a, AruteQuantumSupremacy2019, sungRealizationHighFidelityCZ2021a, jinSuperconductingArchitectureDemonstrating2025}, at the cost of limiting connectivity, additional circuit elements, control lines and calibration requirements.
\\\indent
These limitations and costs could be strongly mitigated with superconducting flux qubit \cite{mooijJosephsonPersistentCurrentQubit1999a, orlandoSuperconductingPersistentcurrentQubit1999a, orgiazziFluxQubitsPlanar2016},  whose anharmonicities are large. Thus, leakage can be suppressed even under strong driving. Moreover, gradiometric designs \cite{paauwTuningGapSuperconducting2009a, schwarzGradiometricFluxQubits2013} suppress the magnetic dipole-dipole interactions leaving only quadrupolar coupling which can be made negligible between sufficiently spaced qubits. 
Operated at their flux sweet spots, where the transition frequency is first-order insensitive to flux noise, they have good coherence properties \cite{bylanderNoiseSpectroscopyDynamical2011, sternFluxQubitsLong2014a, changReproducibilityGapControl2022} and are subject to an approximately transverse qubit–cavity interaction with negligible static $ZZ$ crosstalk. This provides a natural platform for an all-to-all-connected superconducting processor, provided that interactions can be activated selectively and efficiently between arbitrary qubit pairs.
\\\indent
Established microwave-activated gates like FLICFORQ \cite{rigettiProtocolForUniverslGates2005} and cross-resonance \cite{rigettiFullyMicrowaveTunable2010,kirchhoffOptimizedCrossresonanceGate2018}, together with other ac-Stark-shift \cite{ashhabGeneralizedSwitchableCoupling2006, douceCouplingSingleNitrogenvacancy2015, xueTunableInteractionSuperconducting2016, mitchellHardwareEfficientMicrowaveActivatedTunable2021b, weiHamiltonianEngineeringMulticolor2022a} and Floquet-engineered
\cite{songFloquetEngineeringTwo2020,nguyenProgrammableHeisenbergInteractions2024} protocols rely on resonances between the drive and combinations of qubits and cavity transitions, or on operating close to one \cite{xueTunableInteractionSuperconducting2016, mitchellHardwareEfficientMicrowaveActivatedTunable2021b, weiHamiltonianEngineeringMulticolor2022a}.
Their fidelity is consequently limited by the
accuracy of these conditions, and the spectral selectivity they demand tightens crowding constraints as the register grows.
\\\indent
Here, we introduce a possible processor architecture in which multiple flux qubits share a common resonator (see Fig.~\ref{fig1_schemes}a). Within this setting, we propose a baseband-driven adiabatic ZZ gate, which entangles any selected pair of qubits.
The qubits remain DC-biased at their flux sweet spots, while strong local baseband drives, applied through the flux lines at frequencies far below the qubit transitions (Fig.~\ref{fig1_schemes}b--c), adiabatically tilt their quantization axes. In the resulting Floquet frame, the transverse qubit--cavity coupling acquires a longitudinal component that mediates an effective $ZZ$ interaction through the common cavity. Unlike resonantly activated or parametrically engineered gates, the scheme requires neither a resonance with a qubit or cavity transition nor tuning the static qubit frequencies away from their optimal operating points. Pair selectivity is achieved by driving the qubits to be entangled at a matching modulation frequency, while the other qubits remain undriven or are driven at different frequencies. The use of a shared cavity bus, low-frequency adiabatic control and frequency-selective activation provides a route to highly connected superconducting processors.

\section{Adiabatic ZZ interaction}
\label{sec:floquet}
We model the flux qubits as effective two-level systems transversely coupled to a single cavity mode, as shown schematically in Fig.~\ref{fig1_schemes}(b). Using the convention $\hbar=1$ with all energies expressed in angular-frequency units, the Hamiltonian of $N$ flux qubits transversely coupled to a common cavity and locally driven by a transverse time-dependent pulse is
\begin{align}
\label{HamSysNQ}
    H(t) = &{}\ \sum_{k=1}^N H_{k}(t) + \omega_c a^\dagger a + \sum_{k=1}^N g_{k}(a+a^\dagger)\sigma_k^x,
\end{align}
where
\begin{align}
\label{eq_Hk_V}
    H_k(t) &{}= \frac{\Delta_k}{2}\sigma_k^z + \frac{\epsilon_k(t)}{2}\sigma_k^x,
\end{align}
$\Delta_k$ denotes the transition frequency of qubit $k$ at its flux sweet spot, and $\epsilon_k(t)$ is the applied drive of the form 
\begin{equation}
\epsilon_{k}(t) = A_k\sin{\left(\omega_{k}t+\xi_k\right)},
\end{equation}
with $\omega_k$ and $\xi_k$ the drive frequency and phase, respectively. Qubits to be entangled are driven at a common frequency $\omega_j=\omega_k\equiv\omega_d$.
Because $H_k(t)$ is periodic with period $\tau_k=2\pi/\omega_{k}$,  its time-evolution can be cast in Floquet form \cite{Shirley1965,Sambe1973,GandonFloquet2022}. 

\subsection{Floquet-adiabatic frame}
The $\sigma_x$ drive tilts the axis of each qubit in the XZ plane of the Bloch sphere by an angle $\varphi_k(t)=\mathrm{arctan}[\epsilon_k(t)/\Delta_k]$. The instantaneous eigenstates of $H_k(t)$ thus follow a rotation around the Y axis by $\varphi_k(t)$, described by the operator
\begin{equation}
    R^y_k(t) = \exp{\left\{-i\frac{\varphi_k(t)}{2}\sigma_k^y\right\}}.
\end{equation}
Explicitly, these eigenstates are
\begin{align}
\label{eq_adiabatic_states}
    \ket{+}_k\!(t) =  R^y_k(t)\ket{0}_k,\quad
    \ket{-}_k\!(t) = R^y_k(t)\ket{1}_k,
\end{align}
with eigenvalues $ \pm  \eta_k(t)/2$, where
\begin{align}
    \eta_k(t) = \sqrt{\Delta_k^2 + \epsilon_k^2(t)},
\end{align} 
is the instantaneous energy gap.

In the adiabatic regime of slow drive, where the nonadiabatic transition matrix element satisfies
\begin{align}
    &{}\max_{t}\left|\frac{\bra{+}_k\!(t)\partial_tH_{k}(t)\ket{-}_k\!(t)}{\eta_k(t)^2  }\right| 
    = \frac{ \omega_{k}A_k }{ 2\,\Delta_k^2} \ll 1,
\end{align}
or 
$ \omega_{k} \ll 2(\Delta_k^2/A_k),
$
the trajectory of a qubit initially prepared in an instantaneous eigenstate approximately follows the corresponding instantaneous eigenstate while accumulating a dynamical phase. This evolution is described by the propagator
\begin{align}
\label{eq_prop_1Q_before_Floquet}
    &{}U_{k}(t) = R^y_k{}(t)\,\exp{\bigg\{-i\frac{\gamma_k(t)}{2} \sigma_k^z\bigg\}}\,R^y_k(0)^{\dagger},
\end{align}
where $\gamma_k(t) = \int_0^t \eta_k(t')dt'$ is the accumulated dynamical phase. Since $\eta_k(t)$ is periodic in time with period $\tau_k$, the dynamical phase satisfies $\gamma_k(n\tau_k)=n\,\gamma_k(\tau_k)$ after an integer number of drive periods. As the driven qubit traverses a single line in the XZ plane with zero area, there is no contribution from the geometric phase.

To identify the quasienergy and micromotion that determine the effective interaction and the sideband resonances discussed below, we decompose the accumulated phase as
\begin{equation}
\gamma_k(t)=\nu_k\, t + \beta_k(t).
\end{equation}
where
\begin{equation}
\nu_k=\frac{\gamma_k(\tau_k)}{\tau_k}=\frac{1}{\tau_k}\int_0^{\tau_k} dt \sqrt{\Delta_k^2+A_k^2\sin^2(\omega_{k} t +\xi_k)\,},
\end{equation}
is the average phase-accumulation rate, independent of $\xi_k$, and $\beta_k(t)$ is a periodic residual satisfying $\beta_k(n\tau_k)=0$.
The adiabatic unitary evolution operator in Eq.~\eqref{eq_prop_1Q_before_Floquet} can now be expressed as
\begin{align}
\label{eq_eff_V}
    U_{k}(t)&= P_k(t)\,\exp\{-iH_k^F t\}\,P_k(0)^{\dagger},
\end{align}
where 
\begin{equation}
\label{eq_U1Q_floquet_frame}
    P_k(t) = R_k^y(t)\exp{\left\{ -i\frac{\beta_k(t)}{2}\sigma_k^z\right\}},
\end{equation}
is the periodic micromotion operator, and
 \begin{equation}
 H_k^F= \frac{\nu_k}{2}\sigma_k^z.
 \end{equation} 
the Floquet Hamiltonian. The quasienergies are given by the eigenenergies of  $H_k^F$, which are $\pm \nu_k/2$, and the corresponding Floquet states are eigenstates of $H_k^F$ evolved by $P_k(t)$, i.e., $P_k(t)\ket{0}_k$ and $P_k(t)\ket{1}_k$.
 
In the frame of the Floquet states, defined by $P_k(t)$, the single qubit Hamiltonian $H_k(t)$ becomes the effective Floquet Hamiltonian, as
\begin{align}
\label{eq_Z_floquet_dressing}
\tilde{H}_{k}(t) &{}= P_{k}^{\dagger}(t)H_{k}(t)P_{k}(t) - i\,P_{k}^{\dagger}(t)\,\partial_t P_{k}(t) \nonumber\\
&{}\simeq \frac{\eta_k(t)}{2}\sigma_k^z - \frac{\dot{\beta}_k(t)}{2}\sigma_k^z = \frac{\nu_k}{2} \sigma_k^z,
\end{align}
where the last step follows from $\eta_k(t)=\dot{\gamma}_k(t)=\nu_k+\dot{\beta}_k(t)$, and a $-\dot{\varphi}_k(t)\sigma_k^y/2$  term is dropped in the adiabatic approximation. The error arising from this  non-adiabatic term can be corrected with small local single-qubit $X$ rotations (Appendix \ref{app:nonad}).

In this Floquet frame, the single-qubit $\sigma_k^x$ operator gets transformed into
\begin{align}
\label{eq_X_floquet_dressing}
    &{}P_{k}^{\dagger}(t)\sigma_k^xP_{k}(t)=\frac{\epsilon_k(t)}{\eta_k(t)} \sigma_k^z + \frac{\Delta_k}{\eta_k(t)} \sigma_k^x(t),
\end{align}
with 
\begin{align}
\label{eq_sx_rotated}
    \sigma_k^x(t) = \cos[\beta_k(t)]\sigma_k^x - \sin[\beta_k(t)]\sigma_k^y.
\end{align}
It is shown below that the $\sigma_k^z$ component of the transformed $\sigma_k^x$  leads to the effective longitudinal qubit-qubit coupling.

\subsection{Effective qubit-qubit interaction}
\label{sec:effective_qubit_qubit_interaction}
By treating the Hamiltonian (\ref{HamSysNQ}) in the adiabatic frame of the uncoupled qubits, defined by the transformation $P(t) = \prod_{k=1}^N P_k(t)$, the transverse qubit-cavity coupling terms acquire a longitudinal component:
\begin{align}
\label{HamSysNQ_frame_ad_SM}
\tilde{H}(t) \simeq \sum_k\frac{\nu_k}{2}\sigma_k^z +  \omega_c a^\dagger a + V^{\perp}(t) \, + \, V^{\parallel}(t),
\end{align}
with
\begin{align}
\label{eq_Vper_Vpar}
    &{}V^{\perp}(t) = \sum_{k=1}^N \frac{g_{k}\Delta_k}{\eta_k(t)}(a+a^\dagger)\sigma_k^x(t),\nonumber\\
    &{}V^{\parallel}(t) = \sum_{k=1}^N \frac{g_{k}\epsilon_k(t)}{\eta_k(t)}(a+a^\dagger)\sigma_k^z,
\end{align}
with $\sigma_k^x(t)$ given in Eq.~(\ref{eq_sx_rotated}). To obtain an effective ZZ interaction between the driven qubits, the longitudinal qubit–cavity coupling is removed by applying a conditional displacement transformation to the cavity \cite{LangFirsovKineticTheory1963,billangeon_qed_2015},
\begin{align}
\label{D_alpha_adiabatic_frame}
    D[\alpha(t)] =&{}\ e^{\alpha(t) (a^\dagger - a\,)}, \
\alpha(t)=-\sum_{k=1}^N \frac{g_{k}\epsilon_k(t)}{\omega_c\eta_k(t)}\sigma_k^z.
\end{align}
The time dependence of this transformation generates the additional transport term $-iD^\dagger\dot{D}$, which is negligible in the slow-driving regime. Within this 2nd adiabatic approximation, we obtain 
\begin{align}
\label{HamNQ_Displaced}
&{}D^{\dagger}[\alpha(t)]\tilde{H}(t) D[\alpha(t)] \simeq  \sum_k\frac{\nu_k}{2}\sigma_k^z + \omega_c a^\dagger a \nonumber\\
&{} + D^{\dagger}[\alpha(t)]V^{\perp}(t) D[\alpha(t)] +V_{\text{eff}}^{zz}(t),
\end{align}
with the last term representing a non-perturbative effective longitudinal coupling, 
\begin{align}
\label{g_ZZ_eff_NQ}
    &{}V_{\text{eff}}^{zz}(t)=-\sum_{j<k}J_{jk}^{zz}(t)\sigma_j^{z}\sigma_k^{z},\nonumber \\
   &{}J_{jk}^{zz}(t)= \frac{2g_{j}g_{k}}{\omega_c} \frac{\epsilon_j(t)\epsilon_k(t)}{\eta_j(t)\eta_k(t)},
\end{align}
whose strength depends on the drive pulses only and does not rely on resonant transitions. Small dispersive shifts in the quasienergies $\nu_k$ due to coupling with the cavity are neglected (Appendix~\ref{app:transverse}).

For pulses that vanish at the beginning and the end, the displacement and the adiabatic transformations reduce to identities at the end of the gate, eliminating residual
qubit–cavity entanglement and additional qubit rotations. We end up with the pairwise, accumulated longitudinal phase 
\begin{equation}
\label{phaseZZ_jk}
    \theta_{jk}(T) = \int_0^{T} J_{jk}^{zz}(t)\,dt,
\end{equation} 
with $T$ the pulse length. Pair selectivity follows from the Fourier expansion
\begin{align}
\label{w_i_fourier_exp}
\frac{\epsilon_k(t)}{\eta_k(t)}
={}&\kappa_k\sum_{m=0}^{\infty}C_k^{(m)}
\sin[(2m+1)(\omega_{k}t+\xi_k)], 
\end{align}
with $\kappa_k = A_k/\sqrt{\Delta_k^2+A_k^2}$, and where the coefficients are given by
\begin{align}
C_k^{(m)} ={}& \sum_{l= m}^{\infty}\frac{2m+1}{l+m+1}\binom{2l}{l}\binom{2l}{l-m}\left(\frac{\kappa_k}{4}\right)^{2l}\nonumber\\
 \simeq{}& \binom{2m}{m}\left(\frac{\kappa_k}{4}\right)^{2m},
  \label{eq:Cn}
\end{align}
with the last expression being the leading ($l=m$) term. The series converges for arbitrary $A_k/\Delta_k$, with  coefficients decreasing exponentially with $m$. For $\omega_{j} = \omega_{k} = \omega_d$ and $T = 2\pi n / \omega_d $, $n \in \mathbb{N}$, the accumulated phase can be obtained in terms of the Fourier coefficients as
\begin{align}
\label{g_ZZ_ij_matching_freqs}
\theta_{jk}(T) {}&\simeq
\frac{g_{j}g_{k}\kappa_j\kappa_kT}{\omega_c} \nonumber\\
{}&\quad\cdot\sum_{m=0}^{\infty}C_j^{(m)}C_k^{(m)}\cos\!\big[(2m+1)(\xi_j-\xi_k)\big],
\end{align}
while $\theta_{jk}(T)$ vanishes for $\omega_{j}\neq \omega_{k}$ and $T$ a common multiple of both the drive periods, provided that no odd harmonics of the two drives coincide, i.e. $(2m+1)\omega_j \neq (2n+1)\omega_k$ for all $m$, $n$. This excludes frequency ratios $\omega_j/\omega_k$ that are ratios of odd integers. Since the Fourier coefficients decay as $(\kappa/4)^{2m}$, in practice only low-order ratios such as $3$:$1$ must be avoided (see green squares in Fig.~\ref{fig4_parZZ}). Therefore, pair selection  is enabled by tuning the drive frequencies. Unlike resonance-based addressing, the protocol does not require assigning a distinct qubit transition to each gate channel; additional qubits introduce only discrete avoidance conditions in modulation-frequency space.
The relative drive phase provides additional control of the interaction. For simplicity, we use constant-amplitude drives and set $\xi_j = \xi_k=0$ so that the pulses vanish at $t=0$ and $t=T$, ensuring maximal adiabaticity when switching on and off the drive.

To obtain a high-fidelity ZZ entangling gate, the effect of the transverse interaction needs to be suppressed. A Schrieffer--Wolff  transformation of the transverse term $D^{\dagger}[\alpha(t)]V^{\perp}(t) D[\alpha(t)]$ in the Hamiltonian Eq.~\eqref{HamNQ_Displaced} 
yields a residual  flip-flop interaction between qubits, 
\begin{equation}
\label{V_XX_residual}
V_{\text{eff}}^{\perp}(t) \approx \sum_{j<k}\tilde{J}^{\perp}_{jk}(t)\sigma_j^+\sigma_k^- + \text{h.c.}
\end{equation}
where the modulated coupling coefficient $\tilde{J}^{\perp}_{jk}(t)$ is periodic when $\omega_{j} = \omega_{k} = \omega_d$ and carries harmonics at even multiple of the drive frequency (Appendix~\ref{app:sidebands}), with  Fourier expansion 
\begin{align}
\label{g_XX_eff_fourier}
 \tilde{J}^{\perp}_{jk}(t)=\sum_{m=-\infty}^{\infty}\left[\tilde{J}_{jk}^{\perp}\right]_m e^{i2m \omega_d t}.
\end{align}
In the interaction picture, this effective interaction becomes 
\begin{equation} \sum_{j<k}\sum_{m=-\infty}^{\infty}\left[\tilde{J}_{jk}^{\perp}\right]_m e^{i(2m \omega_d +\nu_j-\nu_k)t}\sigma_j^+\sigma_k^- + \text{h.c.}
\end{equation}
Therefore, sideband transitions   occur when
\begin{equation}
\label{condition_to_avoid_sidebands}
   \nu_k - \nu_j=2m\omega_d, \quad m\in \mathbb{Z},
\end{equation}
with $\nu_i[\,\epsilon_i\to 0\,]\to\Delta_i$ in the case one of the qubits ($i=k$ or $i=j$) is not driven. The effect of this modulated interaction can be neglected by choosing qubit driving frequencies far from these resonances. 

\section{Two-qubit gate performance}
\label{sec_Two_qubit_gate_performance}
\subsection{Entangling gate}
\label{sec_subsec_entangling_gate_2Q}
We now numerically simulate the achievable fidelity for the maximally entangling two-qubit gate,
\begin{equation}
\label{eq:target}
G_{ZZ}=\exp\left(i\frac{\pi}{4}\sigma_1^z\sigma_2^z\right),
\end{equation}
in a system with two qubits coupled via a cavity ($N=2$). The adiabatic approximation described above allows  decomposing the two-qubit propagator  into local rotations, a
conditional displacement, and the $ZZ$ rotation:
\begin{align}
\label{effective_propagator_2Q}
  \!  U_{\text{ad}}(t)\simeq\hspace{-3pt} \Big(\prod_{k=1,2}\hspace{-4pt} P_k(t) e^{-i(\nu_kt/2)\sigma_k^z}\Big)D{[\alpha(t)]}e^{i\theta(t)\sigma_1^z\sigma_2^z},
\end{align}
where $\theta(t) \equiv \theta_{12}(t)$  defined in Eq.~\eqref{phaseZZ_jk}. Since $P_k(t)$ and $D{[\alpha(t)]}$ are periodic in time with period  $2\pi/\omega_{k}$, choosing $\omega_{1} = \omega_{2} = 2\pi n/T$, with $n$ an integer, returns these operators to the identities at the final time, resulting in a two-qubit $ZZ$ rotation up to local $Z$ rotations. In addition, it is  shown in Appendix \ref{app:nonad} that the leading effect of the non-adiabatic term are small single-qubit $X$ rotations at $t=0$ and $t=T$. The actual unitary dynamics is
\begin{equation}
    U(t)=R^x_1R_2^xU_{\text{ad}}(t){R^x_1}^{\dagger}{R^x_2}^{\dagger},
\end{equation}
where
\begin{equation}\label{eq:Rx}
  R_k^x=\exp\left(-i\frac{\omega_d A_k}{2\Delta_k^2}\sigma_k^x\right), \quad k=1,2,
\end{equation}
which must be accounted for when infidelity $\sim 10^{-4}$ is required.

From the exact numerical propagator $U_{\mathrm{ex}}(T)$ obtained by propagating $H(t)$ in Eq.~\eqref{HamSysNQ} for $N=2$, the fidelity is evaluated after projection onto the cavity vacuum, correction of the local X rotations, and optimization over final local $Z$ rotations,
\begin{align}
\label{fidelityZZ}
F_{ZZ} = \left|\frac{1}{4} \mathrm{Tr}\left\{ R_1^zR_2^z{R_1^x}^{\dagger}{R_2^x}^{\dagger}\Pi^{(0)}_{c}U_{\mathrm{ex}}(T)\Pi^{(0)}_cR_1^xR_2^xG_{ZZ}^\dagger\right\}\right|^2,
\end{align}
where $\Pi^{(0)}_c=I_1\otimes I_2\otimes\ket{0}\!\bra{0}$ is the projector to the zero-photon subspace. 

\begin{figure}[!t]
    \centering
\includegraphics[width=0.85\linewidth]{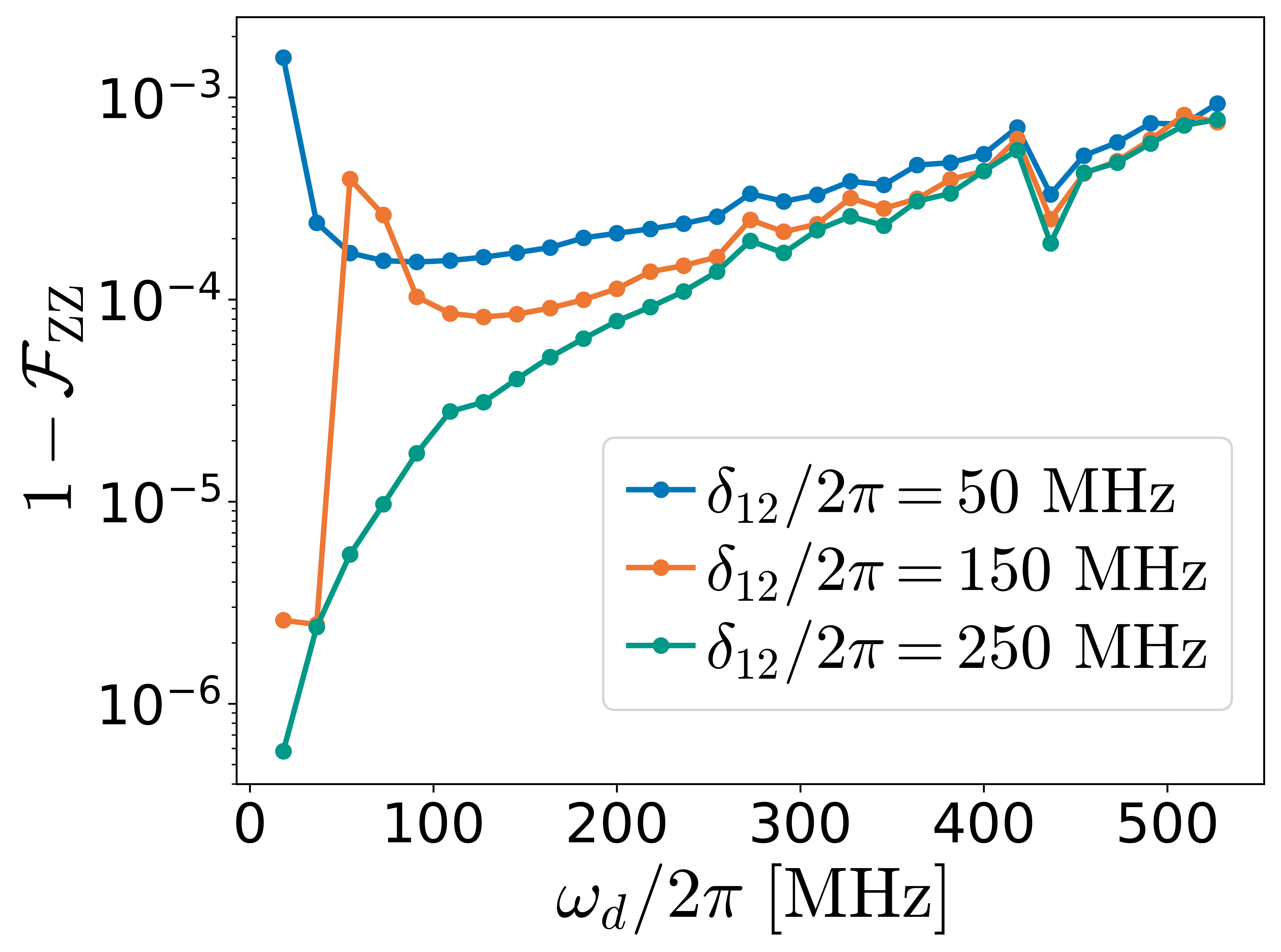}
    \caption{\label{fig2} Adiabatic ZZ entangling gate for $N=2$. The gate error $1 - F_{ZZ}$ is shown as a function of the common drive frequency $\omega_{1}=\omega_{2} = \omega_d = 2\pi n/T$, for $\Delta_1/2\pi= 7.00\ \text{GHz}$ and varying the second qubit frequency $\Delta_2= \Delta_1 + \delta_{12}$ with detuning $\delta_{12}/2\pi = 50,\ 150,\ 250\ \text{MHz}$.  
    The pulse duration is chosen to be $T=55\ \text{ns}$, while the pulse amplitudes and the correcting single-qubit rotations are optimised to maximise the gate fidelity. Other parameters are $\omega_c/2\pi=3.0\ \mathrm{GHz}$ and $g_{1}/2\pi=g_{2}/2\pi=100\ \mathrm{MHz}$.}
\end{figure}

Figure~\ref{fig2} shows the gate error at three qubit detunings, $\delta_{12}/2\pi\equiv(\Delta_2-\Delta_1)/2\pi = 50,\,150,\,250\ \mathrm{MHz}$. For all three, there are broad operating windows with errors $\lesssim 10^{-4}$, reaching $\sim10^{-6}$ at the lowest drive frequencies for $\delta_{12}/2\pi=250\ \mathrm{MHz}$. The slow increase of the error with $\omega_d$ arises from higher-order non-adiabatic terms, which grow with the drive frequency. The peaks appearing for some of the frequencies are sideband resonances of the effective transverse interaction discussed in Sec.~\ref{sec:effective_qubit_qubit_interaction}, which degrade the gate fidelity. Since these resonances are discrete, they are readily avoided by a suitable choice of drive frequency. The overall increase of gate infidelity with decreasing detuning can be attributed to hybridization of the two qubits via a static cavity-mediated XX interaction, an effect explained in more detail in Sec.~\ref{sec:spectators} below.

Identical drives with $A_1=A_2=A$ are used for the numerical simulation in Fig.~\ref{fig2}. The amplitude $A$ fixes the $R_k^x$ rotations in Eq.~\eqref{eq:Rx}. This amplitude and the Z rotations $R_1^z$ and $R_2^z$ are then optimized to minimize the infidelity. For a given duration $T$ the numerically optimized amplitude $A$ is in close agreement with the analytical prediction from Eqs.~\eqref{g_ZZ_eff_NQ} and \eqref{phaseZZ_jk} (Appendix~\ref{app:optamp}), confirming the validity of the adiabatic approximations in the derivation of the ZZ interaction.

\subsection{Robustness to qubit-frequency fluctuations}
The longitudinal nature of the effective entangling operation, Eq.~(\ref{g_ZZ_eff_NQ}), yields an advantage when fluctuations affect one of the targeted qubit frequency.
Since the single-qubit term $\Delta_k\sigma_k^z/2$ and the induced $ZZ$ interaction commute, a frequency fluctuation in $\Delta_k$ produces a  single-qubit  phase error and  a  two-qubit error through its influence on the $ZZ$ interaction strength. 
\\\indent
Starting from the $ZZ$ entangling phase in Eq.~\eqref{g_ZZ_ij_matching_freqs} for the pair $j=1$ and $k=2$, and keeping only the maximal $m=0$ contribution in the Fourier sum,
\begin{align}
\theta\simeq{}&
\frac{g_{1}g_{2}\kappa_1\kappa_2T}{\omega_c}.
\end{align}
The duration $T$ is chosen such that $\theta=\pi/4$ to realize the target $G_{ZZ}$ in Eq.~\eqref{eq:target}. A frequency fluctuation on the second qubit $\Delta_2\to\Delta_2+\delta\Delta_2$ modifies this entangling phase as
\begin{align}
\frac{\delta \theta}{\theta} \simeq -\left(1-\kappa_2^2\right)\frac{\delta\Delta_2}{\Delta_2},
\end{align}
neglecting the weaker dependence of $C_2^{(0)}$ on $\Delta_2$, so that the actual unitary, up to the single qubit phase error, is
\begin{align}
{\cal U}_{ZZ} =&{}\ \exp\left\{i(\theta+\delta \theta)\,\sigma_1^z \sigma_2^z\right\} \nonumber\\
=&{}\ G_{ZZ}\, \exp\left\{-i\frac{\pi}{4}\left(1-\kappa_2^2\right)\frac{\delta\Delta_2}{\Delta_2} \sigma_1^z \sigma_2^z\right\}.
\end{align}
Therefore, the  two-qubit error is 
\begin{equation}
\label{fid_ZZ_freq_fluctuation}
1-\left\vert\frac{\mathrm{tr}\big\{G_{ZZ}^{\dagger}{\cal U}_{ZZ}\big\}}{4}\right\vert^2
\lesssim \left(\frac{\pi}{4}\right)^2\left(\frac{\delta\Delta_2}{\Delta_2}\right)^2,
\end{equation}
which is extremely small because the fluctuation is normalized by the gigahertz qubit splitting. This is important for fault-tolerant quantum computation, as while independent single-qubit errors can be efficiently suppressed, correlated two-qubit errors typically degrade its performance and may require decoding strategies tailored to their correlation structure \cite{Wang2026CorrelatedErrors,Nickerson2019CorrelatedNoise}.
\\\indent
This contrasts with other resonantly activated interactions such as the cross-resonance gate. Here, the unitary under fluctuation of the second qubit frequency, $\Delta_2\rightarrow\Delta_2+\delta\Delta_2$, is given by (Appendix \ref{app:robustness})
\begin{align}
\label{Ucr_t}
&{}U_\text{CR}(T_{\text{CR}})\simeq\exp\left\{{-i\left(\frac{\delta\Delta_2 T_{\text{CR}}}{2}\sigma_2^z + J_{\text{CR}}  T_{\text{CR}}\,\hat\sigma_1^z\sigma_2^x \,\right)}\right\},
\end{align}
where $\hat\sigma_1^z$ is a tilted operator in the XZ plane. Although this frequency fluctuation does not affect the interaction strength $J_{\mathrm{CR}}$, the non-commutativity of the single-qubit $\sigma_2^z$ error term and the CR interaction $\hat\sigma_1^z\sigma_2^x$ leads to an effective $ZY$ spurious term. At $T_{\text{CR}}=\pi/4J_{\mathrm{CR}}$, for example, the unitary operator is
\begin{align}
\label{Ucr_gate_with_fluctuation}
{\cal U}_{\text{CR}}\approx&{}\ 
G_{\text{CR}}
\exp\bigg\{-i\frac{\delta\Delta_2}{4J_{\text{CR}}}\hat\sigma_1^z\sigma_2^y\bigg\},
\end{align}
up to a single-qubit phase error,
where $G_{\mathrm{CR}}=\exp\{-i(\pi/4)\hat\sigma_1^z\sigma_2^x\}$ is the ideal gate.
This yields a two-qubit error scaling as
\begin{equation}
    1-\left\vert\frac{\mathrm{tr}\big\{G_{\mathrm{CR}}^{\dagger}{\cal U}_{\mathrm{CR}}\big\}}{4}\right\vert^2\simeq\left(
\frac{\delta\Delta_2}{4J_{\mathrm{CR}}}\right)^2,
\end{equation}
which is several orders of magnitude higher than the two-qubit error of the ZZ gate, because typically $J_{\text{CR}}\ll\Delta_2$.
For representative values $J_{\text{CR}}/2\pi=10\ \mathrm{MHz}$ and $\Delta_2/2\pi=7\ \mathrm{GHz}$, the frequency-noise-induced correlated error is approximately $5\times10^4$ smaller for the adiabatic $ZZ$ gate, assuming the same frequency fluctuation. This robustness against frequency fluctuations can also provide effective protection against \(1/f\) frequency noise, which is dominated by low-frequency components and can therefore be treated as quasi-static over the relevant timescale \cite{georgeMinimalTimeRobust2025}. 

\section{Multiqubit system}

\subsection{Effect of spectators}
\label{sec:spectators}
\begin{figure}[t]
    \centering
    \includegraphics[width=\linewidth]{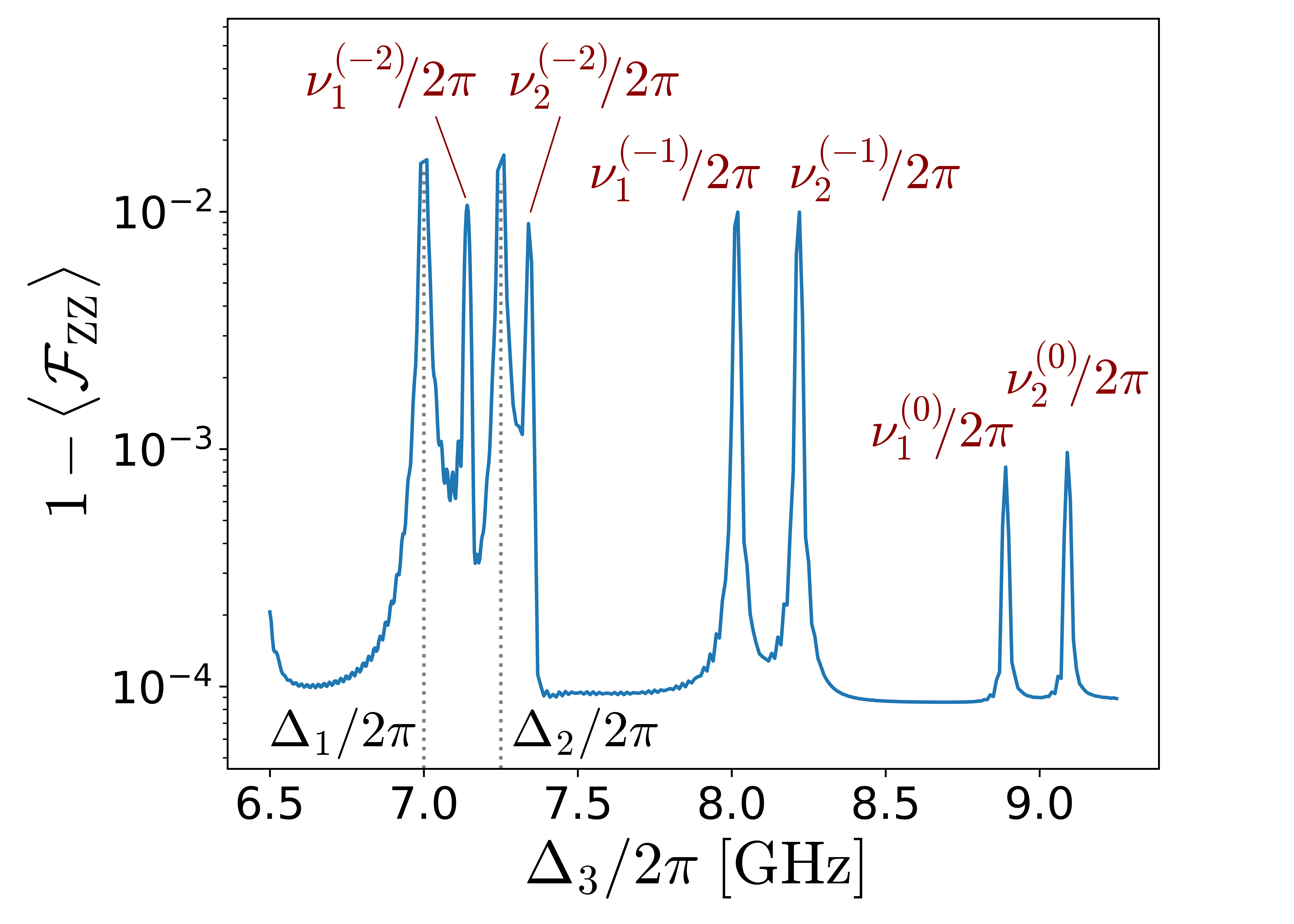}
    \caption{\label{fig_3Q}Optimized average infidelity $1-\braket{\mathcal{F}_{ZZ}}$ for the entangling $G_{ZZ}^{12}$ in a three-qubit register as a function of the spectator qubit frequency $\Delta_3$. Here $\Delta_1/2\pi=7.00\ \mathrm{GHz}$, $(\Delta_2-\Delta_1)/2\pi = 250\ \text{MHz}$, $T=55\ \mathrm{ns}$ and $\omega_d^{12}/2\pi=436.4\ \mathrm{MHz}$. The black dotted lines indicate hybridization at the bare qubit frequencies $\Delta_{1,2}$ via the cavity-mediated flip-flop coupling, while the labelled peaks are the sideband resonances with the  quasienergies $\nu_k^{(m)} = \nu_k + 2m\omega_d$, with $k=1,2$ and $m\in \mathbb{Z}$.  
    Other parameters are $\omega_c/2\pi=3.0\ \mathrm{GHz}$ and $g_{k}/2\pi=100\ \mathrm{MHz}$.}
\end{figure}

We now extend the entangling gate analysis to registers with more than two qubits,
and first assess how an undriven spectator qubit affects the
entangling operation. 
Consider a three-qubit register in which
qubits 1 and 2 are driven at a common frequency $\omega_d$ to
implement an entangling $G_{ZZ}^{12}$ operation, while qubit 3 shares the cavity but is left undriven, $\epsilon_3(t)=0$. The interference mechanism of Eq.~(\ref{g_ZZ_eff_NQ}) guarantees that the effective longitudinal couplings $J^{zz}_{13}$ and $J^{zz}_{23}$ vanish. The residual couplings to the spectator are therefore realised by the cavity-mediated transverse interaction of Eq.~(\ref{V_XX_residual}).
\\\indent
To quantify their impact the dynamics induced by the full Hamiltonian, Eq.~(\ref{HamSysNQ}), for $N=3$, is simulated. Figure~\ref{fig_3Q} shows the average gate infidelity of the target operation, 
$$G_{3Q} = G_{ZZ}^{12}\otimes \mathbb{I}_3,$$ 
up to local correcting $Z$ rotations, as a function of the spectator frequency $\Delta_3$. Similarly to the case $N=2$, correcting single qubit gate rotations are determined by projecting the total propagator $U(t)$ onto the $0$-photon subspace. For a $2$-qubit system, the average fidelity over $SU(4)$ can be approximated by its average over the $2$-qubit Clifford states. For the $3$-qubit case, we heuristically extend the method by averaging each $2$-qubit fidelity over the axial states of the single qubit's Bloch sphere, i.e. over the eigenstates of the Pauli matrices. This should allow us to have an estimate of the gate fidelity over $SU(4)\otimes SU(2)$. We show the result for $\Delta_1/2\pi=7.00\ \mathrm{GHz}$, $\delta_{12}/2\pi=250\ \mathrm{MHz}$, and the same gate duration and drive frequency as in the two-qubit analysis, with $\Delta_3/2\pi$ swept over a $3\ \mathrm{GHz}$ window enclosing the driven pair. 

There are wide regions where the infidelity remains below $10^{-4}$, comparable to the spectator-free result of Fig.~\ref{fig2} at similar parameters, confirming that an idle qubit sharing the cavity does not degrade the gate under generic conditions. However, there are sharp peaks where the fidelity decreases substantially. These peaks are caused by two different types of resonance between the spectator qubit and the driven qubits. 
\\\indent

The first type of resonance is due to the transverse interaction between the spectator and a driven qubit. As discussed in
Sec.~\ref{sec:floquet}, the coupling coefficient between the spectator qubit and the driven qubit $k$ carries
harmonics at even multiples of the drive frequency, Eq.~(\ref{g_XX_eff_fourier}), so the
cavity-mediated flip-flop term
$\sigma_k^+\sigma_3^-+\mathrm{h.c.}$ becomes resonant when the
spectator frequency matches a sideband of the Floquet quasienergies,
\begin{equation}
  \Delta_3 =  \nu_k + 2 m\, \omega_d,
  \qquad k=1,2, \quad m \in \mathbb{Z},
  \label{eq:spectator_sidebands}
\end{equation}
which are the labelled peaks in Fig.~\ref{fig_3Q}. The peak widths are set by
the corresponding sideband coupling strengths, i.e.  the Fourier coefficients
$(\tilde{J}^{\perp}_{k3})_m$ in Eq.~\eqref{g_XX_eff_fourier}.
The substantial displacement of the $\nu_1^{(0)},\,\nu_2^{(0)}$ peaks from the bare frequencies $\Delta_{1}$, $\Delta_{2}$ reflects the strong-drive renormalization of the quasienergies.
\\\indent
The second type of resonance appears at the bare frequencies $\Delta_{1},\Delta_{2}$
(black dotted lines in Fig.~\ref{fig_3Q}). The computational basis is the dressed
basis of the undriven Hamiltonian, which contains the always-on cavity-mediated
flip-flop coupling $J^{\perp}_{k3}$ (Appendix~\ref{app:dress}), with
$J^{\perp}_{k3}/2\pi\simeq1.5$~MHz at $\Delta_3\simeq\Delta_k$, Eq.~\eqref{Jjk_supp}.
When $\Delta_3$ approaches $\Delta_k$ ($k=1,2$), the dressed single-excitation states
hybridize into $|\Psi_\pm\rangle=(|1_k0_3\rangle\pm|0_k1_3\rangle)/\sqrt{2}$, split by
$2J^{\perp}_{k3}$. These resonances are therefore set by the bare detuning
$\Delta_k-\Delta_3$, up to small Lamb shifts, rather than by the quasienergies. The large infidelity originates from the near-degeneracy of this hybridized doublet. 
During the gate, the drive shifts the splitting of qubit $k$ from $\Delta_k$ towards
$\nu_k$, while the spectator is unaffected. Within the doublet this acts as a
detuning term $[(\nu_k-\Delta_k)/2]\,\sigma_k^z$, far exceeding $2J^{\perp}_{k3}$. This term is purely
off-diagonal between the hybridized eigenstates, because
$\langle \Psi_\pm|\sigma_k^z|\Psi_\pm\rangle=0$, whereas 
$\sigma_k^z|\Psi_\pm\rangle=-|\Psi_\mp\rangle$.  The drive therefore directly interconverts the
two computational states. Any pulse with short duration compared with the
flip-flop timescale $\pi/J^{\perp}_{k3}\simeq 340$~ns --- as is 
the case here because $T=55$~ns --- generates strong Rabi flopping within
the doublet that
deviates from the $ZZ$ dynamics.  The error  
falls off as $[J^{\perp}_{k3}/(\Delta_k-\Delta_3)]^2$ with the hybridization
coefficient, which sets the width of these peaks. 

Between these peaks, the spectator has no significant effect on the
gate, and it is itself left unperturbed apart from a dispersive
local $Z$ phase, correctable by a local $Z$ rotation. In addition, the sideband peaks can be shifted by varying the amplitude and the frequency of the drives. Thus, adding spectators to the bus  merely excludes a
discrete, narrow comb of frequencies that can be avoided by tuning the drive parameters.

\subsection{Parallel entangling gates}
\begin{figure}[t]
    \centering
    \includegraphics[width=\linewidth]{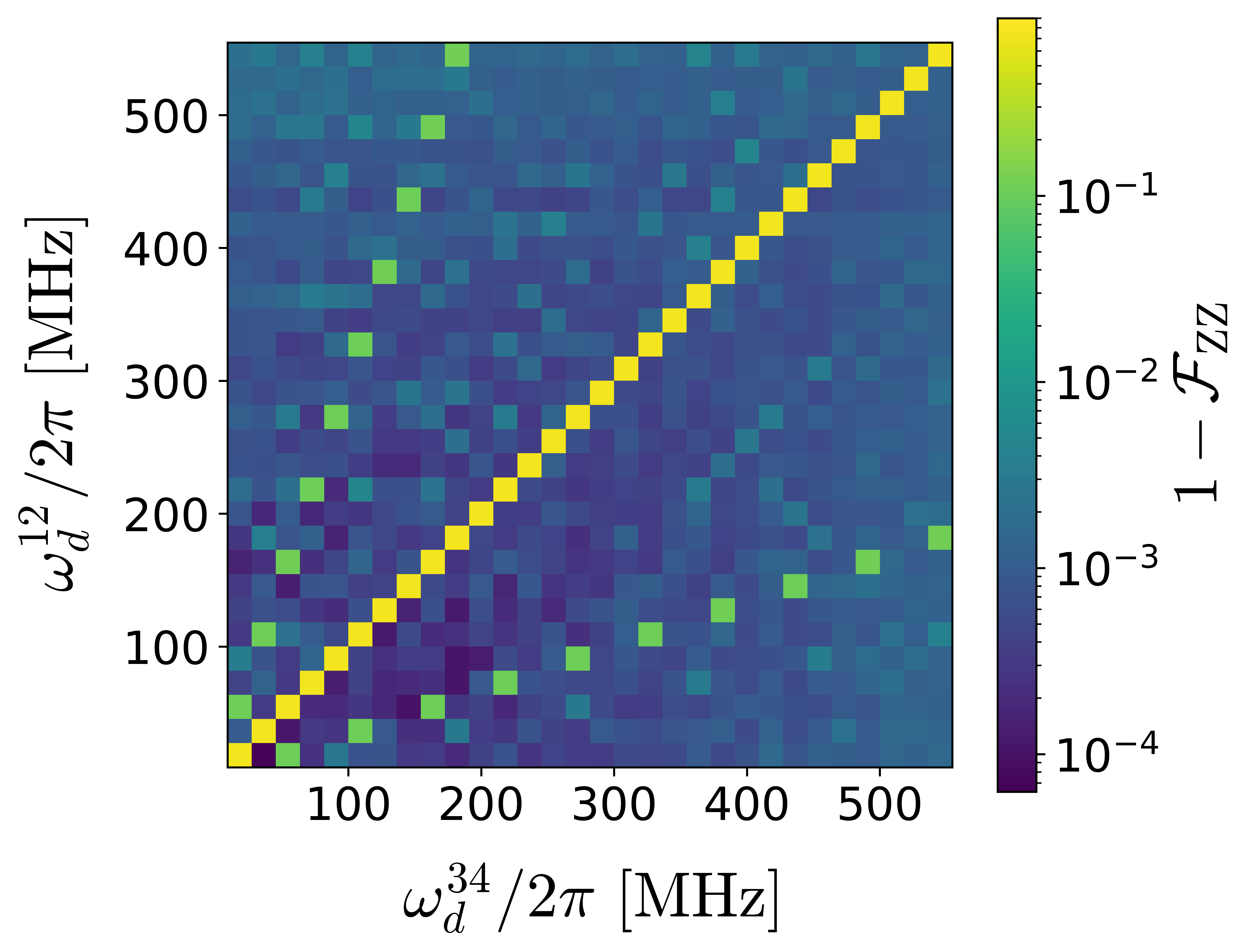}
    \caption{\label{fig4_parZZ}Optimized infidelity $1-\mathcal{F}$ for two parallel entangling gates $G_{ZZ}^{12}\otimes G_{ZZ}^{34}$ in a four-qubit register as a function of $\omega_d^{12}$ and $\omega_d^{34}$. Here $\Delta_1/2\pi=7.00\ \mathrm{GHz}$, $\Delta_k=\Delta_1+(k-1)\delta$, $\delta/2\pi=250\ \mathrm{MHz}$, and $T=55\ \mathrm{ns}$. The pulse amplitudes and final local $Z$ rotations are optimized using the corrected pulse. Other parameters are $\omega_c/2\pi=3.0\ \mathrm{GHz}$ and $g_{k}/2\pi=100\ \mathrm{MHz}$. }
\end{figure}

We show how the frequency-selection mechanism also permits simultaneous gates on disjoint pairs. We consider a four-qubit register and target gate
$$G_{4Q} = G_{ZZ}^{12}\otimes G_{ZZ}^{34}$$
up to correcting local $Z$ rotations. The two members of each pair are driven at a common frequency, $\omega_d^{12}$ or $\omega_d^{34}$, while the two pairs are assigned different frequencies, $\omega_d^{12}\neq\omega_d^{34} $. This distinction is essential because if all four qubits were driven at the same frequency, every pairwise product in Eq.~(\ref{g_ZZ_eff_NQ}) would contain a constant component and additional cross-pair $ZZ$ phases would be generated. Because all the qubits are involved in the dynamics, here we consider again the gate fidelity calculated within the cavity $0$-photon subspace.
\\\indent
Figure~\ref{fig4_parZZ} maps the optimized gate error as a function of $\omega_d^{12}$ and $\omega_d^{34}$ for a register with uniform adjacent detuning $\delta_{k,k+1}/2\pi=250\ \mathrm{MHz}$. The diagonal feature at $\omega_d^{12}=\omega_d^{34}$ directly exposes the loss of pair selectivity when the two channels coincide. Additional narrow features arise from Floquet sidebands involving different dressed-qubit pairs. Away from the equal-frequency line and the sideband conditions, an extended region with errors approaching $10^{-4}$ is obtained.
\\\indent
Parallel operation on a common cavity can therefore be formulated as a frequency-allocation problem. Each simultaneously active pair is assigned a distinct low-frequency channel, while inactive qubits are left undriven. Importantly, a processor does not require a permanently assigned channel for every one of its $N(N-1)/2$ possible pairs. The channels are allocated dynamically only to the gates executed in parallel, and can be reused in subsequent circuit layers.

\section{Discussion and Outlook}
The scheme presented is well suited to qubits that closely approximate ideal two-level systems, and flux qubits provide a natural superconducting implementation. Because the interaction strength is set by the amplitudes and relative phase of the drives rather than by a resonance condition, pair selectivity is available over a broad band of drive frequencies (Fig.~\ref{fig2}). For this architecture, strong driving, up to $A_k\sim\Delta_k$, is enabled by both the high flux-qubit anharmonicity, which can reach tens of gigahertz~\cite{mooijJosephsonPersistentCurrentQubit1999a,orlandoSuperconductingPersistentcurrentQubit1999a,sternFluxQubitsLong2014a} and strongly suppresses leakage out of the computational subspace, and by the highly off-resonant operating regime, $\omega_k\ll\Delta_k$.  
The low drive frequencies have a further practical advantage: they lie outside the stop band of the Bragg filters used to suppress near-resonant spontaneous emission. They thus allow
strong drives to reach the flux qubits without significant attenuation or unrealistically high input powers.

The use of strong AC drives does not compromise the protection against
low-frequency $1/f$ flux noise provided by the sweet spot. A flux fluctuation $\delta\epsilon_k$ enters as $\delta\epsilon_k\sigma_k^x/2$, whose longitudinal
component in the adiabatic frame, $(\epsilon_k/\eta_k)\,\delta\epsilon_k\sigma_k^z/2$,
sets the first-order flux sensitivity of the driven
qubit~\cite{changTunableSuperconductingFlux2023}. This $\epsilon_k/\eta_k$ contains only odd
harmonics of $\omega_k$, and $\int_0^T \epsilon_k/\eta_k\,dt=0$ for gate durations
of an integer number of periods. Quasi-static flux noise is therefore refocused by
the drive itself, which shifts the noise sensitivity of the qubit away from the
low-frequency part of the $1/f$ spectrum, analogously to spin-locked driven
qubits~\cite{Yan2013RotatingFrame}. The residual second-order shift,
$\delta\eta_k\simeq(\Delta_k^2/2\eta_k^3)\,\delta\epsilon_k^2$ with
$\Delta_k^2/2\eta_k^3\le 1/(2\Delta_k)$, does not exceed that of the static sweet
spot. Remarkably, the same harmonic structure underlies the gate: for matched drive
frequencies, the product $J^{zz}_{jk}\propto\epsilon_j\epsilon_k/\eta_j\eta_k$ of two
functions with only odd harmonics contains a non-vanishing DC component. As for
qubit relaxation, the chosen gate duration is sufficiently short that the resulting
gate error is below the thresholds of leading quantum error correction
codes~\cite{Wang2011,Bravyi2024}: for a gradiometer flux qubit with
$T_1=50~\mu$s~\cite{changReproducibilityGapControl2022}, the error for a
$T=55$~ns gate is of order $T/T_1\sim10^{-3}$.

The nearly two-level character of flux qubits also strongly suppresses static $ZZ$ crosstalk, which is essential for scaling to highly connected processors. 
The $ZZ$ crosstalk between a pair of qubits  can be quantified as
$
  \zeta=E_{11}-E_{10}-E_{01}+E_{00},
$
where the energies include shifts induced by the cavity, spectator qubits, and higher qubit levels. These shifts can be estimated perturbatively. As an example, consider a processor containing 20 flux qubits coupled to a common cavity, with qubit frequencies spanning $7$--$8~\mathrm{GHz}$ in increments of $50~\mathrm{MHz}$. For qubit--cavity couplings $g/2\pi=100~\mathrm{MHz}$ and an anharmonicity of $25~\mathrm{GHz}$, the residual $ZZ$ coupling between any qubit pair is approximately $\zeta/4\lesssim 1.67~\mathrm{kHz}$ (Appendix~\ref{app:crosstalk}). Over a gate duration of $55~\mathrm{ns}$, this interaction contributes an infidelity of order $10^{-7}$ and is therefore negligible compared with the other error sources considered here.

\begin{figure}[t]
    \centering
    \includegraphics[width=0.7\linewidth]{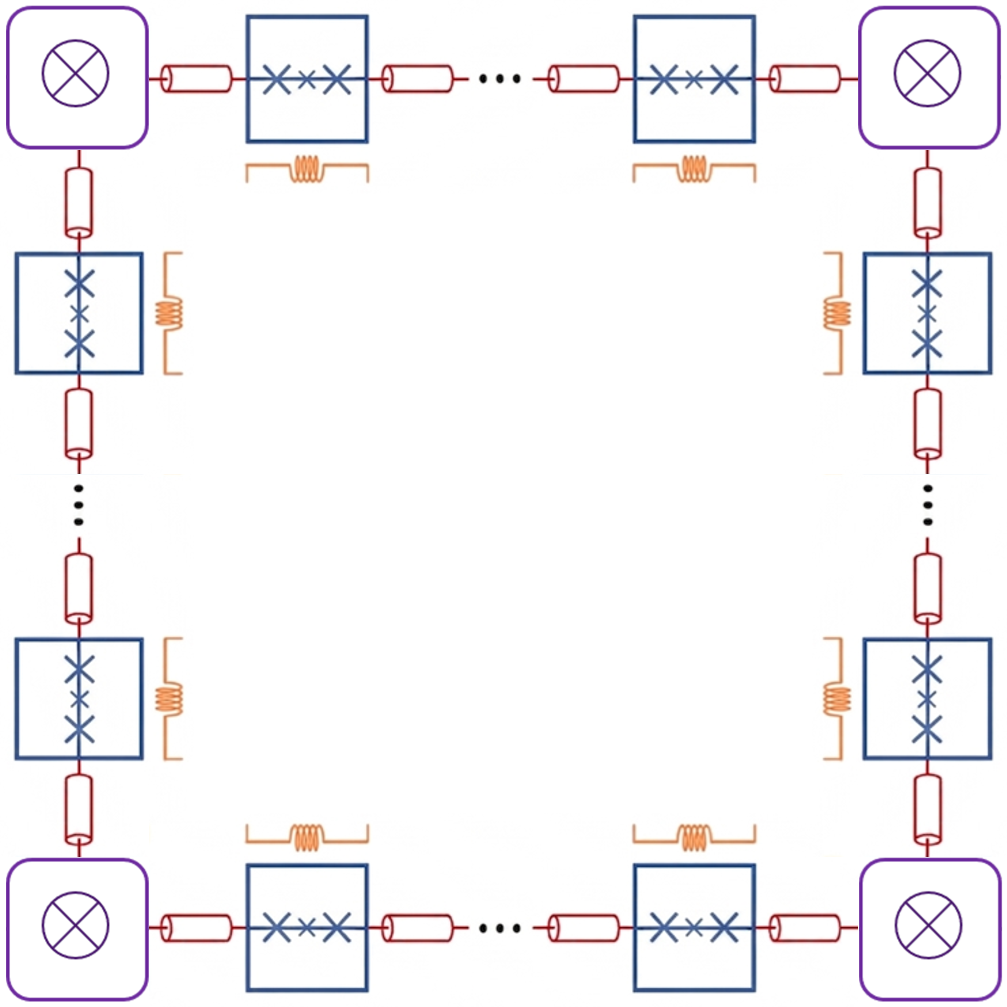}   \caption{\label{fig_multicore}Schematic of a modular multi-core processor. Each core consists of a common resonator coupled to multiple flux qubits, providing all-to-all intra-core connectivity. Neighboring resonators are linked by flux-tunable couplers. When an inter-core coupler is activated, the resonator modes hybridize into delocalized normal modes spanning the two cores, thereby mediating longitudinal interactions between flux qubits located in different cores.}
\end{figure}

Scaling beyond a single common cavity can be achieved through a modular
multi-core architecture. Each multi-qubit register may be treated as a
processor core whose cavity bus is connected to other cores through a
frequency-tunable bosonic coupler, such as a SQUID-tunable superconducting resonator
\cite{palaciosLaloyTunableResonators2008,
sandbergTuningField2008,wangQuantumState2013}. Figure~\ref{fig_multicore} illustrates such a multicore design. Varying the coupler frequency allows the effective interaction between the cavity modes to be activated, tuned, or parked near zero. When activated, the modes hybridize into nonlocal bosonic modes that couple flux qubits in different cores; applying matched low-frequency drives to a selected inter-core pair can then generate the same adiabatic $ZZ$ interaction through these modes (Appendix~\ref{app:multicore}). 
This extends pair-selective long-range gates across different cores, providing a modular route toward substantially larger superconducting quantum processors using flux qubits. Current flux qubit fabrication technologies could in principle support a system comprising up to 20 interconnected cores, with each core comprising 20 flux qubits, yielding a 400-qubit quantum processor.
\\

\section*{Acknowledgements}
M.S. thanks the Helen Diller Quantum Center at the Technion.
E.L. and E.G. acknowledge support from Engineering and Physical Sciences Research Council grant number EP/T001062/1. E.G. acknowledges support from the Engineering and Physical Sciences Research Council grant number EP/Z000505/1.

%

\clearpage
\appendix

\section{Local X rotations due to non-adiabatic corrections}\label{app:nonad}
In this section, we show that the effect of the non-adiabatic term on the qubit dynamics are local single-qubit $X$ rotations at $t=0$ and $t=T$. For notational simplicity, explicit time dependence in variables is dropped in the derivations below, unless necessary.
Each qubit is described by
\begin{equation}
    H_k=\frac{\Delta_k}{2}\sigma^z_k+\frac{\epsilon_k}{2}\sigma^x_k ,
\end{equation}
with a baseband drive $\epsilon_k=A_k\sin(\omega_k t)$, i.e.\ $\xi_k=0$ as in the main text, and instantaneous gap $\eta_k=\sqrt{\Delta_k^2+\epsilon_k^2}$. The gate time is
commensurate with the drive, $T=2\pi n/\omega_k$, so that
$\epsilon_k(0)=\epsilon_k(T)=0$ while $\dot\epsilon_k(0)=\dot\epsilon_k(T)=A_k\omega_k$.
The adiabatic frame is defined by
\begin{equation}
    R^y_k(t)=e^{-i\varphi_k\sigma^y_k/2},
\end{equation}
with tilt angle $\tan\varphi_k=\epsilon_k/\Delta_k$, yielding the single-qubit Hamiltonian
\begin{equation}
  H^{(1)}_k=(R^y_k)^\dagger H_kR^y_k-i(R^y_k)^\dagger\partial_tR^y_k
  =\frac{\eta_k}{2}\sigma^z_k-\frac{\dot\varphi_k}{2}\sigma^y_k,
  \label{eq:H1}
\end{equation}
with $\dot\varphi_k=\Delta_k\dot\epsilon_k/\eta_k^2$. 
Neglecting the transport term $\dot\varphi_k$ here would lead to the time-dynamics described in Eq.~(\ref{eq_prop_1Q_before_Floquet}). Here instead, we retain the term and treat $H^{(1)}_k$ as a slow time-dependent Hamiltonian by itself, and diagonalise it in its instantaneous eigenbasis. This is done with a rotation about $x$,
\begin{equation}
  R_k^x(t)=e^{-i\vartheta_k\sigma^x_k/2},\qquad
  \tan\vartheta_k=\frac{\dot\varphi_k}{\eta_k},
  \label{eq:chi}
\end{equation}
which gives
\begin{align}
  H^{(2)}_k&={R_k^x}^\dagger H^{(1)}_kR_k^x-i{(R_k^x)}^\dagger\partial_t R_k^x
  =\frac{\bar\eta_k}{2}\sigma^z_k-\frac{\dot\vartheta_k}{2}\sigma^x_k,\nonumber \\
  \bar\eta_k&=\sqrt{\eta_k^2+\dot\varphi_k^2}.
  \label{eq:H2}
\end{align}
The structure of $H^{(2)}_k$ is the same as in Eq.~\eqref{eq:H1}, with a transport term that has shrunk from $\dot\varphi_k\sim\omega_k A_k/\Delta_k$ to
$\dot\vartheta_k\sim\omega_k^2A_k/\Delta_k^2$.

Eq.~\eqref{eq:H2} shows that the state actually followed by the qubit
is the superadiabatic state $R^y_kR_k^x|0\rangle$ rather than the adiabatic state
$R^y_k|0\rangle$; the two differ by the tilt $\vartheta_k(t)$ about $x$. In addition, the dynamical
phase is accumulated with $\bar\eta_k$ rather than $\eta_k$, a second-order shift
$\dot\varphi_k^2/2\eta_k$ which is a pure $Z$ phase.

With $|\psi\rangle=R^y_kR_k^x|\psi^{(2)}\rangle$ the lab-frame propagator reads
\begin{equation}
  U_k(t,0)=R^y_k(t)R_k^x(t)\,U^{(2)}_k(t,0)\,{R_k^x}(0)^{\dagger}R^{y}_k(0)^{\dagger}.
  \label{eq:Uframes}
\end{equation}
Neglecting the residual $\dot\vartheta_k$ term in Eq.~\eqref{eq:H2}, whose effect is a
second-order $Z$ shift of relative size $\omega_k^4A_k^2/\Delta_k^6$, the time-evolution in this new frame now reads
\begin{equation}
  U^{(2)}_k(T,0)=\exp\left\{-\frac{i}{2}\left(\int_0^T\bar\eta_k\,dt\right)\sigma^z_k\right\}. 
\end{equation}
At $t=0$ and $t=T$,
$\epsilon_k=0$, so $R^y_k=\openone$ and $\eta_k=\Delta_k$, but $\dot\epsilon_k=A_k\omega_k$,
so the $R_k^x$ tilt does not vanish:
\begin{equation}
  \vartheta_k(0)=\vartheta_k(T)=\arctan\frac{\omega_kA_k}{\Delta_k^2}\equiv\vartheta_k .
  \label{eq:edge}
\end{equation}
Hence $R_k^x(0)=R_k^x(T)\equiv R_k^x$ where
\begin{equation}
    R_k^x=\exp\left(-i\frac{\vartheta_k}{2}\sigma_k^x\right),
\end{equation}
is a fixed rotation and
\begin{equation}
  U_k(T,0)\simeq R_k^x\,e^{-\frac{i}{2}\big(\int_0^T\bar\eta_k\,dt\big)\sigma^z_k}\,{R_k^x}^\dagger .
  \label{eq:Uk}
\end{equation}
The non-adiabatic term thus leaves the main evolution as a pure $Z$ phase with additional small local $R_k^x$ rotations.

\section{Effective qubit-qubit transverse interaction} 
\label{app:transverse}
While in the main text the derivation of the effective longitudinal coupling is presented, here we derive the effective transverse interaction term, Eq.~(\ref{V_XX_residual}), of the $N$ qubit system.

\subsection{Transverse term displacement}
\label{app:transverse-displacement}
We first show that the expansion of the displaced transverse qubit-cavity term in Hamiltonian Eq.~\eqref{HamNQ_Displaced}
\begin{align}
\label{Vperp_Displaced}
D^\dagger[\alpha]V^{\perp} D[\alpha],
\end{align}
where $V^{\perp}$ is given by Eq.~(\ref{eq_Vper_Vpar}),
\begin{align}
    &{}V^{\perp} = \sum_{k=1}^N \lambda_k(a+a^\dagger)\tilde\sigma_k^x; \quad\lambda_k = g_{k}\frac{\Delta_k}{\eta_k};
\end{align}
and $\sigma^x_k(t)$ is defined in Eq.~\eqref{eq_sx_rotated},
\begin{align}
    &{}\tilde\sigma_k^x=R_k^z(\beta_k)^{\dagger}\sigma^x_k R_k^z(\beta_k);\quad 
    R_k^z(\beta_k)=\exp\left\{-i\frac{\beta_k}{2}\sigma^z_k\right\},
\end{align}
leads to a shift of the couplings $\lambda_k$ and far off resonant interaction terms. 

We remind ourselves that $D[\alpha]$ is the Lang--Firsov transformation \cite{LangFirsovKineticTheory1963} in Eq.~(\ref{D_alpha_adiabatic_frame}),
\begin{align}
    D[\alpha] =&{}\ e^{\alpha (a^\dagger - a\,)}, \
\alpha=\sum_{k=1}^N\alpha_k\sigma_k^z,
\end{align}
where $\alpha_k = -g_{k}\epsilon_k/\eta_k\omega_c$.
Since $D[\alpha]$ contains only $\sigma_k^z$ terms, it commutes with the rotations $R_k^z(\beta_k)$ applied in $\tilde\sigma_k^x$, and, introducing the resonator quadrature $p = i(a^\dagger - a)$ and $x = (a^\dagger + a)$, we have that
\begin{align}
    &{}D^\dagger[\alpha] \left(a^\dagger + a\right) \tilde\sigma_k^x\, D[\alpha] \nonumber\\
    &{} = R_k^z(\beta_k)^\dagger\left\{D[\alpha(t)]^\dagger \left(a^\dagger + a\right) \sigma_k^x\, D[\alpha]\right\}R_k^{z}(\beta_k) \nonumber\\
    &{} = R_k^z(\beta_k)^{\dagger}\left\{D[\alpha]^\dagger \left(a^\dagger + a\right) D[\alpha]\right\} \left\{D[\alpha]^\dagger\,  \sigma_k^xD[\alpha]\right\}R_k^{z}(\beta_k) \nonumber\\
    &{} = \,R_k^z(\beta_k)^{\dagger}\Big[a^\dagger + a + 2\sum_j\alpha_j\sigma_j^z \Big] X_k(p)\,R_k^{z}(\beta_k)
    \nonumber\\
    &{} = \,R_k^z(\beta_k)^{\dagger}\Big\{\Big[x + 2\alpha_k\sigma_k^z \Big] X_k(p) \nonumber\\
    &{}\qquad + 2\sum_{j\neq k}\alpha_j\sigma_j^zX_k(p)\Big\}R_k^{z}(\beta_k),
\end{align}
with 
\begin{equation}
    X_k(p) = \cos{[2\alpha_k p]}\,\sigma_k^x - \sin{[2\alpha_k p]}\,\sigma_k^y.
\end{equation}
One can verify that $[x+ 2\alpha_k\sigma_k^z]X_k(p)  = (1/2)\,x\,X_k(p) + \text{h.c.}$, therefore yielding 
\begin{align}
    &{}D[\alpha]^{\dagger} \left(a^\dagger + a\right) \tilde\sigma_k^x\, D[\alpha] \nonumber\\
    &{} = \frac{1}{2}\left(a^\dagger + a\right) \big( \cos{[2\alpha_k p]}\,\tilde\sigma_k^x - \sin{[2\alpha_k p]}\,\tilde\sigma_k^y\big) \nonumber\\
    &{}- \sum_{j\neq k}\alpha_j \Big\{ \cos{[2\alpha_k p]}\,\sigma_j^z\tilde\sigma_k^x - \sin{[2\alpha_k p]}\,\sigma_j^z\tilde\sigma_k^y\Big\}\nonumber\\
    &{}+ \text{h.c.},
\end{align}
where
\begin{equation}
    \tilde\sigma_k^y=R_k^z(\beta_k)^{\dagger}\sigma^y_k R_k^z(\beta_k).
\end{equation}
We now expand to second order in $\alpha_k$ and retain only the terms linear in the bosonic operators:
\begin{align}
    &{}D[\alpha]^{\dagger} \left(a^\dagger + a\right) \tilde\sigma_k^x\, D[\alpha] \nonumber\\
    &{} \simeq \left(a^\dagger + a\right)\tilde\sigma_k^x - 2i\alpha_k\left(a^{\dagger,2} - a^2\right)\tilde\sigma_k^y \nonumber\\
    &{}\quad + \alpha_k^2\left\{a^{\dagger,2}-a^2,\ a^\dagger - a\right\}\tilde\sigma_k^x \nonumber\\
    &{}\quad + 2\sum_{j\neq k}\alpha_j \sigma_j^z\tilde\sigma_k^x \nonumber\\
    &{}\quad - 4i\sum_{j\neq k}\alpha_k\alpha_j(a^\dagger - a)\sigma_j^z\tilde\sigma_k^y \nonumber\\
    &{} \simeq \left(1 - 2\alpha_k^2\right)\left(a^\dagger + a\right)\tilde\sigma_k^x + 2\sum_{j\neq k}\alpha_j \sigma_j^z\tilde\sigma_k^x \nonumber\\
    &{}\quad - 4i\sum_{j\neq k}\alpha_k\alpha_j(a^\dagger - a)\sigma_j^z\tilde\sigma_k^y,
\end{align}
where we have used $\left\{(a^{\dagger})^2-a^2,\ a^\dagger - a\right\} = 2(a^{\dagger})^3 + 2a^3 - 2(a^{\dagger})^2a - 2a^\dagger a^2 - 2\left(a^\dagger + a\right)$.

The net effect of the displacement on the transverse coupling is a renormalization factor, $\lambda_k\to\lambda_k\big(1-2\alpha_k^2\big)$, which can be neglected as $\alpha_k^2=O(g^2_k/\omega_c^2)\ll1$. Also, for typical parameter values in our system, the $\sigma_j^z\tilde\sigma_k^x$  and $(a^{\dagger}-a)\sigma_j^z\tilde\sigma_k^y$  terms are far off resonant and can be discarded. With these approximations, we have that 
$$ D^\dagger[\alpha]V^{\perp} D[\alpha] \approx V^\perp,$$
and the effective Hamiltonian (\ref{HamNQ_Displaced}) is simply:
\begin{align}
\label{HamNQ_eff_expanded_SM}
H_{\text{eff}} \simeq&{} \sum_k\frac{\nu_k}{2}\sigma_k^z + \omega_c a^\dagger a + V^\perp + V_{\text{eff}}^\parallel\,,
\end{align}
where $V_{\text{eff}}^\parallel = -\sum_{j<k}J_{jk}^{zz}\sigma_j^z\sigma_k^z$ is given by Eq.~(\ref{g_ZZ_eff_NQ}).

\subsection{Effective dispersive Hamiltonian}
\label{app:transverse-SW}
At this point, in the dispersive regime of the Floquet and cavity frequencies, i.e. $\omega_c-\nu_k \gg \lambda_k \sim g_{k}$, we can apply a Schrieffer--Wolff (SW) transformation $e^{S}$, with 
\begin{align}
    S=&{}\ \sum_k \bigg\{ \frac{\lambda_k}{\omega_c-\nu_k}\left( a^\dagger\tilde\sigma_k^- - a\tilde\sigma_k^+\right) \nonumber\\
    &{}\  + \frac{\lambda_k}{\omega_c+\nu_k}\left( a^\dagger\tilde\sigma_k^+ - a\tilde\sigma_k^-\right) \bigg\},
\end{align}
and we have defined  $\tilde\sigma^{\pm}_{k}= (\tilde\sigma_k^x\pm i \tilde\sigma_k^y)/2 = e^{\pm i\beta_k}\sigma^\pm_{k}$.
In the adiabatic approximation the term $-i\dot{S}$ is neglected. 
Denote by $H_0$ the uncoupled, time-independent part of Eq.~(\ref{HamNQ_eff_expanded_SM}), $V^{\perp}$ the part containing the transverse $(a^\dagger + a )\tilde\sigma_k^x$ terms, and neglecting terms of the order  $O(g_{k}^3,\,\lambda_k^3)$, the transformed Hamiltonian takes the form
\begin{align}
\label{SW_expansion_H_eff}
    H_{\text{eff}}^{SW} \simeq&{}\ e^{S}\left[H_0 + V^{\perp}  + V_{\text{eff}}^{zz}  \right]e^{-S} \nonumber\\
    \simeq&{}\ H_0 + V_{\text{eff}}^{zz} + \frac{1}{2}[S,V^\perp],
\end{align} 
where we have  substituted $[S,H_0] = - V^{\perp}$.  The commutator can be expressed as
\begin{align}
    &{}\frac{1}{2}[S, V^{\perp}] = - \sum_{k,\,j\neq k} \frac{\omega_c\,\lambda_j\lambda_k}{\omega_c^2 - \nu_j^2}\tilde\sigma_j^x\,\tilde\sigma_k^x \nonumber\\
    &{}\hspace{53pt}  -\sum_k\frac{\nu_k\,\lambda_k^2}{\omega_c^2 - \nu_k^2} \big(1 + 2a^\dagger a\big)\sigma_k^z \nonumber\\
    &{}\hspace{53pt}  -\sum_k\frac{\nu_k\,\lambda_k^2}{\omega_c^2 - \nu_k^2}\big(a^2 + (a^\dagger)^2 \big)\sigma_k^z,
\end{align}
up to a constant term. Thus, when we neglect the two-photon terms and define $\delta\nu_k = 2\nu_k \lambda_k^2/(\nu_k^2 - 
\omega_c^2)$, a dispersive shift of the quasienergy due to coupling with the cavity, we obtain the effective Hamiltonian
\begin{align}
\label{HamNQ_eff_SW_SM}
H_{\text{eff}}^{SW} &\simeq \sum_k\frac{\tilde\nu_k}{2}\sigma_k^z + \bigg(\omega_c + \sum_k\delta\nu_k\,\sigma_k^z\bigg) a^\dagger a\nonumber\\
&+\sum_{j<k}J_{jk}^{\perp}\,\tilde\sigma_j^x\tilde\sigma_k^x 
 -\sum_{j<k}J_{jk}^{zz}\sigma_j^z\sigma_k^z,
\end{align}
where
\begin{equation}
  \tilde \nu_k  =\nu_k+ \delta\nu_k
\end{equation}
is the shifted quasienergy, and
\begin{equation}\label{eq:Jperp}
J_{jk}^{\perp} = \lambda_j\lambda_k\omega_c\left[\frac{1}{\nu_j^2-\omega_c^2}+\frac{1}{\nu_k^2-\omega_c^2}\right]
\end{equation} 
is the effective transverse coupling strength. The DC component of $J_{jk}^{\perp}$ is an always-on coupling, however its effect is suppressed when the qubits are far detuned, $J_{jk}^{\perp} \ll \vert\nu_j - \nu_k\vert$, leaving only the effective $\sigma_j^z\sigma_k^z$ interaction. 

\section{Sidebands transitions}
\label{app:sidebands}
In this section we explore the conditions for  unwanted resonant transitions during the ZZ entangling gate.  In the interaction picture of the free terms, and within the RWA, the Hamiltonian (\ref{HamNQ_eff_SW_SM}) becomes
\begin{align}
\label{HamNQ_eff_SM}
H_{\text{eff},I}^{SW} &= \sum_{j<k}J_{jk}^{\perp}\left( e^{i(\tilde \nu_j-\tilde \nu_k)t}\tilde\sigma_j^+\tilde\sigma_k^- 
+ \text{h.c.} \right)-\sum_{j<k}J_{jk}^{zz}\sigma_j^z\sigma_k^z \nonumber\\
&=\sum_{j<k}\left(\tilde J_{jk}^{\perp}e^{i(\tilde \nu_j-\tilde \nu_k)t}\sigma_j^+\sigma_k^- 
+ \text{h.c.} \right)
-\sum_{j<k}J_{jk}^{zz}\sigma_j^z\sigma_k^z,
\end{align}
with
\begin{equation}\label{eq:Jtilde}
    \tilde J_{jk}^{\perp}=J_{jk}^{\perp}e^{i(\beta_j-\beta_k)},
\end{equation}
where the last step follows from a $\tilde \sigma_j^{+}=e^{i\beta_j}\sigma_j^{+}$ substitution.

Recall that $\lambda_k$, $\alpha_k$ and $\beta_k$ are time-dependent coefficients, whose harmonic content follows from the parity and periodicity of $\eta_k$. Since  $\eta_k(t)$ is a function of $\sin^2(\omega_kt+\xi_k)$, it has period $\tau_k/2$, and so does $\beta_k(t)$. Therefore, $\tilde\lambda_k(t) \equiv\lambda_k(t)\exp\{-i\beta_k(t)\}$ has period $\tau_k/2$ and admits Fourier series in even harmonics $2n\omega_k$, i.e.,
\begin{align}
\label{HamNQ_coefficients_harmonics_SM}
&{} \tilde\lambda_k(t)=   \sum_{n=-\infty}^{\infty} [\tilde\lambda_k]_{n}e^{i2n\omega_{k}t}.
\end{align}
where $[\tilde\lambda_k]_{n}$ are the Fourier coefficients.
As $\tilde J_{jk}^{\perp} \sim \tilde\lambda_j^{*} \tilde\lambda_k$, up to a constant, (see Eqs.~\eqref{eq:Jtilde} and \eqref{eq:Jperp}), it is a product of two Fourier series with even harmonics in $\omega_j$ and $\omega_k$, i.e,
\begin{equation}
    \tilde J^{\perp}_{jk}(t)=\sum_{m,n}[\tilde J^{\perp}_{jk}]_{m}^ne^{i2(n\omega_k-m\omega_j)t}, \quad [J^{\perp}_{jk}]_{m}^n=[\tilde\lambda_j]^{*}_{m}[\tilde\lambda_k]_{n}
\end{equation}
\begin{figure}[!b]
    \centering
\includegraphics[width=0.8\linewidth]{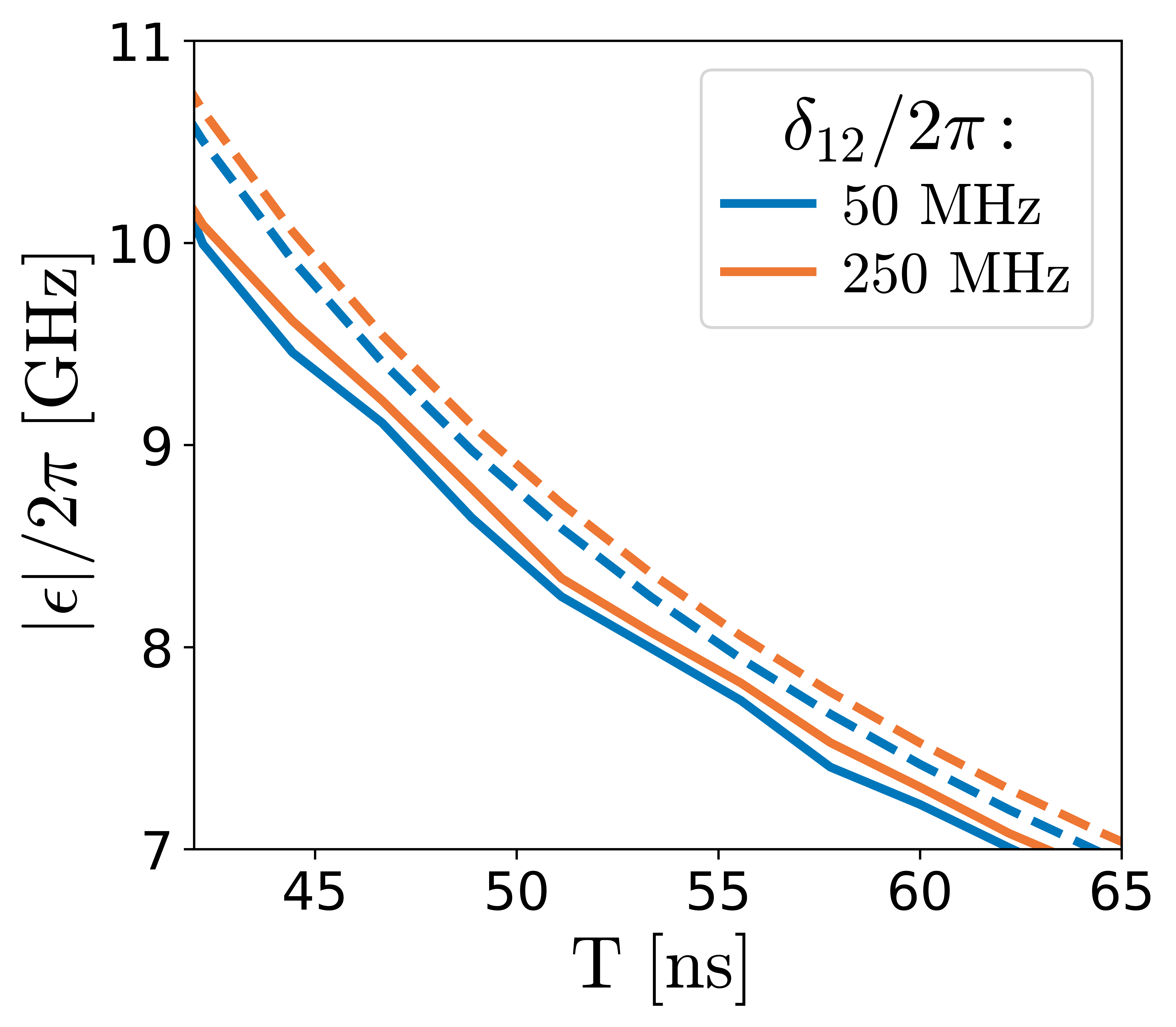}
    \caption{\label{figD}Optimal drive amplitude $A$ for the two-qubit gate in Fig.~\ref{fig2} as a function of $T$ at fixed $\omega_d/2\pi=450\ \mathrm{MHz}$. Solid curves show the numerical optimized results, while dashed curves follow from the analytical effective interaction by imposing $\theta_{12}(T)=\pi/4$.}
\end{figure}

The Hamiltonian Eq.~(\ref{HamNQ_eff_SM}) can now be written as
\begin{align}
\label{HamNQ_eff_harmonics_SM}
&\sum_{j<k}\sum_{m,n}[J^{\perp}_{jk}]_{mn}e^{i2(n\omega_k-m\omega_j)t}e^{i(\tilde \nu_j-\tilde \nu_k)t}\sigma_j^+\sigma_k^- 
+ \text{h.c.}\nonumber \\ 
&-\sum_{j<k}J_{jk}^{zz}\sigma_j^z\sigma_k^z,
\end{align}
The modulated transverse coupling thus gives rise to resonant transition when
\begin{equation}
    2(m\omega_{j}-n\omega_{k}) = \tilde\nu_{j} - \tilde\nu_{k}, \quad m,n \in \mathbb{Z}
\end{equation}
For a selected pair $(k,j)$, the implementation of an entangling operation is bound to the condition $\omega_{k} = \omega_{j} = \omega_{d}$, so the resonant transitions happen at
\begin{equation}
    2q\omega_d=\tilde\nu_{j} - \tilde\nu_{k}, \quad q\in \mathbb{Z}.
\end{equation}
All sideband transitions with $\vert q \vert >0$ can be avoided by using a driving frequency above half energy separation, i.e.,
\begin{equation}
    \omega_{d} > \frac{\vert \tilde\nu_{k} - \tilde\nu_{j}\vert}{2}.
\end{equation}
Typically the dispersive shifts of the quasienergies due to interaction with the cavity are small and  hence $\tilde\nu_k\approx \nu_k$.

For an undriven spectator $l$, $\lambda_l=g_l$ and $\beta_l=0$, so only the $m=0$ harmonic of $\tilde\lambda_l$ survives and $\tilde\nu_l\simeq\Delta_l$. The resonance condition with a driven qubit $k$ then reduces to
\begin{equation}
    \Delta_l = \tilde\nu_k + 2n\,\omega_d, \qquad n\in\mathbb{Z},
\end{equation}
i.e. Eq.~\eqref{eq:spectator_sidebands}, with sideband strength proportional to $|[\tilde\lambda_k]_{-n}|$, i.e.\ to the $e^{2in\omega_d t}$ component of $(\Delta_k/\eta_k)\,e^{i\beta_k}$. Unlike the driven-pair case, these resonances cannot all be removed by increasing $\omega_d$, since the $n=0$ resonance at $\Delta_l\simeq\nu_k$ is always present; they are instead avoided by placing spectators away from the comb $\nu_k+2n\omega_d$. Because $\tilde\lambda_k=(g_k\Delta_k/\eta_k)\,e^{-i\beta_k}$ and $\beta_k$ has a large amplitude at strong driving (of order $A_k^2/(8\Delta_k\omega_d)$ for $A_k \lesssim \Delta_k$), the coefficients $|[\tilde\lambda_k]_n|$ are distributed over several harmonics rather than dominated by $n=0$.

\section{Optimal drive amplitude}\label{app:optamp}
To maximize the fidelity for the two-qubit entangling gate in Sec.~\ref{sec_subsec_entangling_gate_2Q} the amplitude of the two identical drives is optimized. Figure~\ref{figD} shows the numerically optimized values for a given duration $T$, and values obtained from the analytical effective interaction, by numerically integrating Eq.~\eqref{phaseZZ_jk} with the coupling of Eq.~\eqref{g_ZZ_eff_NQ} and imposing $\theta_{12}(T)=\pi/4$. The close agreement validates the adiabatic approximations utilized in the derivation of the ZZ interaction strength.

\section{Robustness comparison with cross-resonance gate}
\label{app:robustness}
In this section, we compare the two-qubit correlated error induced by qubit frequency fluctuation of the adiabatic $ZZ$ gate and the more widely used cross-resonance (CR) gate. The derivation related to the adiabatic $ZZ$ gate is already presented above (Eq.~\eqref{fid_ZZ_freq_fluctuation}), yielding a correlated error of 
\begin{equation}
\label{gate_error_ZZ_freq_fluctuation_SM}
1-F_{\text{ZZ}} \approx \left(\frac{\pi}{4}\right)^2\left(\frac{\delta\Delta_2}{\Delta_2}\right)^2.
\end{equation}
Here we provide more details on the CR gate \cite{rigettiFullyMicrowaveTunable2010,kirchhoffOptimizedCrossresonanceGate2018}, deriving the  effective Hamiltonian  and estimating the leading correlated error.

\subsection{Cross-resonance gate}
\label{sec_SM_subsection_CR}
We start from the Hamiltonian (\ref{HamSysNQ}) for $N=2$ without the drives
\begin{equation}
H=\frac{\Delta_1}{2}\sigma_1^z+\frac{\Delta_2}{2}\sigma_2^z+\omega_c a^\dagger a+\sum_{j=1,2} g_j \sigma_j^x(a+a^\dagger).
\end{equation}
In the dispersive regime where $g_k\ll |\omega_c - \Delta_k|$, applying the Schrieffer-Wolff transformation $e^S$, with
\begin{align}
S = &{}\ \sum_k \bigg\{ \frac{g_{k}}{\omega_c-\Delta_k}\left( a^\dagger\sigma_k^- - a\sigma_k^+\right)  \nonumber\\
&{}+ \frac{g_{k}}{\omega_c+\Delta_k}\left( a^\dagger\sigma_k^+ - a\sigma_k^-\right) \bigg\},
\end{align}
one obtains, up to small dispersive shifts, the effective qubit Hamiltonian with a $XX$ interaction
\begin{align}
H_{q0}&{}\simeq\frac{\Delta_1}{2}\sigma_1^z+\frac{\Delta_2}{2}\sigma_2^z+J_{xx}\sigma_1^x\sigma_2^x, \nonumber\\
J_{xx}&{}=g_1g_2\,\omega_c\left(\sum_{k=1}^2 \frac{1}{\Delta_k^2-\omega_c^2}\right).
\end{align}
With a cross-resonant drive on the first qubit at the predicted frequency of the second qubit, $H^{\mathrm{cr}}_d=A_{\text{CR}}\, \sigma_1^x \cos{\Delta_2 t}$, assuming the real frequency of the 2nd qubit is shifted by  $\delta\Delta_2$, the Hamiltonian becomes
\begin{align}
H_q=&{}\,\frac{\Delta_1}{2}\sigma_1^z+\frac{\Delta_2+\delta\Delta_2}{2}\sigma_2^z \nonumber\\
&{}\qquad\quad +J_{xx}\sigma_1^x\sigma_2^x + A_{\text{CR}}\sigma_1^x\cos{\Delta_2 t}.
\end{align}
In the rotating frame defined by $U_0(t)=\exp\{-i(\Delta_2/2)(\sum_k\sigma_k^z)t\}$ and within the RWA, we obtain 
\begin{align}
\tilde{H}_{q}=&{}\,\frac{\delta}{2}\sigma_1^z+\frac{A_{\text{CR}}}{2} \sigma_1^x+\frac{\delta\Delta_2}{2}\sigma_2^z+\frac{J_{xx}}{2}(\sigma_1^x\sigma_2^x+\sigma_1^y\sigma_2^y),
\end{align}
where $\delta=\Delta_1-\Delta_2$. Introducing the tilted frame
\begin{align}
\hat\sigma_1^z&{}=\sigma_1^z\cos{\Phi}+\sigma_1^x\sin{\Phi}, \nonumber\\
\hat\sigma_1^x&{}=\sigma_1^x\cos{\Phi}-\sigma_1^z\sin{\Phi}, \nonumber\\
\hat\sigma_1^y&{}=\sigma_1^y,
\end{align}
with $\cos{\Phi}=\delta/\delta'$, $\sin{\Phi}=A_{\text{CR}}/\delta'$ and $\delta'=\sqrt{\delta^2+A_{\text{CR}}^2}$, we rewrite
\begin{align}
\tilde{H}_{q}=&{}\ \frac{\delta'}{2}\hat\sigma_1^z+\frac{\delta\Delta_2}{2}\sigma_2^z \nonumber\\
&{}\qquad +\frac{J_{xx}}{2}(\cos{\Phi}\,\hat\sigma_1^x\sigma_2^x+\hat\sigma_1^y\sigma_2^y) \nonumber\\
&{}\qquad +\frac{J_{xx}}{2}\sin{\Phi}\,\hat\sigma_1^z\sigma_2^x.
\end{align}
In the doubly rotating frame defined by $U_1(t)=e^{-i\delta' \hat\sigma_1^z t/2}$, the effective $ZX$ Hamiltonian arises within the RWA,
\begin{align}
H_{\text{CR}} =&{}\, U_1^\dagger(t)\tilde{H}_{q}\,U_1(t) - iU_1^\dagger(t)\partial_tU_1(t) 
\nonumber\\
=&{}\,\frac{\delta\Delta_2}{2} \sigma_2^z+J_{\text{CR}}\hat\sigma_1^z\sigma_2^x, \quad J_{\text{CR}}=\frac{J_{xx}}{2}\sin{\Phi}.
\end{align}
This effective Hamiltonian generates the time-evolution operator Eq.~(\ref{Ucr_t}).
In the ideal case $\delta\Delta_2=0$, evolution for a duration of $T_{\text{CR}}=\pi/(4J_{\text{CR}})$ realizes the target gate 
\begin{equation}
\label{target_CR_SM}
    U_{\text{CR}}(T_{\text{CR}})= G_{\text{CR}} \equiv e^{-i(\pi/4)\hat\sigma_1^z\sigma_2^x}.
\end{equation} 

\subsection{CR gate error due to frequency fluctuation}
\label{sec_SM_subsection_CR_error}
We derive here the expression for the CR time-evolution operator in the presence of fluctuations of the target qubit frequency, Eq.~(\ref{Ucr_gate_with_fluctuation}). With  $\delta\Delta_2 \neq 0$, the actual unitary $\mathcal{U}(T_{\text{CR}})$ is given by
\begin{align}
U_{\text{CR}}(T_{\text{CR}})&{}=e^{-iH_{\text{CR}} T_{\text{CR}}} \nonumber\\
&{}=\exp\left[{-i\left(\frac{\delta\Delta_2 T_{\text{CR}}}{2}\sigma_2^z +\frac{\pi}{4}\hat\sigma_1^z\sigma_2^x\right)}\right],
\end{align}
which, using the product formula derived below, can be decomposed as
\begin{align}\label{eq:ucr_product}
&{}U_{\text{CR}}(T_{\text{CR}})\approx
\exp\bigg\{-i\frac{\pi}{4}\hat\sigma_1^z\sigma_2^x\bigg\}\,\exp\bigg\{-i\frac{\delta\Delta_2}{4J_{\text{CR}}}\sigma_2^z\bigg\}\,\nonumber\\
&{}\qquad\qquad\times
\exp\bigg\{-i\frac{\delta\Delta_2}{4J_{\text{CR}}}\hat\sigma_1^z\sigma_2^y\bigg\}+
O\!\left(\frac{\delta \Delta_2^2}{J_{\text{CR}}^2}\right) \nonumber\\
&{}\approx G_{\text{CR}}\exp\bigg\{-i\frac{\delta\Delta_2}{4J_{\text{CR}}}\sigma_2^z\bigg\}
\exp\bigg\{-i\frac{\delta\Delta_2}{4J_{\text{CR}}}\hat\sigma_1^z\sigma_2^y\bigg\},
\end{align}
yielding 
\begin{equation}
 {\cal U}_{\text{CR}}=  G_{\text{CR}} \exp\bigg\{-i\frac{\delta\Delta_2}{4J_{\text{CR}}}\hat\sigma_1^z\sigma_2^y\bigg\}.
\end{equation} 
 when the  single-qubit error is corrected. The two-qubit gate error is thus
\begin{equation}
1-\left\vert\frac{\mathrm{Tr}\big\{G_{\text{CR}}^{\dagger}{\mathcal{U}_{\text{CR}}}\big\}}{4}\right\vert^2\approx \left(\frac{\delta\Delta_2}{4J_{\text{CR}}}\right)^2.
\end{equation}
The CR gate is therefore much less robust than the adiabatic $ZZ$ gate, since the fluctuation is normalized by the small coupling $J_{\text{CR}}$ rather than by the qubit frequency, like in (\ref{gate_error_ZZ_freq_fluctuation_SM}). Assuming $J_{\text{CR}}/2\pi = 10$ MHz and $\Delta_2/2\pi= 7$ GHz, the longitudinal gate is $\sim 5\times10^4$ times more resilient against frequency noise.

\subsection{Derivation of the product formula}
We derive here the product formula used in the section above. Let's consider $\mathcal{U}=e^{-i\alpha B-i\beta A}$ with $A=\sigma_1^z\sigma_2^x$, $B=\sigma_2^z$, and $\alpha,\beta\in\mathbb{R}$. We note that $A$ and $B$ anticommute and $AB=-iC$, with $C=\sigma_1^z\sigma_2^y$. Therefore we have that
\begin{equation}
e^{is\beta A}B\, e^{-is\beta A}
=
B\cos(2s\beta)+C\sin(2s\beta).
\end{equation}
The Baker-Campbell-Hausdorff expansion to first order in $\alpha$ yields
\begin{align}
&{}e^{-i\alpha B-i\beta A} \nonumber\\
&{} = e^{-i\beta A}
\exp\left[
-i\alpha
\int_0^1
e^{is\beta A}B\, e^{-is\beta A}\,ds
\right]
+
O(\alpha^2)\nonumber\\
&{}\approx e^{-i\beta A} 
\exp\left[-
i\frac{\alpha}{2\beta}
\Big( B\sin(2\beta)-C\cos(2\beta)+C \Big)
\right].
\end{align}
For our case, $\beta=\pi/4$ and
\begin{align}
&{}\exp\bigg\{-i\alpha \sigma_2^z-i\frac{\pi}{4}\hat\sigma_1^z\sigma_2^x\bigg\} \approx
\exp\bigg\{-i\frac{\pi}{4}\hat\sigma_1^z\sigma_2^x\bigg\}\nonumber\\
&{}\times
\exp\bigg\{
-i\frac{2\alpha}{\pi}
\left(
\sigma_2^z+\hat\sigma_1^z\sigma_2^y
\right)
\bigg\}
\end{align}
or, as a product of Pauli exponentials to first order,
\begin{align}
&{}\exp\bigg\{-i\alpha \sigma_2^z-i\frac{\pi}{4}\hat\sigma_1^z\sigma_2^x\bigg\} \approx
\exp\bigg\{-i\frac{\pi}{4}\hat\sigma_1^z\sigma_2^x\bigg\}\nonumber\\
&{}\times
\exp\bigg\{-
i\frac{2\alpha}{\pi}
\sigma_2^z
\bigg\}\exp\bigg\{-
i\frac{2\alpha}{\pi}\hat\sigma_1^z\sigma_2^y
\bigg\}+
O(\alpha^2).
\end{align}
with $\alpha=\pi\delta\Delta_2/(8J_{\text{CR}})$ we obtain Eq.~\eqref{eq:ucr_product}.

\section{Static Hamiltonian}
Even in the undriven register, the qubits are dressed by their interaction with the cavity. The basis of these dressed states is the relevant computational basis in experiments and is also used in all numerical simulations in this work. In this basis residual $ZZ$ crosstalk between qubits remains. While in Sec.~\ref{app:dress} we obtain the effective dispersive basis for a multi-qubit system, a detailed analysis of the crosstalk, including the one originating from higher qubit levels, is presented in Sec.~\ref{app:crosstalk}.

\subsection{The dressed basis}\label{app:dress} 
Consider the system Hamiltonian Eq.~(\ref{HamSysNQ}) without the drives,
\begin{align}
\label{HamSysNQ_undriven_SM}
	& H =  H_0+V,\nonumber \\
    &H_0=\sum_{k=1}^N \frac{\Delta_k}{2}\sigma_k^z +\omega_c a^\dagger a; \quad V= \sum_{k=1}^N g_{k}(a+a^\dagger)\sigma_k^x.
\end{align}
In the dispersive regime where $g_k\ll \vert\Delta_k-\omega_c\vert$, the dressed basis can be obtained from successive SW transformations.
When the cavity is eliminated dispersively by the SW transformation
$e^{S}$ with
\begin{align}
    S&=\sum_k \bigg\{ \frac{g_k}{\delta_{kc}}\left(   a\sigma_k^+ -a^\dagger\sigma_k^-\right)  + \frac{g_k}{\Sigma_{kc}}\left( a^\dagger\sigma_k^+ - a\sigma_k^-\right) \bigg\}
\end{align}\vspace{-15pt}\\\noindent
and
\begin{equation}
    \delta_{kc}=\Delta_k-\omega_c, \quad \Sigma_{kc}=\Delta_k+\omega_c,
\end{equation}
each  qubit frequency is shifted to
\begin{equation}
\tilde\Delta_k=\Delta_k +g_{k}^2\left( \frac{1}{\delta_{kc}}+\frac{1}{\Sigma_{kc}}\right),
\end{equation}
 while a pairwise effective $XX$ interaction arises with coupling strength
\begin{equation}
\label{Jjk_supp}
J_{jk}^{\perp}=\frac{g_{j}g_{k}}{2}\left(\frac{1}{\delta_{jc}}+\frac{1}{\delta_{kc}}-\frac{1}{\Sigma_{jc}}-\frac{1}{\Sigma_{kc}}\right).
\end{equation}
The qubit register is then described by the Hamiltonian, to 2nd order in $g_k$,
\begin{equation}
\label{H_NQ_SW_undriven_SM}
    H_1\simeq  \sum_{k=1}^N \frac{\tilde\Delta_k}{2}\sigma_k^z + \omega_c a^\dagger a + \sum_{j<k} J_{jk}^{\perp}\sigma_j^x\sigma_k^x.
\end{equation} 
Assuming that the qubits are sufficiently detuned so that the $XX$ interaction is off-resonance, i.e., $J_{jk}^{\perp}\ll\vert \Delta_j-\Delta_k\vert$, this interaction can be  eliminated by a 2nd SW transformation  $e^{S_{XX}}$
with
\begin{align}
\label{SW_transformation_undriven_H_SM}
{S}_{XX}=-\sum_{j<k}\bigg\{& \frac{J_{jk}^{\perp}}{\delta_{jk}}\left(\sigma_j^+\sigma_k^--\sigma_j^-\sigma_k^+\right)  \nonumber \\&+\frac{J_{jk}^{\perp}}{\Sigma_{jk}}\left(\sigma_j^+\sigma_k^+-\sigma_j^-\sigma_k^-\right)\bigg\},
\end{align}
where $\delta_{jk}=\Delta_j-\Delta_k$ and $\Sigma_{jk}=\Delta_j+\Delta_k$, yielding the dressed qubit frequencies 
\begin{equation}
\tilde{\tilde\Delta}_j=\tilde\Delta_j+\sum_{k\neq j}(J_{jk}^{\perp})^2\left(\frac{1}{\delta_{jk}}+\frac{1}{\Sigma_{jk}}\right),
\end{equation}
to 4th order in $g_k$. 
In addition, the second order term of this SW expansion gives rise to three-body effective $ZXX$ interaction
\begin{align}
    \frac{1}{2}&{}\bigg[S_{XX},\sum_{j<k}J_{jk}^{\perp}\sigma_j^x\sigma_k^x\bigg]=\sum_{j}\sum_{\substack{k<l \\ k,l\neq j}} \mu_{j;kl}\sigma_j^z\sigma_k^x\sigma_l^x,
\end{align}
where  
\begin{equation}
    \mu_{j;kl} =\frac{1}{2}J_{jk}^{\perp}J_{jl}^{\perp} \left(\frac{1}{\delta_{jk}}+\frac{1}{\Sigma_{jk}} + \frac{1}{\delta_{jl}}+\frac{1}{\Sigma_{jl}}\right),
\end{equation}
which can be further eliminated by a third SW transformation $e^{S_{ZXX}}$, where
\begin{align}
S_{ZXX} = \sum_{j} \sum_{\substack{k<l \\ k,l\neq j}} &{}\, \mu_{j;kl}\,\sigma_j^z \Bigg[\frac{1}{\delta_{kl}}\left( \sigma_k^+\sigma_l^- - \sigma_k^-\sigma_l^+\right) \nonumber\\
&{}
+\frac{1}{\Sigma_{kl}}\left(\sigma_k^+\sigma_l^+ - \sigma_k^-\sigma_l^-\right)\Bigg].
\end{align}
The dressed basis can thus be obtained from the basis of the non-interacting Hamiltonian $H_0$ by the total unitary transformation
\begin{equation}
    e^{S_{ZXX}}e^{S_{XX}}e^S.
\end{equation}
In this frame each qubit has a well defined shifted frequency, with a residual longitudinal $\sigma_j^z \sigma_k^z$ coupling on the order of $O(g_k^6)$.  This represents the $ZZ$ crosstalk caused by the interaction between a qubit and other qubits in the cavity, which we analyse in more detail below.

\subsection{Residual static ZZ crosstalk}
\label{app:crosstalk}
Here we proceed with the evaluation of the static ZZ crosstalk present in the undriven register using both perturbation theory and exact numerical diagonalization. The crosstalk discussed in section~\ref{app:dress} does not take into account virtual transitions which have been adiabatically eliminated before the basis transformation Eq.~(\ref{SW_transformation_undriven_H_SM}), or virtual transitions involving higher qubit levels. Thus, we proceed with the evaluation of the residual crosstalk using perturbation theory. The crosstalk between a pair of qubits is given by 
\begin{equation}
\label{zeta_def_SM}
    \zeta_{jk} = (E_{11} - E_{01} ) - (E_{10} - E_{00}),
\end{equation}
where $E_{s}$ are the  system energies when the pair $(j,k)$ are in the states $\{00,01,10,11\}$, with all the other qubits and cavity in the ground state. This quantity then yields an effective static $J_{jk,\,\text{cr.}}^{zz}\sigma_j^z\sigma_k^z$ interaction, with 
\begin{equation}
\label{JZZ_static_def_SM}
    J_{jk,\,\text{cr.}}^{zz} = \zeta_{jk}/4.
\end{equation}
Treating the coupling term $V$ in
Hamiltonian~(\ref{HamSysNQ_undriven_SM}) as a perturbation with $g_{k}$ as the small parameter, we can use fourth-order perturbation theory
\cite{krishnanApproximateFourthorderPerturbation1978} to find the correction
terms to the energies,
\begin{equation}
    E_s = \sum_{n=0}^4 E_s^{(n)},
\end{equation}
with
\begin{align}
\label{energy_corrections_pert_theory_SM}
E_s^{(1)} &{}= V_{ss} \equiv 0;\nonumber\\
E_s^{(2)} &{}= \sum_{r\neq s} \frac{ |V_{sr}|^2 }{\delta E_{sr} }, \nonumber\\
E_s^{(3)} &{}= \sum_{r,q \neq s} \frac{ V_{sr}V_{rq}V_{qs} }{ \delta E_{sr} \delta E_{sq}}\equiv 0,\nonumber\\
E_s^{(4)} &{}= \sum_{r,q,w \neq s} \frac{ V_{sr}V_{rq}V_{qw}V_{ws} }{ \delta E_{sr} \delta E_{sq} \delta E_{sw} } - \sum_{r,q \neq s} \frac{ |V_{sr}|^2 |V_{sq}|^2 }{\delta E_{sr}^2 \delta E_{sq} },
\end{align}
where $V_{sr} = \bra{s}V\ket{r}$, $\delta E_{sr} = E_s^{(0)} - E_r^{(0)}$. Here, $\ket{r},\ket{q},\ket{w}$ denote the eigenstates of the unperturbed Hamiltonian. The odd orders vanish because $V$ changes the photon number by one. The processes that contribute to Eq.~(\ref{zeta_def_SM}) are those in which
each of the two qubits is acted upon exactly twice, i.e.\ each of them emits and
reabsorbs one virtual photon. Processes in which a single qubit is acted upon
four times produce a one-qubit energy shift of the form
$c_j\sigma^z_{j}+c_k\sigma_k^z$, which cancels identically in the
combination (\ref{zeta_def_SM}). At fourth order, no spectator qubit of the register can
contribute, the leading spectator correction to $\zeta$ is
$\mathcal{O}(g_{k}^6)$ (see below).

Including the anharmonicity at the third level of each qubit, i.e. $E_{\ket{2}_k} = 3\Delta_k/2 + \chi_k$, with the coupling connecting the second and third level being $g^{(1\to2)}_{k}=g'_{k}$, the bare energies of the states relevant here are
\begin{align}
\label{bare_energies_SM}
&{}E_{(0,0,n_c)}^{(0)} = -\frac{\Delta_j}{2}-\frac{\Delta_k}{2} + n_c\omega_c, \nonumber\\
&{}E_{(1,0,n_c)}^{(0)} = \frac{\Delta_j}{2}-\frac{\Delta_k}{2} + n_c\omega_c, \nonumber\\
&{}E_{(0,1,n_c)}^{(0)} = -\frac{\Delta_j}{2}+\frac{\Delta_k}{2} + n_c\omega_c, \nonumber\\
&{}E_{(1,1,n_c)}^{(0)} = \frac{\Delta_j}{2}+\frac{\Delta_k}{2} + n_c\omega_c, \nonumber\\
&{}E_{(2,0,n_c)}^{(0)} = \frac{3\Delta_j}{2}+\chi_j-\frac{\Delta_k}{2} + n_c\omega_c, \nonumber\\
&{}E_{(0,2,n_c)}^{(0)} = -\frac{\Delta_j}{2}+\frac{3\Delta_k}{2}+\chi_k + n_c\omega_c, \nonumber\\
&{}E_{(2,1,n_c)}^{(0)} = \frac{3\Delta_j}{2}+\chi_j+\frac{\Delta_k}{2} + n_c\omega_c, \nonumber\\
&{}E_{(1,2,n_c)}^{(0)} = \frac{\Delta_j}{2}+\frac{3\Delta_k}{2}+\chi_k + n_c\omega_c.
\end{align}
At second order each of the four computational states is shifted by
\begin{align}
\label{second_order_SM}
&{}E_{(0,0,0)}^{(2)}  = -\sum_{l=j,k}\frac{ g_{l}^2 }{\Delta_l + \omega_c }, \nonumber\\
&{}E_{(1,0,0)}^{(2)}  = \frac{ g_{j}^2 }{\Delta_j - \omega_c}
                             -\frac{ g_{k}^2 }{\Delta_k + \omega_c }
                             -\frac{ g'^2_{j} }{\Delta_j +\chi_j+ \omega_c}
                             , \nonumber\\\nonumber\\
&{}E_{(0,1,0)}^{(2)}  = \frac{ g_{k}^2 }{\Delta_k - \omega_c}
                             -\frac{ g_{j}^2 }{\Delta_j + \omega_c }
                             -\frac{ g'^2_{k} }{\Delta_k +\chi_k+ \omega_c}
                             , \nonumber\\\nonumber\\
&{}E_{(1,1,0)}^{(2)}  = \sum_{l=j,k}\frac{ g_{l}^2 }{\Delta_l - \omega_c}
                             -\sum_{l=j,k}\frac{ g'^2_{l} }{\Delta_l +\chi_l+ \omega_c}.
\end{align}
The second order energy corrections are merely single-qubit shifts and lead to no $ZZ$ crosstalk, $\zeta^{(2)} = 0$, thus a fourth order correction is required. At fourth order the terms contributing to the energy corrections (\ref{energy_corrections_pert_theory_SM}) increase the qubits excitations by at most two when only one qubit is involved in the virtual transitions, and by one excitation when two qubits are involved. Keeping only the two-qubit contributions (the one-qubit ones cancel in
Eq.~(\ref{zeta_def_SM})), we write:
\begin{widetext}
\begin{align}
\label{E4_00_SM}
E_{(0,0,0)}^{(4)} =&{}\ -\frac{2\,g_{j}^2g_{k}^2}{(\Delta_j+\omega_c)(\Delta_k+\omega_c)}
\left[\frac{1}{\omega_c}
+\frac{(\Delta_j+\Delta_k+2\omega_c)(\Delta_j+\Delta_k+\omega_c)}
{(\Delta_j+\omega_c)(\Delta_k+\omega_c)(\Delta_j+\Delta_k)}\right],
\\[4pt]
\label{E4_10_SM}
E_{(1,0,0)}^{(4)} =&{}\ -\frac{2\,g_{j}^2g_{k}^2}{(\omega_c-\Delta_j)(\Delta_k+\omega_c)}
\left[\frac{1}{\omega_c}
+\frac{(\Delta_k-\Delta_j+2\omega_c)(\Delta_k-\Delta_j+\omega_c)}
{(\omega_c-\Delta_j)(\Delta_k+\omega_c)(\Delta_k-\Delta_j)}\right] \nonumber\\
&{}\ -\frac{2\,g'^2_{j}g_{k}^2}{(\Delta_j+\chi_j+\omega_c)(\Delta_k+\omega_c)}
\left[\frac{1}{\omega_c}
+\frac{(\Delta_j+\chi_j+\Delta_k+2\omega_c)(\Delta_j+\chi_j+\Delta_k+\omega_c)}
{(\Delta_j+\chi_j+\omega_c)(\Delta_k+\omega_c)(\Delta_j+\chi_j+\Delta_k)}\right],
\\[4pt]
\label{E4_01_SM}
E_{(0,1,0)}^{(4)} =&{}\ -\frac{2\,g_{j}^2g_{k}^2}{(\Delta_j+\omega_c)(\omega_c-\Delta_k)}
\left[\frac{1}{\omega_c}
+\frac{(\Delta_j-\Delta_k+2\omega_c)(\Delta_j-\Delta_k+\omega_c)}
{(\Delta_j+\omega_c)(\omega_c-\Delta_k)(\Delta_j-\Delta_k)}\right] \nonumber\\
&{}\ -\frac{2\,g_{j}^2g'^2_{k}}{(\Delta_j+\omega_c)(\Delta_k+\chi_k+\omega_c)}
\left[\frac{1}{\omega_c}
+\frac{(\Delta_j+\Delta_k+\chi_k+2\omega_c)(\Delta_j+\Delta_k+\chi_k+\omega_c)}
{(\Delta_j+\omega_c)(\Delta_k+\chi_k+\omega_c)(\Delta_j+\Delta_k+\chi_k)}\right],
\\[4pt]
\label{E4_11_SM}
E_{(1,1,0)}^{(4)} =&{}\ -\frac{2\,g_{j}^2g_{k}^2}{(\omega_c-\Delta_j)(\omega_c-\Delta_k)}
\left[\frac{1}{\omega_c}
+\frac{(2\omega_c-\Delta_j-\Delta_k)(\omega_c-\Delta_j-\Delta_k)}
{(\omega_c-\Delta_j)(\omega_c-\Delta_k)\,(-\Delta_j-\Delta_k)}\right] \nonumber\\
&{}\ -\frac{2\,g_{j}^2g'^2_{k}}{(\omega_c-\Delta_j)(\Delta_k+\chi_k+\omega_c)}
\left[\frac{1}{\omega_c}
+\frac{(\Delta_k+\chi_k-\Delta_j+2\omega_c)(\Delta_k+\chi_k-\Delta_j+\omega_c)}
{(\omega_c-\Delta_j)(\Delta_k+\chi_k+\omega_c)(\Delta_k+\chi_k-\Delta_j)}\right] \nonumber\\
&{}\ -\frac{2\,g'^2_{j}g_{k}^2}{(\Delta_j+\chi_j+\omega_c)(\omega_c-\Delta_k)}
\left[\frac{1}{\omega_c}
+\frac{(\Delta_j+\chi_j-\Delta_k+2\omega_c)(\Delta_j+\chi_j-\Delta_k+\omega_c)}
{(\Delta_j+\chi_j+\omega_c)(\omega_c-\Delta_k)(\Delta_j+\chi_j-\Delta_k)}\right] \nonumber\\
&{}\ -\frac{2\,g'^2_{j}g'^2_{k}}{(\Delta_j+\chi_j+\omega_c)(\Delta_k+\chi_k+\omega_c)}
\left[\frac{1}{\omega_c}
+\frac{(\Delta_j+\chi_j+\Delta_k+\chi_k+2\omega_c)(\Delta_j+\chi_j+\Delta_k+\chi_k+\omega_c)}
{(\Delta_j+\chi_j+\omega_c)(\Delta_k+\chi_k+\omega_c)(\Delta_j+\chi_j+\Delta_k+\chi_k)}\right],
\end{align}
\end{widetext}
and finally
\begin{equation}
\label{zeta4_SM}
    \zeta^{(4)}_{jk}
    = E_{(1,1,0)}^{(4)} + E_{(0,0,0)}^{(4)}
    - E_{(0,1,0)}^{(4)} - E_{(1,0,0)}^{(4)}.
\end{equation}
In the two-level limit $\chi_{j},\chi_k\to\infty$ all terms containing $g'_{j}$ or $g'_{k}$ vanish, and Eqs.~(\ref{E4_00_SM})--(\ref{zeta4_SM}) collapse to the closed form
\begin{align}
\label{zeta4_two_level_SM}
&{}\zeta^{(4)}_{jk}\Big|_{\chi\to\infty} \nonumber\\
&{}= -\frac{ 8g_{j}^2g_{k}^2\,\Delta_j\Delta_k \left( 
\Delta_j^2\Delta_k^2 
-3\Delta_j^2\omega_c^2
-3\Delta_k^2\omega_c^2
+5\omega_c^4
\right)
}{
\omega_c\,
\delta_{jc}^2
\delta_{kc}^2
\Sigma_{jc}^2
\Sigma_{kc}^2
},
\end{align}
with $\delta_{kc}=\Delta_k-\omega_c$ and $\Sigma_{kc}=\Delta_k+\omega_c$. For finite anharmonicity no comparable simplification occurs, and Eqs.~(\ref{E4_00_SM})--(\ref{zeta4_SM}) are already the closed form.

\begin{figure}[t]
    \centering
    \includegraphics[width=0.9\linewidth]{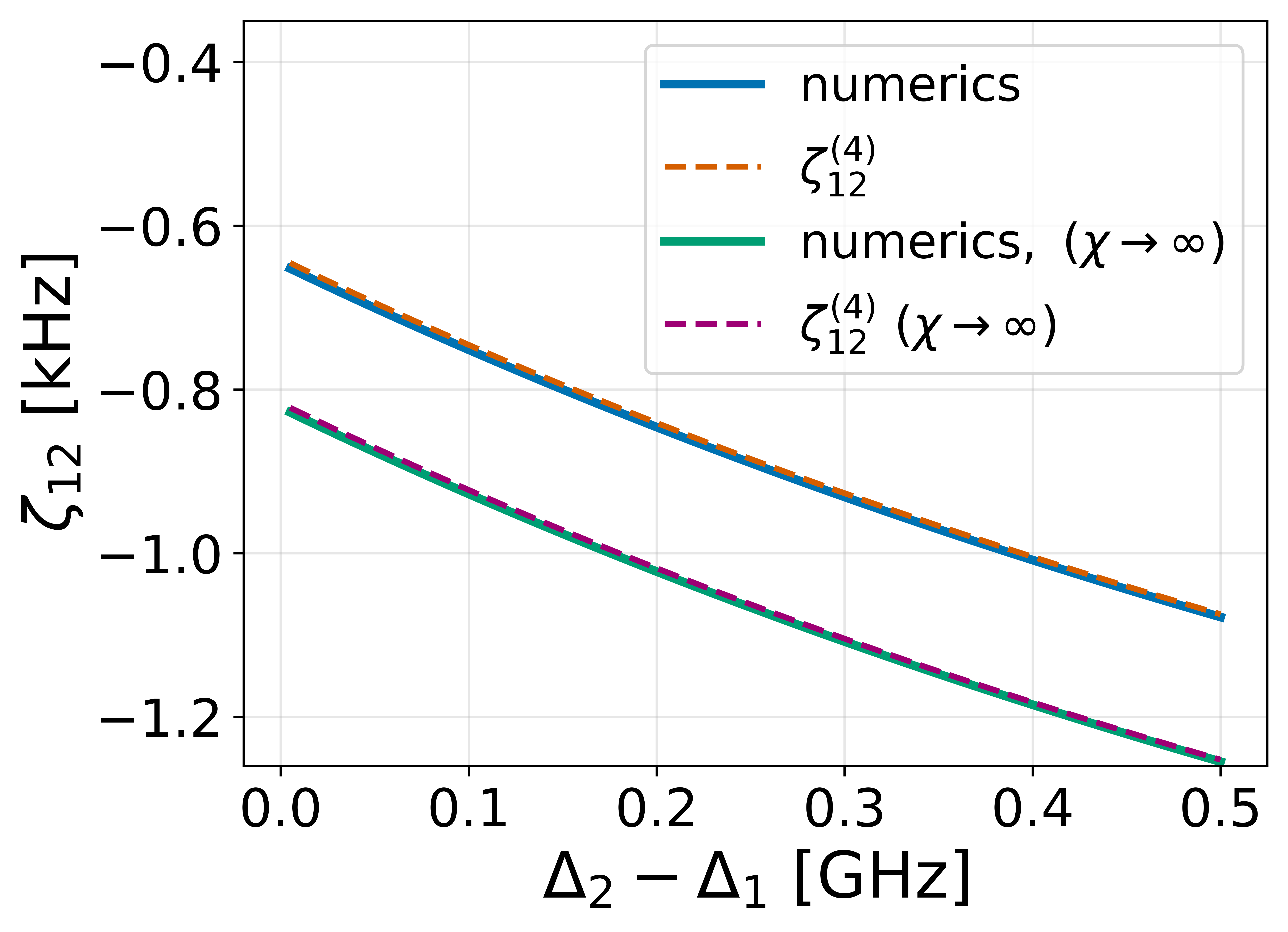}
    \caption{\label{fig_zeta2Q_SM}Residual $ZZ$ crosstalk for 2 qubits obtained both numerically and using Eq.~(\ref{zeta4_SM}) at finite anharmonicity $\chi_j=\chi_k=2\pi\times25~\mathrm{GHz}$, compared with the two-level limit, Eq.~(\ref{zeta4_two_level_SM}). Parameters: $\Delta_1/2\pi=7.0~\mathrm{GHz}$, $\omega_c/2\pi=3.0~\mathrm{GHz}$, $g_{j}/2\pi=g_{k}/2\pi=  100~\mathrm{MHz}$, $g'_{k}=0.48\,g_{k}$.}
\end{figure}

The two-level limit is not quantitatively accurate even for the large anharmonicities of flux qubits. The reason is that the denominators of the primed terms are controlled by $\Delta_k+\chi_k+\omega_c$, so that the relevant comparison is between $\chi_k$ and the qubit-cavity detuning rather than between $\chi_k$ and the qubit-qubit detuning $\delta_{jk}$. For the parameters used here, $\omega_c/2\pi=3.0~\mathrm{GHz}$, $g_{k}/2\pi=100~\mathrm{MHz}$ and setting $\chi_k/2\pi \sim 25~\mathrm{GHz}$, we have $g_{k}' \sim 0.48\,g_k$ across the $7$--$8~\mathrm{GHz}$ band, which partially cancels the two-level result (\ref{zeta4_two_level_SM}), as shown in Fig.~\ref{fig_zeta2Q_SM}.

A further contribution arises from the presence of spectator qubits. It is absent at fourth order, as it can be noted from Eqs.~(\ref{E4_00_SM})--(\ref{zeta4_SM}), and first appears at sixth order in $g_{k}$. It is convenient to estimate it after the adiabatic elimination of the cavity, in terms of the effective exchange $J_{jk}^\perp$ of Eq.~(\ref{Jjk_supp}), i.e. treating the effective qubit-qubit coupling,
\begin{equation}
 V_{qq} \sim \sum_{j<k} J_{jk}^{\perp}\,\sigma_j^x\sigma_k^x,
\end{equation}
perturbatively with respect to $H_0$. The term generates no $ZZ$ at second order, the four shifts cancelling pairwise in Eq.~(\ref{zeta_def_SM}); a nonzero result first appears at third order, whenever $V_{qq}^3$ flips every qubit an even number of times. Summing over the six time orderings and over all spectators, 
\begin{equation}
\label{zeta_spectator_SM}
\tilde\zeta_{jk}^{(6)} = \sum_{l\neq j,k}\frac{16J_{jk}^{\perp}J_{jl}^{\perp}J_{lk}^{\perp}\,\Delta_j\Delta_k}{\delta_{jl}\Sigma_{jl}\,\delta_{kl}\Sigma_{kl}
},
\end{equation}
with $\delta_{jl}=\Delta_j-\Delta_l$ and $\Sigma_{jl}=\Delta_j+\Delta_l$. This result should be read as the leading estimate of the sixth-order three-body coupling rather than as its complete expression, since we have neglected the cavity-mediated processes of the same order.\\
The total residual crosstalk between the pair $(j,k)$ is therefore
\begin{equation}
\label{zeta_total_SM}
    \zeta_{jk} = \underbrace{\zeta^{(4)}_{jk}}_{\mathcal{O}(g_{k}^4)}
    + \underbrace{\tilde\zeta^{(6)}_{jk}}_{\mathcal{O}(g_{k}^6)} + \dots ,
\end{equation}
where the superscripts denote the order in the qubit--cavity coupling $g_k$.
\begin{figure}[t]
    \centering
    \includegraphics[width=0.9\linewidth]{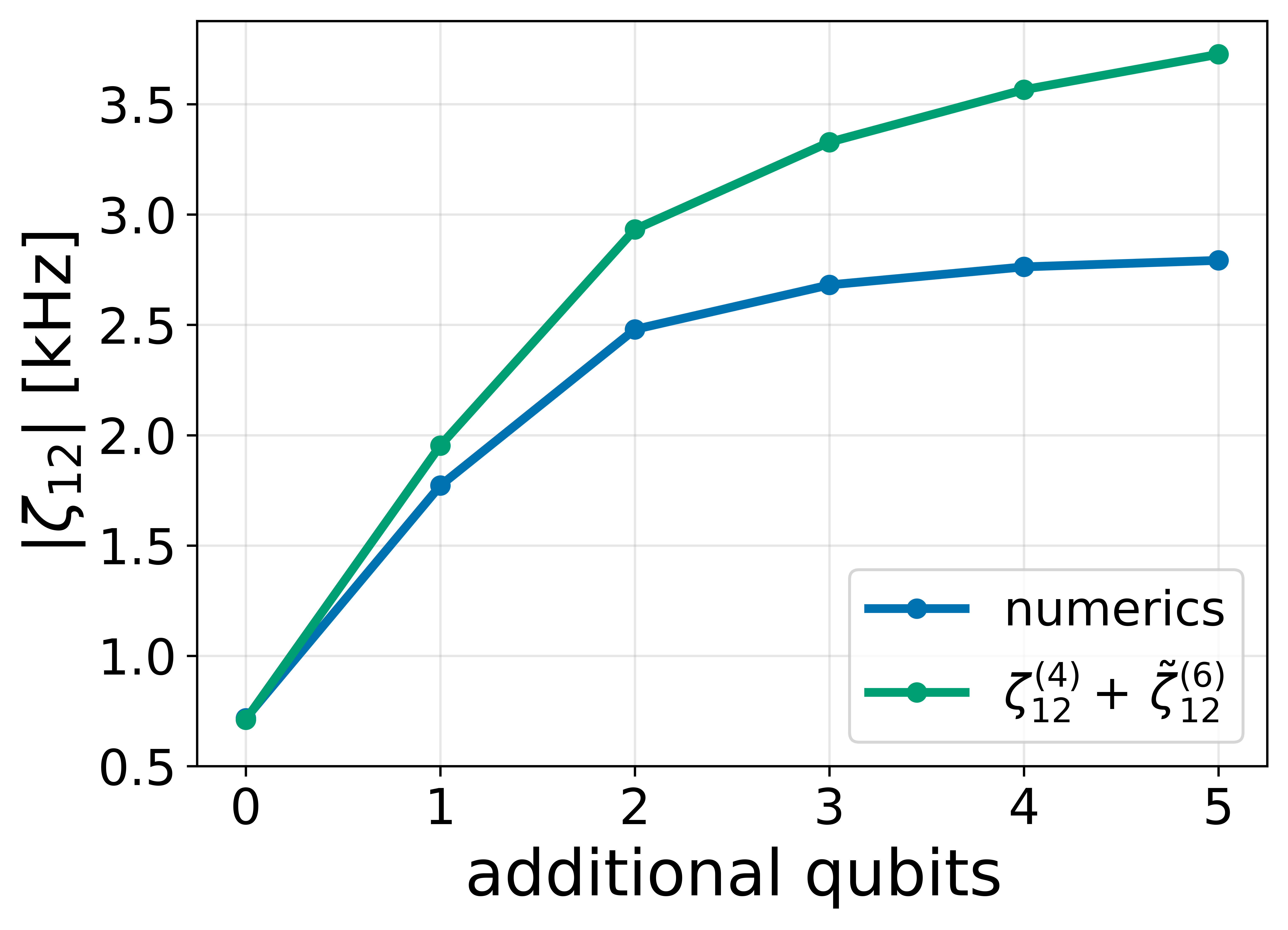}\caption{\label{fig_speccross}Residual $ZZ$ crosstalk between two qubits with $\Delta_1/2\pi = 7.00\ \text{GHz}$ and $\Delta_2 - \Delta_1 = 2\pi \times 50\ \text{MHz}$, when spectator qubits at $\Delta_k = \Delta_2 + k\,\delta$, with $\delta/2\pi = 50\ \text{MHz}$, $k\geq 1$ are added. Other parameters: $\chi_k/2\pi \sim 25\ \text{GHz}$, $g_k'/g_k \sim 0.48$, and $g_k/2\pi = 100\ \text{MHz}$ for all the qubits $k=1,\dots 5$; $\omega_c/2\pi= 3.0\ \text{GHz}$.}
\end{figure}

For the $20$-qubit register considered in the discussion, with frequencies spanning $7$--$8~\mathrm{GHz}$ in steps of $50~\mathrm{MHz}$, Eq.~(\ref{zeta_total_SM}) gives  a worst-case value of $\zeta/2\pi\sim 6.67~\mathrm{kHz}$ over all pairs. Over a gate duration $T=55~\mathrm{ns}$ this residual coupling accumulates a phase of $2\pi(\zeta\, T/4)\sim 6\times10^{-4}$ rad, i.e.\ an infidelity of order $3\times10^{-7}$.

\section{Multi-core architecture with interconnected processors}
\label{app:multicore}
The discussed architecture is inherently scalable through a modular multi-core design in which cavity buses belonging to different processor cores are linked by tunable coupling resonators. The coupling between the cores is controlled by tuning the frequency of the coupling resonator. As an illustrative example, consider two cavity buses with frequencies $\omega_1^c$ and $\omega_2^c$, respectively. These cavities are coupled both directly to one another and indirectly through a frequency-tunable coupling resonator with frequency $\omega_r$. The system Hamiltonian may be written as
\begin{equation}
\begin{aligned}
H={}&
\omega_1^c a_1^\dagger a_1
+\omega_2^c a_2^\dagger a_2
+\omega_r a_r^\dagger a_r \\
&+\Lambda_{1r}
\left(a_1^\dagger a_r+a_1a_r^\dagger\right)
+\Lambda_{2r}
\left(a_2^\dagger a_r+a_2a_r^\dagger\right) \\
&+\Lambda_{12}
\left(a_1^\dagger a_2+a_1a_2^\dagger\right).
\end{aligned}
\end{equation}
Here, $\Lambda_{1r}$, $\Lambda_{2r}$, and $\Lambda_{12}$ denote the corresponding coupling strengths.

Applying a Schrieffer--Wolff transformation $e^{-S}$, where
\begin{equation}
S=
\frac{\Lambda_{1r}}{\omega_1^c-\omega_r}
\left(a_1^\dagger a_r-a_1a_r^\dagger\right)
+
\frac{\Lambda_{2r}}{\omega_2^c-\omega_r}
\left(a_2^\dagger a_r-a_2a_r^\dagger\right),
\end{equation}
yields an effective interaction between the two cavity buses of the form
\begin{equation}
V_{\mathrm{eff}}
=
\Lambda_{\mathrm{eff}}
\left(a_1^\dagger a_2+a_1a_2^\dagger\right),
\end{equation}
with effective coupling strength
\begin{equation}
\Lambda_{\mathrm{eff}}
=
\frac{\Lambda_{1r}\Lambda_{2r}}{2}
\left(
\frac{1}{\omega_1^c-\omega_r}
+
\frac{1}{\omega_2^c-\omega_r}
\right)
+\Lambda_{12}.
\end{equation}

The interaction strength between the two bus resonators can be tuned by varying the coupling-resonator frequency $\omega_r$. Importantly, the interaction may be completely suppressed by parking the coupling resonator at a frequency for which $\Lambda_{\mathrm{eff}}=0$. This provides a controllable mechanism for selectively turning the inter-core coupling on or off.

For superconducting implementations with representative parameters
$
\Lambda_{1r}/2\pi\sim\Lambda_{2r}/2\pi\sim500~\mathrm{MHz}$, $
\omega_1^c/2\pi\sim\omega_2^c/2\pi\sim 3~\mathrm{GHz},
$
and a coupling resonator frequency
$
\omega_r/2\pi\sim 3.5~\mathrm{GHz},
$
the effective inter-core coupling strength is on the order of
$
\Lambda_{\mathrm{eff}}/2\pi \sim -500~\mathrm{MHz},
$
assuming that $\Lambda_{12}$ is negligible on this scale.

With this effective coupling, the modes of the two bus resonators hybridize into two nonlocal modes, thereby enabling long-range interactions between computational qubits located on different processor cores. Assuming $\omega_1^c\approx \omega_2^c=\omega_c$,  the two hybridized eigenmodes have frequencies
\begin{equation}
\omega_c^{\pm}=\omega_c\pm \Lambda_{\rm{eff}},
\end{equation}
and annihilation operators $c_\pm=(a_1\pm a_2)/\sqrt{2}$.
As these modes have support on both resonators, the qubits residing on different processor cores are now  coupled to the shared non-local modes, which mediate the  effective $ZZ$ interaction when the periodic adiabatic drives are applied on the qubits.

Consider, for example, two  qubits with frequencies $\Delta_1$ and $\Delta_2$, driven by adiabatic drives $\epsilon_1(t)$ and $\epsilon_2(t)$ with amplitudes $A_1$ and $A_2$, and coupled to two local cavity buses with strength $g_1$ and $g_2$, respectively. Assuming $\Lambda_{\rm{eff}}\ll \omega_c$, the strength of the effective inter-processor $ZZ$ interaction is
\begin{equation}
J_{12,\,\text{eff.}}^{\text{inter}}=\frac{\theta_{12}^{\text{inter}}(T)}{T}
\approx
\frac{\Lambda_{\mathrm{eff}}}{\omega_c}\frac{g_1g_2\kappa_1\kappa_2}{\omega_c}\sum_{m}C_1^{(m)}C_2^{(m)},
\end{equation}
which differs from the intra-processor effective coupling (\ref{g_ZZ_ij_matching_freqs}) by a factor of 
$(\Lambda_{\mathrm{eff}}/\omega_c)$ approximately, and is obtained by summing the contributions of the two hybridized nonlocal modes. This effective coupling enables the implementation of a $ZZ$ entangling gate  between computational qubits residing on different processor cores. 

For representative parameters used in the main text, 
$A_1\sim A_2\sim2\pi\times8~\mathrm{GHz}$, $g_1\sim g_2\sim2\pi\times100~\mathrm{MHz}$,
$\omega_1^c\sim\omega_2^c\sim2\pi\times3~\mathrm{GHz}$, $\Delta_1\sim\Delta_2\sim2\pi\times7~\mathrm{GHz}$, and an effective inter-core cavities coupling $\Lambda_{\mathrm{eff}}\sim -2\pi\times 500~\mathrm{MHz}$, the resulting effective inter-core $ZZ$ interaction strength is 
\begin{equation}
    \big|J_{12,\,\text{eff.}}^{\text{inter}}\big| \sim  2\pi\times 0.38~\mathrm{MHz}.
\end{equation}
An inter-core $\pi/2$ $ZZ$ rotation between the  qubits may therefore be implemented with a duration of approximately
\begin{equation}
T\sim\frac{\pi}{4|J_{12,\,\text{eff.}}^{\text{inter}}|}\sim330~\mathrm{ns}.
\end{equation}

\begin{thebibliography}{49}%
\makeatletter
\providecommand \@ifxundefined [1]{%
 \@ifx{#1\undefined}
}%
\providecommand \@ifnum [1]{%
 \ifnum #1\expandafter \@firstoftwo
 \else \expandafter \@secondoftwo
 \fi
}%
\providecommand \@ifx [1]{%
 \ifx #1\expandafter \@firstoftwo
 \else \expandafter \@secondoftwo
 \fi
}%
\providecommand \natexlab [1]{#1}%
\providecommand \enquote  [1]{``#1''}%
\providecommand \bibnamefont  [1]{#1}%
\providecommand \bibfnamefont [1]{#1}%
\providecommand \citenamefont [1]{#1}%
\providecommand \href@noop [0]{\@secondoftwo}%
\providecommand \href [0]{\begingroup \@sanitize@url \@href}%
\providecommand \@href[1]{\@@startlink{#1}\@@href}%
\providecommand \@@href[1]{\endgroup#1\@@endlink}%
\providecommand \@sanitize@url [0]{\catcode `\\12\catcode `\$12\catcode `\&12\catcode `\#12\catcode `\^12\catcode `\_12\catcode `\%12\relax}%
\providecommand \@@startlink[1]{}%
\providecommand \@@endlink[0]{}%
\providecommand \url  [0]{\begingroup\@sanitize@url \@url }%
\providecommand \@url [1]{\endgroup\@href {#1}{\urlprefix }}%
\providecommand \urlprefix  [0]{URL }%
\providecommand \Eprint [0]{\href }%
\providecommand \doibase [0]{https://doi.org/}%
\providecommand \selectlanguage [0]{\@gobble}%
\providecommand \bibinfo  [0]{\@secondoftwo}%
\providecommand \bibfield  [0]{\@secondoftwo}%
\providecommand \translation [1]{[#1]}%
\providecommand \BibitemOpen [0]{}%
\providecommand \bibitemStop [0]{}%
\providecommand \bibitemNoStop [0]{.\EOS\space}%
\providecommand \EOS [0]{\spacefactor3000\relax}%
\providecommand \BibitemShut  [1]{\csname bibitem#1\endcsname}%
\let\auto@bib@innerbib\@empty
\bibitem [{\citenamefont {Herbert}(2020)}]{herbert2020depth}%
  \BibitemOpen
  \bibfield  {author} {\bibinfo {author} {\bibfnamefont {S.}~\bibnamefont {Herbert}},\ }\bibfield  {title} {\bibinfo {title} {On the depth overhead incurred when running quantum algorithms on near-term quantum computers with limited qubit connectivity},\ }\href {https://doi.org/10.26421/QIC20.9-10-5} {\bibfield  {journal} {\bibinfo  {journal} {Quantum Information and Computation}\ }\textbf {\bibinfo {volume} {20}},\ \bibinfo {pages} {787} (\bibinfo {year} {2020})}\BibitemShut {NoStop}%
\bibitem [{\citenamefont {Gottesman}(2014)}]{gottesmanConstantOverhead2014}%
  \BibitemOpen
  \bibfield  {author} {\bibinfo {author} {\bibfnamefont {D.}~\bibnamefont {Gottesman}},\ }\bibfield  {title} {\bibinfo {title} {Fault-tolerant quantum computation with constant overhead},\ }\href {https://doi.org/10.26421/QIC14.15-16-5} {\bibfield  {journal} {\bibinfo  {journal} {Quantum Information and Computation}\ }\textbf {\bibinfo {volume} {14}},\ \bibinfo {pages} {1338} (\bibinfo {year} {2014})}\BibitemShut {NoStop}%
\bibitem [{\citenamefont {Breuckmann}\ and\ \citenamefont {Eberhardt}(2021)}]{breuckmannQuantumLDPCCodes2021}%
  \BibitemOpen
  \bibfield  {author} {\bibinfo {author} {\bibfnamefont {N.~P.}\ \bibnamefont {Breuckmann}}\ and\ \bibinfo {author} {\bibfnamefont {J.~N.}\ \bibnamefont {Eberhardt}},\ }\bibfield  {title} {\bibinfo {title} {Quantum low-density parity-check codes},\ }\href {https://doi.org/10.1103/PRXQuantum.2.040101} {\bibfield  {journal} {\bibinfo  {journal} {PRX Quantum}\ }\textbf {\bibinfo {volume} {2}},\ \bibinfo {pages} {040101} (\bibinfo {year} {2021})}\BibitemShut {NoStop}%
\bibitem [{\citenamefont {Panteleev}\ and\ \citenamefont {Kalachev}(2022)}]{panteleevAsymptoticallyGood2022}%
  \BibitemOpen
  \bibfield  {author} {\bibinfo {author} {\bibfnamefont {P.}~\bibnamefont {Panteleev}}\ and\ \bibinfo {author} {\bibfnamefont {G.}~\bibnamefont {Kalachev}},\ }\bibfield  {title} {\bibinfo {title} {Asymptotically good quantum and locally testable classical ldpc codes},\ }in\ \href {https://doi.org/10.1145/3519935.3520017} {\emph {\bibinfo {booktitle} {Proceedings of the 54th Annual ACM SIGACT Symposium on Theory of Computing (STOC 2022)}}}\ (\bibinfo {year} {2022})\ pp.\ \bibinfo {pages} {375--388}\BibitemShut {NoStop}%
\bibitem [{\citenamefont {Cain}\ \emph {et~al.}(2026)\citenamefont {Cain}, \citenamefont {Xu}, \citenamefont {King}, \citenamefont {Picard}, \citenamefont {Levine}, \citenamefont {Endres}, \citenamefont {Preskill}, \citenamefont {Huang},\ and\ \citenamefont {Bluvstein}}]{cain2026shor}%
  \BibitemOpen
  \bibfield  {author} {\bibinfo {author} {\bibfnamefont {M.}~\bibnamefont {Cain}}, \bibinfo {author} {\bibfnamefont {Q.}~\bibnamefont {Xu}}, \bibinfo {author} {\bibfnamefont {R.}~\bibnamefont {King}}, \bibinfo {author} {\bibfnamefont {L.~R.~B.}\ \bibnamefont {Picard}}, \bibinfo {author} {\bibfnamefont {H.}~\bibnamefont {Levine}}, \bibinfo {author} {\bibfnamefont {M.}~\bibnamefont {Endres}}, \bibinfo {author} {\bibfnamefont {J.}~\bibnamefont {Preskill}}, \bibinfo {author} {\bibfnamefont {H.-Y.}\ \bibnamefont {Huang}},\ and\ \bibinfo {author} {\bibfnamefont {D.}~\bibnamefont {Bluvstein}},\ }\href {https://arxiv.org/abs/2603.28627} {\bibinfo {title} {Shor's algorithm is possible with as few as 10,000 reconfigurable atomic qubits}} (\bibinfo {year} {2026}),\ \Eprint {https://arxiv.org/abs/2603.28627} {arXiv:2603.28627 [quant-ph]} \BibitemShut {NoStop}%
\bibitem [{\citenamefont {Zhao}\ \emph {et~al.}(2026)\citenamefont {Zhao}, \citenamefont {Duckering}, \citenamefont {Gu}, \citenamefont {Maskara},\ and\ \citenamefont {Zhou}}]{zhao2026ultra}%
  \BibitemOpen
  \bibfield  {author} {\bibinfo {author} {\bibfnamefont {C.}~\bibnamefont {Zhao}}, \bibinfo {author} {\bibfnamefont {C.}~\bibnamefont {Duckering}}, \bibinfo {author} {\bibfnamefont {A.}~\bibnamefont {Gu}}, \bibinfo {author} {\bibfnamefont {N.}~\bibnamefont {Maskara}},\ and\ \bibinfo {author} {\bibfnamefont {H.}~\bibnamefont {Zhou}},\ }\href {https://arxiv.org/abs/2604.16209} {\bibinfo {title} {Towards ultra-high-rate quantum error correction with reconfigurable atom arrays}} (\bibinfo {year} {2026}),\ \Eprint {https://arxiv.org/abs/2604.16209} {arXiv:2604.16209 [quant-ph]} \BibitemShut {NoStop}%
\bibitem [{\citenamefont {Koch}\ \emph {et~al.}(2007)\citenamefont {Koch}, \citenamefont {Yu}, \citenamefont {Gambetta}, \citenamefont {Houck}, \citenamefont {Schuster}, \citenamefont {Majer}, \citenamefont {Blais}, \citenamefont {Devoret}, \citenamefont {Girvin},\ and\ \citenamefont {Schoelkopf}}]{Koch2007}%
  \BibitemOpen
  \bibfield  {author} {\bibinfo {author} {\bibfnamefont {J.}~\bibnamefont {Koch}}, \bibinfo {author} {\bibfnamefont {T.~M.}\ \bibnamefont {Yu}}, \bibinfo {author} {\bibfnamefont {J.}~\bibnamefont {Gambetta}}, \bibinfo {author} {\bibfnamefont {A.~A.}\ \bibnamefont {Houck}}, \bibinfo {author} {\bibfnamefont {D.~I.}\ \bibnamefont {Schuster}}, \bibinfo {author} {\bibfnamefont {J.}~\bibnamefont {Majer}}, \bibinfo {author} {\bibfnamefont {A.}~\bibnamefont {Blais}}, \bibinfo {author} {\bibfnamefont {M.~H.}\ \bibnamefont {Devoret}}, \bibinfo {author} {\bibfnamefont {S.~M.}\ \bibnamefont {Girvin}},\ and\ \bibinfo {author} {\bibfnamefont {R.~J.}\ \bibnamefont {Schoelkopf}},\ }\bibfield  {title} {\bibinfo {title} {Charge-insensitive qubit design derived from the {Cooper-Pair Box}},\ }\href {https://doi.org/10.1103/PhysRevA.76.042319} {\bibfield  {journal} {\bibinfo  {journal} {Physical Review A}\ }\textbf {\bibinfo {volume} {76}},\ \bibinfo {pages} {042319} (\bibinfo {year} {2007})}\BibitemShut {NoStop}%
\bibitem [{\citenamefont {Schutjens}\ \emph {et~al.}(2013)\citenamefont {Schutjens}, \citenamefont {Abu~Dagga}, \citenamefont {Egger},\ and\ \citenamefont {Wilhelm}}]{Schutjens2013}%
  \BibitemOpen
  \bibfield  {author} {\bibinfo {author} {\bibfnamefont {R.}~\bibnamefont {Schutjens}}, \bibinfo {author} {\bibfnamefont {F.}~\bibnamefont {Abu~Dagga}}, \bibinfo {author} {\bibfnamefont {D.~J.}\ \bibnamefont {Egger}},\ and\ \bibinfo {author} {\bibfnamefont {F.~K.}\ \bibnamefont {Wilhelm}},\ }\bibfield  {title} {\bibinfo {title} {Single-qubit gates in frequency-crowded transmon systems},\ }\href {https://doi.org/10.1103/PhysRevA.88.052330} {\bibfield  {journal} {\bibinfo  {journal} {Physical Review A}\ }\textbf {\bibinfo {volume} {88}},\ \bibinfo {pages} {052330} (\bibinfo {year} {2013})}\BibitemShut {NoStop}%
\bibitem [{\citenamefont {Zhao}\ \emph {et~al.}(2021)\citenamefont {Zhao}, \citenamefont {Lan}, \citenamefont {Xu}, \citenamefont {Xue}, \citenamefont {Blank}, \citenamefont {Tan}, \citenamefont {Yu},\ and\ \citenamefont {Yu}}]{zhaoSuppressionStaticZZ2021}%
  \BibitemOpen
  \bibfield  {author} {\bibinfo {author} {\bibfnamefont {P.}~\bibnamefont {Zhao}}, \bibinfo {author} {\bibfnamefont {D.}~\bibnamefont {Lan}}, \bibinfo {author} {\bibfnamefont {P.}~\bibnamefont {Xu}}, \bibinfo {author} {\bibfnamefont {G.}~\bibnamefont {Xue}}, \bibinfo {author} {\bibfnamefont {M.}~\bibnamefont {Blank}}, \bibinfo {author} {\bibfnamefont {X.}~\bibnamefont {Tan}}, \bibinfo {author} {\bibfnamefont {H.}~\bibnamefont {Yu}},\ and\ \bibinfo {author} {\bibfnamefont {Y.}~\bibnamefont {Yu}},\ }\bibfield  {title} {\bibinfo {title} {Suppression of static {$ZZ$} interaction in an all-transmon quantum processor},\ }\href {https://doi.org/10.1103/PhysRevApplied.16.024037} {\bibfield  {journal} {\bibinfo  {journal} {Physical Review Applied}\ }\textbf {\bibinfo {volume} {16}},\ \bibinfo {pages} {024037} (\bibinfo {year} {2021})}\BibitemShut {NoStop}%
\bibitem [{\citenamefont {Kandala}\ \emph {et~al.}(2021)\citenamefont {Kandala}, \citenamefont {Wei}, \citenamefont {Srinivasan}, \citenamefont {Magesan}, \citenamefont {Carnevale}, \citenamefont {Keefe}, \citenamefont {Klaus}, \citenamefont {Dial},\ and\ \citenamefont {McKay}}]{kandalaHighFidelityCNOT2021}%
  \BibitemOpen
  \bibfield  {author} {\bibinfo {author} {\bibfnamefont {A.}~\bibnamefont {Kandala}}, \bibinfo {author} {\bibfnamefont {K.~X.}\ \bibnamefont {Wei}}, \bibinfo {author} {\bibfnamefont {S.}~\bibnamefont {Srinivasan}}, \bibinfo {author} {\bibfnamefont {E.}~\bibnamefont {Magesan}}, \bibinfo {author} {\bibfnamefont {S.}~\bibnamefont {Carnevale}}, \bibinfo {author} {\bibfnamefont {G.~A.}\ \bibnamefont {Keefe}}, \bibinfo {author} {\bibfnamefont {D.}~\bibnamefont {Klaus}}, \bibinfo {author} {\bibfnamefont {O.}~\bibnamefont {Dial}},\ and\ \bibinfo {author} {\bibfnamefont {D.~C.}\ \bibnamefont {McKay}},\ }\bibfield  {title} {\bibinfo {title} {Demonstration of a high-fidelity {CNOT} gate for fixed-frequency transmons with engineered {$ZZ$} suppression},\ }\href {https://doi.org/10.1103/PhysRevLett.127.130501} {\bibfield  {journal} {\bibinfo  {journal} {Physical Review Letters}\ }\textbf {\bibinfo {volume} {127}},\ \bibinfo {pages} {130501} (\bibinfo {year} {2021})}\BibitemShut {NoStop}%
\bibitem [{\citenamefont {Magesan}\ and\ \citenamefont {Gambetta}(2020)}]{magesanEffectiveHamiltonian2020}%
  \BibitemOpen
  \bibfield  {author} {\bibinfo {author} {\bibfnamefont {E.}~\bibnamefont {Magesan}}\ and\ \bibinfo {author} {\bibfnamefont {J.~M.}\ \bibnamefont {Gambetta}},\ }\bibfield  {title} {\bibinfo {title} {Effective hamiltonian models of the cross-resonance gate},\ }\href {https://doi.org/10.1103/PhysRevA.101.052308} {\bibfield  {journal} {\bibinfo  {journal} {Physical Review A}\ }\textbf {\bibinfo {volume} {101}},\ \bibinfo {pages} {052308} (\bibinfo {year} {2020})}\BibitemShut {NoStop}%
\bibitem [{\citenamefont {Yan}\ \emph {et~al.}(2018)\citenamefont {Yan}, \citenamefont {Krantz}, \citenamefont {Sung}, \citenamefont {Kjaergaard}, \citenamefont {Campbell}, \citenamefont {Wang}, \citenamefont {Orlando}, \citenamefont {Gustavsson},\ and\ \citenamefont {Oliver}}]{yanTunableCouplingScheme2018a}%
  \BibitemOpen
  \bibfield  {author} {\bibinfo {author} {\bibfnamefont {F.}~\bibnamefont {Yan}}, \bibinfo {author} {\bibfnamefont {P.}~\bibnamefont {Krantz}}, \bibinfo {author} {\bibfnamefont {Y.}~\bibnamefont {Sung}}, \bibinfo {author} {\bibfnamefont {M.}~\bibnamefont {Kjaergaard}}, \bibinfo {author} {\bibfnamefont {D.}~\bibnamefont {Campbell}}, \bibinfo {author} {\bibfnamefont {J.~I.~J.}\ \bibnamefont {Wang}}, \bibinfo {author} {\bibfnamefont {T.~P.}\ \bibnamefont {Orlando}}, \bibinfo {author} {\bibfnamefont {S.}~\bibnamefont {Gustavsson}},\ and\ \bibinfo {author} {\bibfnamefont {W.~D.}\ \bibnamefont {Oliver}},\ }\bibfield  {title} {\bibinfo {title} {A tunable coupling scheme for implementing high-fidelity two-qubit gates},\ }\href {https://doi.org/10.1103/PhysRevApplied.10.054062} {\bibfield  {journal} {\bibinfo  {journal} {Phys. Rev. Applied}\ }\textbf {\bibinfo {volume} {10}},\ \bibinfo {pages} {054062} (\bibinfo {year} {2018})},\ \Eprint {https://arxiv.org/abs/1803.09813} {arXiv:1803.09813 [quant-ph]} \BibitemShut
  {NoStop}%
\bibitem [{\citenamefont {Arute}\ \emph {et~al.}(2019)\citenamefont {Arute} \emph {et~al.}}]{AruteQuantumSupremacy2019}%
  \BibitemOpen
  \bibfield  {author} {\bibinfo {author} {\bibfnamefont {F.}~\bibnamefont {Arute}} \emph {et~al.},\ }\bibfield  {title} {\bibinfo {title} {Quantum supremacy using a programmable superconducting processor},\ }\href {https://doi.org/10.1038/s41586-019-1666-5} {\bibfield  {journal} {\bibinfo  {journal} {Nature}\ }\textbf {\bibinfo {volume} {574}},\ \bibinfo {pages} {505–510} (\bibinfo {year} {2019})}\BibitemShut {NoStop}%
\bibitem [{\citenamefont {Sung}\ \emph {et~al.}(2021)\citenamefont {Sung}, \citenamefont {Ding}, \citenamefont {Braum{\"u}ller}, \citenamefont {Veps{\"a}l{\"a}inen}, \citenamefont {Kannan}, \citenamefont {Kjaergaard}, \citenamefont {Greene}, \citenamefont {Samach}, \citenamefont {McNally}, \citenamefont {Kim}, \citenamefont {Melville}, \citenamefont {Niedzielski}, \citenamefont {Schwartz}, \citenamefont {Yoder}, \citenamefont {Orlando}, \citenamefont {Gustavsson},\ and\ \citenamefont {Oliver}}]{sungRealizationHighFidelityCZ2021a}%
  \BibitemOpen
  \bibfield  {author} {\bibinfo {author} {\bibfnamefont {Y.}~\bibnamefont {Sung}}, \bibinfo {author} {\bibfnamefont {L.}~\bibnamefont {Ding}}, \bibinfo {author} {\bibfnamefont {J.}~\bibnamefont {Braum{\"u}ller}}, \bibinfo {author} {\bibfnamefont {A.}~\bibnamefont {Veps{\"a}l{\"a}inen}}, \bibinfo {author} {\bibfnamefont {B.}~\bibnamefont {Kannan}}, \bibinfo {author} {\bibfnamefont {M.}~\bibnamefont {Kjaergaard}}, \bibinfo {author} {\bibfnamefont {A.}~\bibnamefont {Greene}}, \bibinfo {author} {\bibfnamefont {G.~O.}\ \bibnamefont {Samach}}, \bibinfo {author} {\bibfnamefont {C.}~\bibnamefont {McNally}}, \bibinfo {author} {\bibfnamefont {D.}~\bibnamefont {Kim}}, \bibinfo {author} {\bibfnamefont {A.}~\bibnamefont {Melville}}, \bibinfo {author} {\bibfnamefont {B.~M.}\ \bibnamefont {Niedzielski}}, \bibinfo {author} {\bibfnamefont {M.~E.}\ \bibnamefont {Schwartz}}, \bibinfo {author} {\bibfnamefont {J.~L.}\ \bibnamefont {Yoder}}, \bibinfo {author} {\bibfnamefont {T.~P.}\ \bibnamefont {Orlando}}, \bibinfo {author}
  {\bibfnamefont {S.}~\bibnamefont {Gustavsson}},\ and\ \bibinfo {author} {\bibfnamefont {W.~D.}\ \bibnamefont {Oliver}},\ }\bibfield  {title} {\bibinfo {title} {Realization of {{High-Fidelity CZ}} and {{Z Z}} -{{Free iSWAP Gates}} with a {{Tunable Coupler}}},\ }\href {https://doi.org/10.1103/PhysRevX.11.021058} {\bibfield  {journal} {\bibinfo  {journal} {Phys. Rev. X}\ }\textbf {\bibinfo {volume} {11}},\ \bibinfo {pages} {021058} (\bibinfo {year} {2021})}\BibitemShut {NoStop}%
\bibitem [{\citenamefont {Jin}\ \emph {et~al.}(2025)\citenamefont {Jin}, \citenamefont {Parrott}, \citenamefont {Cicak}, \citenamefont {Kotler}, \citenamefont {Lecocq}, \citenamefont {Teufel}, \citenamefont {Aumentado}, \citenamefont {Kapit},\ and\ \citenamefont {Simmonds}}]{jinSuperconductingArchitectureDemonstrating2025}%
  \BibitemOpen
  \bibfield  {author} {\bibinfo {author} {\bibfnamefont {X.}~\bibnamefont {Jin}}, \bibinfo {author} {\bibfnamefont {Z.}~\bibnamefont {Parrott}}, \bibinfo {author} {\bibfnamefont {K.}~\bibnamefont {Cicak}}, \bibinfo {author} {\bibfnamefont {S.}~\bibnamefont {Kotler}}, \bibinfo {author} {\bibfnamefont {F.}~\bibnamefont {Lecocq}}, \bibinfo {author} {\bibfnamefont {J.}~\bibnamefont {Teufel}}, \bibinfo {author} {\bibfnamefont {J.}~\bibnamefont {Aumentado}}, \bibinfo {author} {\bibfnamefont {E.}~\bibnamefont {Kapit}},\ and\ \bibinfo {author} {\bibfnamefont {R.}~\bibnamefont {Simmonds}},\ }\bibfield  {title} {\bibinfo {title} {Superconducting architecture demonstrating fast, tunable high-fidelity {{{\emph{CZ}}}} gates with parametric control of {{{\emph{ZZ}}}} coupling},\ }\href {https://doi.org/10.1103/kmls-lgp5} {\bibfield  {journal} {\bibinfo  {journal} {Phys. Rev. Applied}\ }\textbf {\bibinfo {volume} {24}},\ \bibinfo {pages} {064026} (\bibinfo {year} {2025})}\BibitemShut {NoStop}%
\bibitem [{\citenamefont {Mooij}\ \emph {et~al.}(1999)\citenamefont {Mooij}, \citenamefont {Orlando}, \citenamefont {Levitov}, \citenamefont {Tian}, \citenamefont {{van der Wal}},\ and\ \citenamefont {Lloyd}}]{mooijJosephsonPersistentCurrentQubit1999a}%
  \BibitemOpen
  \bibfield  {author} {\bibinfo {author} {\bibfnamefont {J.~E.}\ \bibnamefont {Mooij}}, \bibinfo {author} {\bibfnamefont {T.~P.}\ \bibnamefont {Orlando}}, \bibinfo {author} {\bibfnamefont {L.}~\bibnamefont {Levitov}}, \bibinfo {author} {\bibfnamefont {L.}~\bibnamefont {Tian}}, \bibinfo {author} {\bibfnamefont {C.~H.}\ \bibnamefont {{van der Wal}}},\ and\ \bibinfo {author} {\bibfnamefont {S.}~\bibnamefont {Lloyd}},\ }\bibfield  {title} {\bibinfo {title} {Josephson {{Persistent-Current Qubit}}},\ }\href {https://doi.org/10.1126/science.285.5430.1036} {\bibfield  {journal} {\bibinfo  {journal} {Science}\ }\textbf {\bibinfo {volume} {285}},\ \bibinfo {pages} {1036} (\bibinfo {year} {1999})}\BibitemShut {NoStop}%
\bibitem [{\citenamefont {Orlando}\ \emph {et~al.}(1999)\citenamefont {Orlando}, \citenamefont {Mooij}, \citenamefont {Tian}, \citenamefont {{van der Wal}}, \citenamefont {Levitov}, \citenamefont {Lloyd},\ and\ \citenamefont {Mazo}}]{orlandoSuperconductingPersistentcurrentQubit1999a}%
  \BibitemOpen
  \bibfield  {author} {\bibinfo {author} {\bibfnamefont {T.~P.}\ \bibnamefont {Orlando}}, \bibinfo {author} {\bibfnamefont {J.~E.}\ \bibnamefont {Mooij}}, \bibinfo {author} {\bibfnamefont {L.}~\bibnamefont {Tian}}, \bibinfo {author} {\bibfnamefont {C.~H.}\ \bibnamefont {{van der Wal}}}, \bibinfo {author} {\bibfnamefont {L.~S.}\ \bibnamefont {Levitov}}, \bibinfo {author} {\bibfnamefont {S.}~\bibnamefont {Lloyd}},\ and\ \bibinfo {author} {\bibfnamefont {J.~J.}\ \bibnamefont {Mazo}},\ }\bibfield  {title} {\bibinfo {title} {Superconducting persistent-current qubit},\ }\href {https://doi.org/10.1103/PhysRevB.60.15398} {\bibfield  {journal} {\bibinfo  {journal} {Phys. Rev. B}\ }\textbf {\bibinfo {volume} {60}},\ \bibinfo {pages} {15398} (\bibinfo {year} {1999})}\BibitemShut {NoStop}%
\bibitem [{\citenamefont {Orgiazzi}\ \emph {et~al.}(2016)\citenamefont {Orgiazzi}, \citenamefont {Deng}, \citenamefont {Layden}, \citenamefont {Marchildon}, \citenamefont {Kitapli}, \citenamefont {Shen}, \citenamefont {Bal}, \citenamefont {Ong},\ and\ \citenamefont {Lupascu}}]{orgiazziFluxQubitsPlanar2016}%
  \BibitemOpen
  \bibfield  {author} {\bibinfo {author} {\bibfnamefont {J.-L.}\ \bibnamefont {Orgiazzi}}, \bibinfo {author} {\bibfnamefont {C.}~\bibnamefont {Deng}}, \bibinfo {author} {\bibfnamefont {D.}~\bibnamefont {Layden}}, \bibinfo {author} {\bibfnamefont {R.}~\bibnamefont {Marchildon}}, \bibinfo {author} {\bibfnamefont {F.}~\bibnamefont {Kitapli}}, \bibinfo {author} {\bibfnamefont {F.}~\bibnamefont {Shen}}, \bibinfo {author} {\bibfnamefont {M.}~\bibnamefont {Bal}}, \bibinfo {author} {\bibfnamefont {F.~R.}\ \bibnamefont {Ong}},\ and\ \bibinfo {author} {\bibfnamefont {A.}~\bibnamefont {Lupascu}},\ }\bibfield  {title} {\bibinfo {title} {Flux qubits in a planar circuit quantum electrodynamics architecture: {{Quantum}} control and decoherence},\ }\href {https://doi.org/10.1103/PhysRevB.93.104518} {\bibfield  {journal} {\bibinfo  {journal} {Phys. Rev. B}\ }\textbf {\bibinfo {volume} {93}},\ \bibinfo {pages} {104518} (\bibinfo {year} {2016})}\BibitemShut {NoStop}%
\bibitem [{\citenamefont {Paauw}\ \emph {et~al.}(2009)\citenamefont {Paauw}, \citenamefont {Fedorov}, \citenamefont {Harmans},\ and\ \citenamefont {Mooij}}]{paauwTuningGapSuperconducting2009a}%
  \BibitemOpen
  \bibfield  {author} {\bibinfo {author} {\bibfnamefont {F.~G.}\ \bibnamefont {Paauw}}, \bibinfo {author} {\bibfnamefont {A.}~\bibnamefont {Fedorov}}, \bibinfo {author} {\bibfnamefont {C.~J. P.~M.}\ \bibnamefont {Harmans}},\ and\ \bibinfo {author} {\bibfnamefont {J.~E.}\ \bibnamefont {Mooij}},\ }\bibfield  {title} {\bibinfo {title} {Tuning the {{Gap}} of a {{Superconducting Flux Qubit}}},\ }\href {https://doi.org/10.1103/PhysRevLett.102.090501} {\bibfield  {journal} {\bibinfo  {journal} {Phys. Rev. Lett.}\ }\textbf {\bibinfo {volume} {102}},\ \bibinfo {pages} {090501} (\bibinfo {year} {2009})}\BibitemShut {NoStop}%
\bibitem [{\citenamefont {Schwarz}\ \emph {et~al.}(2013)\citenamefont {Schwarz}, \citenamefont {Goetz}, \citenamefont {Jiang}, \citenamefont {Niemczyk}, \citenamefont {Deppe}, \citenamefont {Marx},\ and\ \citenamefont {Gross}}]{schwarzGradiometricFluxQubits2013}%
  \BibitemOpen
  \bibfield  {author} {\bibinfo {author} {\bibfnamefont {M.~J.}\ \bibnamefont {Schwarz}}, \bibinfo {author} {\bibfnamefont {J.}~\bibnamefont {Goetz}}, \bibinfo {author} {\bibfnamefont {Z.}~\bibnamefont {Jiang}}, \bibinfo {author} {\bibfnamefont {T.}~\bibnamefont {Niemczyk}}, \bibinfo {author} {\bibfnamefont {F.}~\bibnamefont {Deppe}}, \bibinfo {author} {\bibfnamefont {A.}~\bibnamefont {Marx}},\ and\ \bibinfo {author} {\bibfnamefont {R.}~\bibnamefont {Gross}},\ }\bibfield  {title} {\bibinfo {title} {Gradiometric flux qubits with a tunable gap},\ }\href {https://doi.org/10.1088/1367-2630/15/4/045001} {\bibfield  {journal} {\bibinfo  {journal} {New J. Phys.}\ }\textbf {\bibinfo {volume} {15}},\ \bibinfo {pages} {045001} (\bibinfo {year} {2013})}\BibitemShut {NoStop}%
\bibitem [{\citenamefont {Bylander}\ \emph {et~al.}(2011)\citenamefont {Bylander}, \citenamefont {Gustavsson}, \citenamefont {Yan}, \citenamefont {Yoshihara}, \citenamefont {Harrabi}, \citenamefont {Fitch}, \citenamefont {Cory}, \citenamefont {Nakamura}, \citenamefont {Tsai},\ and\ \citenamefont {Oliver}}]{bylanderNoiseSpectroscopyDynamical2011}%
  \BibitemOpen
  \bibfield  {author} {\bibinfo {author} {\bibfnamefont {J.}~\bibnamefont {Bylander}}, \bibinfo {author} {\bibfnamefont {S.}~\bibnamefont {Gustavsson}}, \bibinfo {author} {\bibfnamefont {F.}~\bibnamefont {Yan}}, \bibinfo {author} {\bibfnamefont {F.}~\bibnamefont {Yoshihara}}, \bibinfo {author} {\bibfnamefont {K.}~\bibnamefont {Harrabi}}, \bibinfo {author} {\bibfnamefont {G.}~\bibnamefont {Fitch}}, \bibinfo {author} {\bibfnamefont {D.~G.}\ \bibnamefont {Cory}}, \bibinfo {author} {\bibfnamefont {Y.}~\bibnamefont {Nakamura}}, \bibinfo {author} {\bibfnamefont {J.-S.}\ \bibnamefont {Tsai}},\ and\ \bibinfo {author} {\bibfnamefont {W.~D.}\ \bibnamefont {Oliver}},\ }\bibfield  {title} {\bibinfo {title} {Noise spectroscopy through dynamical decoupling with a superconducting flux qubit},\ }\href {https://doi.org/10.1038/nphys1994} {\bibfield  {journal} {\bibinfo  {journal} {Nature Phys}\ }\textbf {\bibinfo {volume} {7}},\ \bibinfo {pages} {565} (\bibinfo {year} {2011})}\BibitemShut {NoStop}%
\bibitem [{\citenamefont {Stern}\ \emph {et~al.}(2014)\citenamefont {Stern}, \citenamefont {Catelani}, \citenamefont {Kubo}, \citenamefont {Grezes}, \citenamefont {Bienfait}, \citenamefont {Vion}, \citenamefont {Esteve},\ and\ \citenamefont {Bertet}}]{sternFluxQubitsLong2014a}%
  \BibitemOpen
  \bibfield  {author} {\bibinfo {author} {\bibfnamefont {M.}~\bibnamefont {Stern}}, \bibinfo {author} {\bibfnamefont {G.}~\bibnamefont {Catelani}}, \bibinfo {author} {\bibfnamefont {Y.}~\bibnamefont {Kubo}}, \bibinfo {author} {\bibfnamefont {C.}~\bibnamefont {Grezes}}, \bibinfo {author} {\bibfnamefont {A.}~\bibnamefont {Bienfait}}, \bibinfo {author} {\bibfnamefont {D.}~\bibnamefont {Vion}}, \bibinfo {author} {\bibfnamefont {D.}~\bibnamefont {Esteve}},\ and\ \bibinfo {author} {\bibfnamefont {P.}~\bibnamefont {Bertet}},\ }\bibfield  {title} {\bibinfo {title} {Flux {{Qubits}} with {{Long Coherence Times}} for {{Hybrid Quantum Circuits}}},\ }\href {https://doi.org/10.1103/PhysRevLett.113.123601} {\bibfield  {journal} {\bibinfo  {journal} {Phys. Rev. Lett.}\ }\textbf {\bibinfo {volume} {113}},\ \bibinfo {pages} {123601} (\bibinfo {year} {2014})}\BibitemShut {NoStop}%
\bibitem [{\citenamefont {Chang}\ \emph {et~al.}(2022)\citenamefont {Chang}, \citenamefont {Holzman}, \citenamefont {Cohen}, \citenamefont {Johnson}, \citenamefont {Jamieson},\ and\ \citenamefont {Stern}}]{changReproducibilityGapControl2022}%
  \BibitemOpen
  \bibfield  {author} {\bibinfo {author} {\bibfnamefont {T.}~\bibnamefont {Chang}}, \bibinfo {author} {\bibfnamefont {I.}~\bibnamefont {Holzman}}, \bibinfo {author} {\bibfnamefont {T.}~\bibnamefont {Cohen}}, \bibinfo {author} {\bibfnamefont {B.~C.}\ \bibnamefont {Johnson}}, \bibinfo {author} {\bibfnamefont {D.~N.}\ \bibnamefont {Jamieson}},\ and\ \bibinfo {author} {\bibfnamefont {M.}~\bibnamefont {Stern}},\ }\bibfield  {title} {\bibinfo {title} {Reproducibility and {{Gap Control}} of {{Superconducting Flux Qubits}}},\ }\href {https://doi.org/10.1103/PhysRevApplied.18.064062} {\bibfield  {journal} {\bibinfo  {journal} {Phys. Rev. Appl.}\ }\textbf {\bibinfo {volume} {18}},\ \bibinfo {pages} {064062} (\bibinfo {year} {2022})}\BibitemShut {NoStop}%
\bibitem [{\citenamefont {Rigetti}\ \emph {et~al.}(2005)\citenamefont {Rigetti}, \citenamefont {Blais},\ and\ \citenamefont {Devoret}}]{rigettiProtocolForUniverslGates2005}%
  \BibitemOpen
  \bibfield  {author} {\bibinfo {author} {\bibfnamefont {C.}~\bibnamefont {Rigetti}}, \bibinfo {author} {\bibfnamefont {A.}~\bibnamefont {Blais}},\ and\ \bibinfo {author} {\bibfnamefont {M.}~\bibnamefont {Devoret}},\ }\bibfield  {title} {\bibinfo {title} {Protocol for universal gates in optimally biased superconducting qubits},\ }\href {https://doi.org/10.1103/PhysRevLett.94.240502} {\bibfield  {journal} {\bibinfo  {journal} {Phys. Rev. Lett.}\ }\textbf {\bibinfo {volume} {94}},\ \bibinfo {pages} {240502} (\bibinfo {year} {2005})}\BibitemShut {NoStop}%
\bibitem [{\citenamefont {Rigetti}\ and\ \citenamefont {Devoret}(2010)}]{rigettiFullyMicrowaveTunable2010}%
  \BibitemOpen
  \bibfield  {author} {\bibinfo {author} {\bibfnamefont {C.}~\bibnamefont {Rigetti}}\ and\ \bibinfo {author} {\bibfnamefont {M.}~\bibnamefont {Devoret}},\ }\bibfield  {title} {\bibinfo {title} {Fully microwave-tunable universal gates in superconducting qubits with linear couplings and fixed transition frequencies},\ }\href {https://doi.org/10.1103/PhysRevB.81.134507} {\bibfield  {journal} {\bibinfo  {journal} {Phys. Rev. B}\ }\textbf {\bibinfo {volume} {81}},\ \bibinfo {pages} {134507} (\bibinfo {year} {2010})}\BibitemShut {NoStop}%
\bibitem [{\citenamefont {Kirchhoff}\ \emph {et~al.}(2018)\citenamefont {Kirchhoff}, \citenamefont {Ke{\ss}ler}, \citenamefont {Liebermann}, \citenamefont {Ass{\'e}mat}, \citenamefont {Machnes}, \citenamefont {Motzoi},\ and\ \citenamefont {Wilhelm}}]{kirchhoffOptimizedCrossresonanceGate2018}%
  \BibitemOpen
  \bibfield  {author} {\bibinfo {author} {\bibfnamefont {S.}~\bibnamefont {Kirchhoff}}, \bibinfo {author} {\bibfnamefont {T.}~\bibnamefont {Ke{\ss}ler}}, \bibinfo {author} {\bibfnamefont {P.~J.}\ \bibnamefont {Liebermann}}, \bibinfo {author} {\bibfnamefont {E.}~\bibnamefont {Ass{\'e}mat}}, \bibinfo {author} {\bibfnamefont {S.}~\bibnamefont {Machnes}}, \bibinfo {author} {\bibfnamefont {F.}~\bibnamefont {Motzoi}},\ and\ \bibinfo {author} {\bibfnamefont {F.~K.}\ \bibnamefont {Wilhelm}},\ }\bibfield  {title} {\bibinfo {title} {Optimized cross-resonance gate for coupled transmon systems},\ }\href {https://doi.org/10.1103/PhysRevA.97.042348} {\bibfield  {journal} {\bibinfo  {journal} {Phys. Rev. A}\ }\textbf {\bibinfo {volume} {97}},\ \bibinfo {pages} {042348} (\bibinfo {year} {2018})}\BibitemShut {NoStop}%
\bibitem [{\citenamefont {Ashhab}\ \emph {et~al.}(2006)\citenamefont {Ashhab}, \citenamefont {Matsuo}, \citenamefont {Hatakenaka},\ and\ \citenamefont {Nori}}]{ashhabGeneralizedSwitchableCoupling2006}%
  \BibitemOpen
  \bibfield  {author} {\bibinfo {author} {\bibfnamefont {S.}~\bibnamefont {Ashhab}}, \bibinfo {author} {\bibfnamefont {S.}~\bibnamefont {Matsuo}}, \bibinfo {author} {\bibfnamefont {N.}~\bibnamefont {Hatakenaka}},\ and\ \bibinfo {author} {\bibfnamefont {F.}~\bibnamefont {Nori}},\ }\bibfield  {title} {\bibinfo {title} {Generalized switchable coupling for superconducting qubits using double resonance},\ }\href {https://doi.org/10.1103/PhysRevB.74.184504} {\bibfield  {journal} {\bibinfo  {journal} {Phys. Rev. B}\ }\textbf {\bibinfo {volume} {74}},\ \bibinfo {pages} {184504} (\bibinfo {year} {2006})}\BibitemShut {NoStop}%
\bibitem [{\citenamefont {Douce}\ \emph {et~al.}(2015)\citenamefont {Douce}, \citenamefont {Stern}, \citenamefont {Zagury}, \citenamefont {Bertet},\ and\ \citenamefont {Milman}}]{douceCouplingSingleNitrogenvacancy2015}%
  \BibitemOpen
  \bibfield  {author} {\bibinfo {author} {\bibfnamefont {T.}~\bibnamefont {Douce}}, \bibinfo {author} {\bibfnamefont {M.}~\bibnamefont {Stern}}, \bibinfo {author} {\bibfnamefont {N.}~\bibnamefont {Zagury}}, \bibinfo {author} {\bibfnamefont {P.}~\bibnamefont {Bertet}},\ and\ \bibinfo {author} {\bibfnamefont {P.}~\bibnamefont {Milman}},\ }\bibfield  {title} {\bibinfo {title} {Coupling a single nitrogen-vacancy center to a superconducting flux qubit in the far-off-resonance regime},\ }\href {https://doi.org/10.1103/PhysRevA.92.052335} {\bibfield  {journal} {\bibinfo  {journal} {Phys. Rev. A}\ }\textbf {\bibinfo {volume} {92}},\ \bibinfo {pages} {052335} (\bibinfo {year} {2015})}\BibitemShut {NoStop}%
\bibitem [{\citenamefont {Xue}\ \emph {et~al.}(2016)\citenamefont {Xue}, \citenamefont {Li}, \citenamefont {Zhou}, \citenamefont {Gao},\ and\ \citenamefont {Zhang}}]{xueTunableInteractionSuperconducting2016}%
  \BibitemOpen
  \bibfield  {author} {\bibinfo {author} {\bibfnamefont {Z.-Y.}\ \bibnamefont {Xue}}, \bibinfo {author} {\bibfnamefont {Y.-F.}\ \bibnamefont {Li}}, \bibinfo {author} {\bibfnamefont {J.}~\bibnamefont {Zhou}}, \bibinfo {author} {\bibfnamefont {Y.-M.}\ \bibnamefont {Gao}},\ and\ \bibinfo {author} {\bibfnamefont {G.}~\bibnamefont {Zhang}},\ }\bibfield  {title} {\bibinfo {title} {Tunable interaction of superconducting flux qubits in circuit {{QED}}},\ }\href {https://doi.org/10.1007/s11128-015-1185-8} {\bibfield  {journal} {\bibinfo  {journal} {Quantum Inf Process}\ }\textbf {\bibinfo {volume} {15}},\ \bibinfo {pages} {721} (\bibinfo {year} {2016})}\BibitemShut {NoStop}%
\bibitem [{\citenamefont {Mitchell}\ \emph {et~al.}(2021)\citenamefont {Mitchell}, \citenamefont {Naik}, \citenamefont {Morvan}, \citenamefont {Hashim}, \citenamefont {Kreikebaum}, \citenamefont {Marinelli}, \citenamefont {Lavrijsen}, \citenamefont {Nowrouzi}, \citenamefont {Santiago},\ and\ \citenamefont {Siddiqi}}]{mitchellHardwareEfficientMicrowaveActivatedTunable2021b}%
  \BibitemOpen
  \bibfield  {author} {\bibinfo {author} {\bibfnamefont {B.~K.}\ \bibnamefont {Mitchell}}, \bibinfo {author} {\bibfnamefont {R.~K.}\ \bibnamefont {Naik}}, \bibinfo {author} {\bibfnamefont {A.}~\bibnamefont {Morvan}}, \bibinfo {author} {\bibfnamefont {A.}~\bibnamefont {Hashim}}, \bibinfo {author} {\bibfnamefont {J.~M.}\ \bibnamefont {Kreikebaum}}, \bibinfo {author} {\bibfnamefont {B.}~\bibnamefont {Marinelli}}, \bibinfo {author} {\bibfnamefont {W.}~\bibnamefont {Lavrijsen}}, \bibinfo {author} {\bibfnamefont {K.}~\bibnamefont {Nowrouzi}}, \bibinfo {author} {\bibfnamefont {D.~I.}\ \bibnamefont {Santiago}},\ and\ \bibinfo {author} {\bibfnamefont {I.}~\bibnamefont {Siddiqi}},\ }\bibfield  {title} {\bibinfo {title} {Hardware-{{Efficient Microwave-Activated Tunable Coupling}} between {{Superconducting Qubits}}},\ }\href {https://doi.org/10.1103/PhysRevLett.127.200502} {\bibfield  {journal} {\bibinfo  {journal} {Phys. Rev. Lett.}\ }\textbf {\bibinfo {volume} {127}},\ \bibinfo {pages} {200502} (\bibinfo {year}
  {2021})}\BibitemShut {NoStop}%
\bibitem [{\citenamefont {Wei}\ \emph {et~al.}(2022)\citenamefont {Wei}, \citenamefont {Magesan}, \citenamefont {Lauer}, \citenamefont {Srinivasan}, \citenamefont {Bogorin}, \citenamefont {Carnevale}, \citenamefont {Keefe}, \citenamefont {Kim}, \citenamefont {Klaus}, \citenamefont {Landers}, \citenamefont {Sundaresan}, \citenamefont {Wang}, \citenamefont {Zhang}, \citenamefont {Steffen}, \citenamefont {Dial}, \citenamefont {McKay},\ and\ \citenamefont {Kandala}}]{weiHamiltonianEngineeringMulticolor2022a}%
  \BibitemOpen
  \bibfield  {author} {\bibinfo {author} {\bibfnamefont {K.~X.}\ \bibnamefont {Wei}}, \bibinfo {author} {\bibfnamefont {E.}~\bibnamefont {Magesan}}, \bibinfo {author} {\bibfnamefont {I.}~\bibnamefont {Lauer}}, \bibinfo {author} {\bibfnamefont {S.}~\bibnamefont {Srinivasan}}, \bibinfo {author} {\bibfnamefont {D.~F.}\ \bibnamefont {Bogorin}}, \bibinfo {author} {\bibfnamefont {S.}~\bibnamefont {Carnevale}}, \bibinfo {author} {\bibfnamefont {G.~A.}\ \bibnamefont {Keefe}}, \bibinfo {author} {\bibfnamefont {Y.}~\bibnamefont {Kim}}, \bibinfo {author} {\bibfnamefont {D.}~\bibnamefont {Klaus}}, \bibinfo {author} {\bibfnamefont {W.}~\bibnamefont {Landers}}, \bibinfo {author} {\bibfnamefont {N.}~\bibnamefont {Sundaresan}}, \bibinfo {author} {\bibfnamefont {C.}~\bibnamefont {Wang}}, \bibinfo {author} {\bibfnamefont {E.~J.}\ \bibnamefont {Zhang}}, \bibinfo {author} {\bibfnamefont {M.}~\bibnamefont {Steffen}}, \bibinfo {author} {\bibfnamefont {O.~E.}\ \bibnamefont {Dial}}, \bibinfo {author} {\bibfnamefont {D.~C.}\
  \bibnamefont {McKay}},\ and\ \bibinfo {author} {\bibfnamefont {A.}~\bibnamefont {Kandala}},\ }\bibfield  {title} {\bibinfo {title} {Hamiltonian {{Engineering}} with {{Multicolor Drives}} for {{Fast Entangling Gates}} and {{Quantum Crosstalk Cancellation}}},\ }\href {https://doi.org/10.1103/PhysRevLett.129.060501} {\bibfield  {journal} {\bibinfo  {journal} {Phys. Rev. Lett.}\ }\textbf {\bibinfo {volume} {129}},\ \bibinfo {pages} {060501} (\bibinfo {year} {2022})}\BibitemShut {NoStop}%
\bibitem [{\citenamefont {Song}\ \emph {et~al.}(2020)\citenamefont {Song}, \citenamefont {You}, \citenamefont {Xu}, \citenamefont {Yang},\ and\ \citenamefont {An}}]{songFloquetEngineeringTwo2020}%
  \BibitemOpen
  \bibfield  {author} {\bibinfo {author} {\bibfnamefont {W.-L.}\ \bibnamefont {Song}}, \bibinfo {author} {\bibfnamefont {J.-B.}\ \bibnamefont {You}}, \bibinfo {author} {\bibfnamefont {J.}~\bibnamefont {Xu}}, \bibinfo {author} {\bibfnamefont {W.}~\bibnamefont {Yang}},\ and\ \bibinfo {author} {\bibfnamefont {J.-H.}\ \bibnamefont {An}},\ }\bibfield  {title} {\bibinfo {title} {Floquet {{Engineering}} of {{Two Weakly Coupled Superconducting Flux Qubits}}},\ }\href {https://doi.org/10.1103/PhysRevApplied.14.054049} {\bibfield  {journal} {\bibinfo  {journal} {Phys. Rev. Applied}\ }\textbf {\bibinfo {volume} {14}},\ \bibinfo {pages} {054049} (\bibinfo {year} {2020})}\BibitemShut {NoStop}%
\bibitem [{\citenamefont {Nguyen}\ \emph {et~al.}(2024)\citenamefont {Nguyen}, \citenamefont {Kim}, \citenamefont {Hashim}, \citenamefont {Goss}, \citenamefont {Marinelli}, \citenamefont {Bhandari}, \citenamefont {Das}, \citenamefont {Naik}, \citenamefont {Kreikebaum}, \citenamefont {Jordan}, \citenamefont {Santiago},\ and\ \citenamefont {Siddiqi}}]{nguyenProgrammableHeisenbergInteractions2024}%
  \BibitemOpen
  \bibfield  {author} {\bibinfo {author} {\bibfnamefont {L.~B.}\ \bibnamefont {Nguyen}}, \bibinfo {author} {\bibfnamefont {Y.}~\bibnamefont {Kim}}, \bibinfo {author} {\bibfnamefont {A.}~\bibnamefont {Hashim}}, \bibinfo {author} {\bibfnamefont {N.}~\bibnamefont {Goss}}, \bibinfo {author} {\bibfnamefont {B.}~\bibnamefont {Marinelli}}, \bibinfo {author} {\bibfnamefont {B.}~\bibnamefont {Bhandari}}, \bibinfo {author} {\bibfnamefont {D.}~\bibnamefont {Das}}, \bibinfo {author} {\bibfnamefont {R.~K.}\ \bibnamefont {Naik}}, \bibinfo {author} {\bibfnamefont {J.~M.}\ \bibnamefont {Kreikebaum}}, \bibinfo {author} {\bibfnamefont {A.~N.}\ \bibnamefont {Jordan}}, \bibinfo {author} {\bibfnamefont {D.~I.}\ \bibnamefont {Santiago}},\ and\ \bibinfo {author} {\bibfnamefont {I.}~\bibnamefont {Siddiqi}},\ }\bibfield  {title} {\bibinfo {title} {Programmable {{Heisenberg}} interactions between {{Floquet}} qubits},\ }\href {https://doi.org/10.1038/s41567-023-02326-7} {\bibfield  {journal} {\bibinfo  {journal} {Nat. Phys.}\ }\textbf
  {\bibinfo {volume} {20}},\ \bibinfo {pages} {240} (\bibinfo {year} {2024})}\BibitemShut {NoStop}%
\bibitem [{\citenamefont {Shirley}(1965)}]{Shirley1965}%
  \BibitemOpen
  \bibfield  {author} {\bibinfo {author} {\bibfnamefont {J.~H.}\ \bibnamefont {Shirley}},\ }\bibfield  {title} {\bibinfo {title} {Solution of the schrödinger equation with a hamiltonian periodic in time},\ }\href {https://doi.org/10.1103/PhysRev.138.B979} {\bibfield  {journal} {\bibinfo  {journal} {Physical Review}\ }\textbf {\bibinfo {volume} {138}},\ \bibinfo {pages} {B979} (\bibinfo {year} {1965})}\BibitemShut {NoStop}%
\bibitem [{\citenamefont {Sambe}(1973)}]{Sambe1973}%
  \BibitemOpen
  \bibfield  {author} {\bibinfo {author} {\bibfnamefont {H.}~\bibnamefont {Sambe}},\ }\bibfield  {title} {\bibinfo {title} {Steady states and quasienergies of a quantum-mechanical system in an oscillating field},\ }\href {https://doi.org/10.1103/PhysRevA.7.2203} {\bibfield  {journal} {\bibinfo  {journal} {Physical Review A}\ }\textbf {\bibinfo {volume} {7}},\ \bibinfo {pages} {2203} (\bibinfo {year} {1973})}\BibitemShut {NoStop}%
\bibitem [{\citenamefont {Gandon}\ \emph {et~al.}(2022)\citenamefont {Gandon}, \citenamefont {Le~Calonnec}, \citenamefont {Shillito}, \citenamefont {Petrescu},\ and\ \citenamefont {Blais}}]{GandonFloquet2022}%
  \BibitemOpen
  \bibfield  {author} {\bibinfo {author} {\bibfnamefont {A.}~\bibnamefont {Gandon}}, \bibinfo {author} {\bibfnamefont {C.}~\bibnamefont {Le~Calonnec}}, \bibinfo {author} {\bibfnamefont {R.}~\bibnamefont {Shillito}}, \bibinfo {author} {\bibfnamefont {A.}~\bibnamefont {Petrescu}},\ and\ \bibinfo {author} {\bibfnamefont {A.}~\bibnamefont {Blais}},\ }\bibfield  {title} {\bibinfo {title} {Engineering, control, and longitudinal readout of floquet qubits},\ }\href {https://doi.org/10.1103/PhysRevApplied.17.064006} {\bibfield  {journal} {\bibinfo  {journal} {Phys. Rev. Appl.}\ }\textbf {\bibinfo {volume} {17}},\ \bibinfo {pages} {064006} (\bibinfo {year} {2022})}\BibitemShut {NoStop}%
\bibitem [{\citenamefont {Lang}\ and\ \citenamefont {Firsov}(1963)}]{LangFirsovKineticTheory1963}%
  \BibitemOpen
  \bibfield  {author} {\bibinfo {author} {\bibfnamefont {I.~G.}\ \bibnamefont {Lang}}\ and\ \bibinfo {author} {\bibfnamefont {Y.~A.}\ \bibnamefont {Firsov}},\ }\bibfield  {title} {\bibinfo {title} {Kinetic theory of semiconductors with low mobility},\ }\href@noop {} {\bibfield  {journal} {\bibinfo  {journal} {Sov. Phys. JETP}\ }\textbf {\bibinfo {volume} {16}},\ \bibinfo {pages} {1301} (\bibinfo {year} {1963})}\BibitemShut {NoStop}%
\bibitem [{\citenamefont {Billangeon}\ \emph {et~al.}(2015)\citenamefont {Billangeon}, \citenamefont {Tsai},\ and\ \citenamefont {Nakamura}}]{billangeon_qed_2015}%
  \BibitemOpen
  \bibfield  {author} {\bibinfo {author} {\bibfnamefont {P.-M.}\ \bibnamefont {Billangeon}}, \bibinfo {author} {\bibfnamefont {J.~S.}\ \bibnamefont {Tsai}},\ and\ \bibinfo {author} {\bibfnamefont {Y.}~\bibnamefont {Nakamura}},\ }\bibfield  {title} {\bibinfo {title} {Circuit-{QED}-based scalable architectures for quantum information processing with superconducting qubits},\ }\href {https://doi.org/10.1103/PhysRevB.91.094517} {\bibfield  {journal} {\bibinfo  {journal} {Physical Review B}\ }\textbf {\bibinfo {volume} {91}},\ \bibinfo {pages} {094517} (\bibinfo {year} {2015})}\BibitemShut {NoStop}%
\bibitem [{\citenamefont {Wang}\ \emph {et~al.}(2026)\citenamefont {Wang}, \citenamefont {Xia}, \citenamefont {Yan},\ and\ \citenamefont {Wang}}]{Wang2026CorrelatedErrors}%
  \BibitemOpen
  \bibfield  {author} {\bibinfo {author} {\bibfnamefont {S.}~\bibnamefont {Wang}}, \bibinfo {author} {\bibfnamefont {Z.}~\bibnamefont {Xia}}, \bibinfo {author} {\bibfnamefont {Y.}~\bibnamefont {Yan}},\ and\ \bibinfo {author} {\bibfnamefont {X.-B.}\ \bibnamefont {Wang}},\ }\bibfield  {title} {\bibinfo {title} {Threshold of surface code under nearest-neighbor correlated errors via an exact statistical mechanical mapping},\ }\href {https://doi.org/10.1038/s41534-026-01276-1} {\bibfield  {journal} {\bibinfo  {journal} {npj Quantum Information}\ }\textbf {\bibinfo {volume} {12}},\ \bibinfo {pages} {134} (\bibinfo {year} {2026})}\BibitemShut {NoStop}%
\bibitem [{\citenamefont {Nickerson}\ and\ \citenamefont {Brown}(2019)}]{Nickerson2019CorrelatedNoise}%
  \BibitemOpen
  \bibfield  {author} {\bibinfo {author} {\bibfnamefont {N.~H.}\ \bibnamefont {Nickerson}}\ and\ \bibinfo {author} {\bibfnamefont {B.~J.}\ \bibnamefont {Brown}},\ }\bibfield  {title} {\bibinfo {title} {Analysing correlated noise on the surface code using adaptive decoding algorithms},\ }\href {https://doi.org/10.22331/q-2019-04-08-131} {\bibfield  {journal} {\bibinfo  {journal} {Quantum}\ }\textbf {\bibinfo {volume} {3}},\ \bibinfo {pages} {131} (\bibinfo {year} {2019})}\BibitemShut {NoStop}%
\bibitem [{\citenamefont {George}\ \emph {et~al.}(2025)\citenamefont {George}, \citenamefont {Allen}, \citenamefont {Kosut},\ and\ \citenamefont {Ginossar}}]{georgeMinimalTimeRobust2025}%
  \BibitemOpen
  \bibfield  {author} {\bibinfo {author} {\bibfnamefont {N.}~\bibnamefont {George}}, \bibinfo {author} {\bibfnamefont {J.~L.}\ \bibnamefont {Allen}}, \bibinfo {author} {\bibfnamefont {R.}~\bibnamefont {Kosut}},\ and\ \bibinfo {author} {\bibfnamefont {E.}~\bibnamefont {Ginossar}},\ }\bibfield  {title} {\bibinfo {title} {Minimal time robust control for two superconducting qubits},\ }\href {https://doi.org/10.1038/s41598-025-32747-8} {\bibfield  {journal} {\bibinfo  {journal} {Scientific Reports}\ }\textbf {\bibinfo {volume} {16}},\ \bibinfo {pages} {2773} (\bibinfo {year} {2025})}\BibitemShut {NoStop}%
\bibitem [{\citenamefont {Chang}\ \emph {et~al.}(2023)\citenamefont {Chang}, \citenamefont {Cohen}, \citenamefont {Holzman}, \citenamefont {Catelani},\ and\ \citenamefont {Stern}}]{changTunableSuperconductingFlux2023}%
  \BibitemOpen
  \bibfield  {author} {\bibinfo {author} {\bibfnamefont {T.}~\bibnamefont {Chang}}, \bibinfo {author} {\bibfnamefont {T.}~\bibnamefont {Cohen}}, \bibinfo {author} {\bibfnamefont {I.}~\bibnamefont {Holzman}}, \bibinfo {author} {\bibfnamefont {G.}~\bibnamefont {Catelani}},\ and\ \bibinfo {author} {\bibfnamefont {M.}~\bibnamefont {Stern}},\ }\bibfield  {title} {\bibinfo {title} {Tunable {{Superconducting Flux Qubits}} with {{Long Coherence Times}}},\ }\href {https://doi.org/10.1103/PhysRevApplied.19.024066} {\bibfield  {journal} {\bibinfo  {journal} {Phys. Rev. Appl.}\ }\textbf {\bibinfo {volume} {19}},\ \bibinfo {pages} {024066} (\bibinfo {year} {2023})}\BibitemShut {NoStop}%
\bibitem [{\citenamefont {Yan}\ \emph {et~al.}(2013)\citenamefont {Yan}, \citenamefont {Gustavsson}, \citenamefont {Bylander}, \citenamefont {Jin}, \citenamefont {Yoshihara}, \citenamefont {Cory}, \citenamefont {Nakamura}, \citenamefont {Orlando},\ and\ \citenamefont {Oliver}}]{Yan2013RotatingFrame}%
  \BibitemOpen
  \bibfield  {author} {\bibinfo {author} {\bibfnamefont {F.}~\bibnamefont {Yan}}, \bibinfo {author} {\bibfnamefont {S.}~\bibnamefont {Gustavsson}}, \bibinfo {author} {\bibfnamefont {J.}~\bibnamefont {Bylander}}, \bibinfo {author} {\bibfnamefont {X.}~\bibnamefont {Jin}}, \bibinfo {author} {\bibfnamefont {F.}~\bibnamefont {Yoshihara}}, \bibinfo {author} {\bibfnamefont {D.~G.}\ \bibnamefont {Cory}}, \bibinfo {author} {\bibfnamefont {Y.}~\bibnamefont {Nakamura}}, \bibinfo {author} {\bibfnamefont {T.~P.}\ \bibnamefont {Orlando}},\ and\ \bibinfo {author} {\bibfnamefont {W.~D.}\ \bibnamefont {Oliver}},\ }\bibfield  {title} {\bibinfo {title} {Rotating-frame relaxation as a noise spectrum analyser of a superconducting qubit undergoing driven evolution},\ }\href {https://doi.org/10.1038/ncomms3337} {\bibfield  {journal} {\bibinfo  {journal} {Nature Communications}\ }\textbf {\bibinfo {volume} {4}},\ \bibinfo {pages} {2337} (\bibinfo {year} {2013})}\BibitemShut {NoStop}%
\bibitem [{\citenamefont {Wang}\ \emph {et~al.}(2011)\citenamefont {Wang}, \citenamefont {Fowler},\ and\ \citenamefont {Hollenberg}}]{Wang2011}%
  \BibitemOpen
  \bibfield  {author} {\bibinfo {author} {\bibfnamefont {D.~S.}\ \bibnamefont {Wang}}, \bibinfo {author} {\bibfnamefont {A.~G.}\ \bibnamefont {Fowler}},\ and\ \bibinfo {author} {\bibfnamefont {L.~C.~L.}\ \bibnamefont {Hollenberg}},\ }\bibfield  {title} {\bibinfo {title} {Surface code quantum computing with error rates over 1\%},\ }\href {https://doi.org/10.1103/PhysRevA.83.020302} {\bibfield  {journal} {\bibinfo  {journal} {Physical Review A}\ }\textbf {\bibinfo {volume} {83}},\ \bibinfo {pages} {020302(R)} (\bibinfo {year} {2011})}\BibitemShut {NoStop}%
\bibitem [{\citenamefont {Bravyi}\ \emph {et~al.}(2024)\citenamefont {Bravyi}, \citenamefont {Cross}, \citenamefont {Gambetta}, \citenamefont {Maslov}, \citenamefont {Rall},\ and\ \citenamefont {Yoder}}]{Bravyi2024}%
  \BibitemOpen
  \bibfield  {author} {\bibinfo {author} {\bibfnamefont {S.}~\bibnamefont {Bravyi}}, \bibinfo {author} {\bibfnamefont {A.~W.}\ \bibnamefont {Cross}}, \bibinfo {author} {\bibfnamefont {J.~M.}\ \bibnamefont {Gambetta}}, \bibinfo {author} {\bibfnamefont {D.}~\bibnamefont {Maslov}}, \bibinfo {author} {\bibfnamefont {P.}~\bibnamefont {Rall}},\ and\ \bibinfo {author} {\bibfnamefont {T.~J.}\ \bibnamefont {Yoder}},\ }\bibfield  {title} {\bibinfo {title} {High-threshold and low-overhead fault-tolerant quantum memory},\ }\href {https://doi.org/10.1038/s41586-024-07107-7} {\bibfield  {journal} {\bibinfo  {journal} {Nature}\ }\textbf {\bibinfo {volume} {627}},\ \bibinfo {pages} {778} (\bibinfo {year} {2024})},\ \Eprint {https://arxiv.org/abs/2308.07915} {arXiv:2308.07915} \BibitemShut {NoStop}%
\bibitem [{\citenamefont {Palacios-Laloy}\ \emph {et~al.}(2008)\citenamefont {Palacios-Laloy}, \citenamefont {Nguyen}, \citenamefont {Mallet}, \citenamefont {Bertet}, \citenamefont {Vion},\ and\ \citenamefont {Esteve}}]{palaciosLaloyTunableResonators2008}%
  \BibitemOpen
  \bibfield  {author} {\bibinfo {author} {\bibfnamefont {A.}~\bibnamefont {Palacios-Laloy}}, \bibinfo {author} {\bibfnamefont {F.}~\bibnamefont {Nguyen}}, \bibinfo {author} {\bibfnamefont {F.}~\bibnamefont {Mallet}}, \bibinfo {author} {\bibfnamefont {P.}~\bibnamefont {Bertet}}, \bibinfo {author} {\bibfnamefont {D.}~\bibnamefont {Vion}},\ and\ \bibinfo {author} {\bibfnamefont {D.}~\bibnamefont {Esteve}},\ }\bibfield  {title} {\bibinfo {title} {Tunable resonators for quantum circuits},\ }\href {https://doi.org/10.1007/s10909-008-9774-x} {\bibfield  {journal} {\bibinfo  {journal} {Journal of Low Temperature Physics}\ }\textbf {\bibinfo {volume} {151}},\ \bibinfo {pages} {1034} (\bibinfo {year} {2008})}\BibitemShut {NoStop}%
\bibitem [{\citenamefont {Sandberg}\ \emph {et~al.}(2008)\citenamefont {Sandberg}, \citenamefont {Wilson}, \citenamefont {Persson}, \citenamefont {Bauch}, \citenamefont {Johansson}, \citenamefont {Shumeiko}, \citenamefont {Duty},\ and\ \citenamefont {Delsing}}]{sandbergTuningField2008}%
  \BibitemOpen
  \bibfield  {author} {\bibinfo {author} {\bibfnamefont {M.}~\bibnamefont {Sandberg}}, \bibinfo {author} {\bibfnamefont {C.~M.}\ \bibnamefont {Wilson}}, \bibinfo {author} {\bibfnamefont {F.}~\bibnamefont {Persson}}, \bibinfo {author} {\bibfnamefont {T.}~\bibnamefont {Bauch}}, \bibinfo {author} {\bibfnamefont {G.}~\bibnamefont {Johansson}}, \bibinfo {author} {\bibfnamefont {V.}~\bibnamefont {Shumeiko}}, \bibinfo {author} {\bibfnamefont {T.}~\bibnamefont {Duty}},\ and\ \bibinfo {author} {\bibfnamefont {P.}~\bibnamefont {Delsing}},\ }\bibfield  {title} {\bibinfo {title} {Tuning the field in a microwave resonator faster than the photon lifetime},\ }\href {https://doi.org/10.1063/1.2929367} {\bibfield  {journal} {\bibinfo  {journal} {Applied Physics Letters}\ }\textbf {\bibinfo {volume} {92}},\ \bibinfo {pages} {203501} (\bibinfo {year} {2008})}\BibitemShut {NoStop}%
\bibitem [{\citenamefont {Wang}\ \emph {et~al.}(2013)\citenamefont {Wang}, \citenamefont {Zhong}, \citenamefont {He}, \citenamefont {Wang}, \citenamefont {Martinis}, \citenamefont {Cleland},\ and\ \citenamefont {Xie}}]{wangQuantumState2013}%
  \BibitemOpen
  \bibfield  {author} {\bibinfo {author} {\bibfnamefont {Z.~L.}\ \bibnamefont {Wang}}, \bibinfo {author} {\bibfnamefont {Y.~P.}\ \bibnamefont {Zhong}}, \bibinfo {author} {\bibfnamefont {L.~J.}\ \bibnamefont {He}}, \bibinfo {author} {\bibfnamefont {H.}~\bibnamefont {Wang}}, \bibinfo {author} {\bibfnamefont {J.~M.}\ \bibnamefont {Martinis}}, \bibinfo {author} {\bibfnamefont {A.~N.}\ \bibnamefont {Cleland}},\ and\ \bibinfo {author} {\bibfnamefont {Q.~W.}\ \bibnamefont {Xie}},\ }\bibfield  {title} {\bibinfo {title} {Quantum state characterization of a fast tunable superconducting resonator},\ }\href {https://doi.org/10.1063/1.4802893} {\bibfield  {journal} {\bibinfo  {journal} {Applied Physics Letters}\ }\textbf {\bibinfo {volume} {102}},\ \bibinfo {pages} {163503} (\bibinfo {year} {2013})}\BibitemShut {NoStop}%
\bibitem [{\citenamefont {Krishnan}\ and\ \citenamefont {Pople}(1978)}]{krishnanApproximateFourthorderPerturbation1978}%
  \BibitemOpen
  \bibfield  {author} {\bibinfo {author} {\bibfnamefont {R.}~\bibnamefont {Krishnan}}\ and\ \bibinfo {author} {\bibfnamefont {J.~A.}\ \bibnamefont {Pople}},\ }\bibfield  {title} {\bibinfo {title} {Approximate fourth-order perturbation theory of the electron correlation energy},\ }\href {https://doi.org/10.1002/qua.560140109} {\bibfield  {journal} {\bibinfo  {journal} {Int. J. Quantum Chem.}\ }\textbf {\bibinfo {volume} {14}},\ \bibinfo {pages} {91} (\bibinfo {year} {1978})}\BibitemShut {NoStop}%
\end{thebibliography}
\end{document}